\documentclass[12pt]{article}
\pdfoutput=1

\usepackage[english]{babel}
\usepackage{graphicx}
\usepackage{amsmath}
\usepackage{amssymb}
\usepackage{multirow}
\usepackage{url}
\usepackage{tikz}
\usepackage{braket}
\usetikzlibrary{decorations.markings}
\usepackage{subcaption}
\usepackage[utf8]{inputenc}

\usepackage{xargs, ifthen}
\usepackage[breakwords]{truncate}

\newcommand\tomega{\tilde{\omega}}
\usepackage{hyperref}
\usepackage{calrsfs}
\DeclareMathAlphabet{\pazocal}{OMS}{zplm}{m}{n}

\allowdisplaybreaks
\def\eq#1{(\ref{#1})}
\def\s[#1\s]{\begin{align}\begin{split}#1\end{split}\end{align}}
\def\[#1\]{\begin{align}#1\end{align}}
\begin{document}

\begin{titlepage}

\title{
\hfill\parbox{4cm}{ \normalsize YITP-18-38}\\ 
\vspace{1cm} 
Canonical tensor model through data analysis\\
-- Dimensions, topologies, and geometries --}

\author{
Taigen Kawano$^{1}$\footnote{taigen.kawano@yukawa.kyoto-u.ac.jp},
Dennis Obster$^{1,2}$\footnote{dobster@science.ru.nl}, 
Naoki Sasakura$^1$\footnote{sasakura@yukawa.kyoto-u.ac.jp}
\\
$^1${\small{\it Yukawa Institute for Theoretical Physics, Kyoto University,}}
\\ {\small{\it  Kitashirakawa, Sakyo-ku, Kyoto 606-8502, Japan}}
\\
$^2${\small{\it Institute for Mathematics, Astrophysics and Particle Physics, Radboud University,}}
\\ {\small{\it Heyendaalseweg 135, 6525 AJ Nijmegen,The Netherlands}}
}

\date{\today}

\maketitle

\begin{abstract}
The canonical tensor model (CTM) is a tensor model in Hamilton formalism and is studied as a model for gravity 
in both classical and quantum frameworks.
Its dynamical variables are a canonical conjugate pair of real symmetric three-index tensors, and 
a question in this model was how to extract spacetime pictures from the tensors. 
We give such an extraction procedure by using two techniques widely known in data analysis. 
One is the tensor-rank (or CP, etc.) decomposition, which is 
a certain generalization of the singular value decomposition of a matrix and 
decomposes a tensor into a number of vectors. By regarding the vectors as points forming a space,
topological properties can be extracted by using the other data analysis technique 
called persistent homology, and geometries by virtual diffusion processes over points. 
Thus, time evolutions of the tensors 
in the CTM can be interpreted as topological and geometric evolutions of spaces. 
We have performed some initial investigations of 
the classical equation of motion of the CTM in terms of these techniques for a homogeneous fuzzy circle
and homogeneous two- and three-dimensional fuzzy spheres as spaces, 
and have obtained agreement with the general relativistic system obtained previously 
in a formal continuum limit of the CTM. It is also demonstrated by some concrete examples that 
the procedure is general for any dimensions and topologies, showing 
the generality of the CTM.
\end{abstract}

\end{titlepage}

%\listoftodos
%\clearpage
 
\section{Introduction}
How to formulate a consistent theory for quantum gravity is one of the major problems in fundamental physics. While general relativity 
and quantum mechanics are believed to be the correct theories in their applied areas in physics, 
quantization of general relativity using standard (perturbative) quantum field theoretical 
method is hard due to non-renormalizable divergences from small 
scale quantum 
fluctuations~\cite{GOROFF1986709}.\footnote{However, see for instance \cite{Eichhorn:2017egq} 
on the recent developments in the asymptotic safety program~\cite{Reuter:2012id}.} 
One very promising direction to solve the issue is to formulate spacetime and matter fields 
in a more fundamental way than continuous spacetime and point-like objects. %D: What is meant by point-like objects? Objects are not point-like in QFT right? Maybe pointlike interaction vertices is what is meant?
In such attempts, spacetime 
is considered to be an emergent entity generated by the dynamics of more fundamental degrees of freedom.  
  
Among the various approaches to quantum gravity in line with the thoughts above, tensor models are of much interest.
They were originally proposed~\cite{Ambjorn:1990ge,Sasakura:1990fs,Godfrey:1990dt} 
as a generalization of random matrix models, which were successful for describing
two-dimensional quantum gravity, with the hope of obtaining consistent theories for quantum gravity in dimensions higher than two.   
While the original models suffer from some difficulties in computability,\footnote{For the original models there are 
no so-called $1/N$ expansions, as do exist for the matrix models. Recently, introducing a traceless condition~\cite{Klebanov:2017nlk} or a pair of symmetric tensors~\cite{Gurau:2017qya} has been proposed 
as possible resolutions.}
improved models called colored tensor models were introduced \cite{Gurau:2009tw}, that enabled various analytical computations in 
what is called $1/N$ expansions \cite{Gurau:2013cbh}.
The results seem to show that the emergent spaces in the colored tensor models are like branched 
polymers \cite{Gurau:2013cbh,Bonzom:2011zz}, 2D quantum gravity, or mixtures 
\cite{Bonzom:2015axa,Lionni:2017xvn}, 
far from macroscopic spacetimes or our actual spacetime.

On the other hand, there is a model of quantum gravity with a causal structure, 
called causal dynamical triangulation, that successfully generates macroscopic 
spacetimes \cite{Ambjorn:2004qm,Ambjorn:2012jv}. 
This is in contrast with the corresponding  
Euclidean model, called dynamical triangulation, which is not successful in this 
regard \cite{Ambjorn:2013eha, Coumbe:2014nea}.
This fact suggests the importance of causal treatment in quantum gravity, and 
one of the authors of this paper proposed a new type of tensor model, which we call
canonical tensor model (CTM) \cite{Sasakura:2011sq,Sasakura:2012fb}.
This is formulated as a first-class constraint system in Hamilton formalism with a canonical conjugate pair of 
real symmetric three-index tensors as its dynamical variables.\footnote{For a concise review of the CTM, see for instance
the review section in \cite{Obster:2017dhx}.}
Its first-class constraint algebra closely resembles that of the ADM formalism of general relativity,
and there indeed exists a formal continuum limit in which they agree~\cite{Sasakura:2015pxa}.
There are also other remarkable connections to general relativity: The $N=1$ 
CTM\footnote{$N$ denotes the dimension of the 
vector space associated to the tensor indices.
In other words, each index takes values from $\{1,2,\ldots,N\}$.} agrees with the mini-superspace treatment of 
general relativity~\cite{Sasakura:2014gia},
and the classical CTM in the formal continuum limit agrees with a general relativistic coupled system of gravity, 
a scalar field, and higher spin fields in the Hamilton-Jacobi formalism~\cite{Chen:2016ate}.  

The formal continuum limit above is obtained by a formal replacement of the 
discrete values of the indices of the tensors, $a=1,2,\ldots,N$, to a continuous one, $x\in {\mathbb R}^D$.
Therefore, this formal continuum limit is assuming a classical continuous spacetime from the beginning and does not 
tell anything about how such a space may emerge from the (quantum) dynamics of the theory.  
A clue to the last question has been obtained in our previous paper \cite{Obster:2017dhx}: 
The wave function of the quantum CTM has strong peaks at values of the tensors symmetric under Lie-groups.  
Since we know that various symmetries are associated to our spacetime, this result
is encouraging. 
Then the next question which naturally arises is how to interpret such preferred values of the tensors as spacetimes.

The first step to answer this question would be to establish the correspondence between tensors and spacetimes.
To this end, we introduce two well-known techniques in data analysis to the CTM, 
and formulate a systematic procedure to extract topological and geometric properties of spaces held by the tensors. 
The first technique is called tensor-rank decomposition (or CP, etc.)
\cite{SAPM:SAPM192761164,Carroll1970,harshman70,comon:hal-00923279}. 
This is a certain generalization of the singular value decomposition of a matrix, and 
decomposes a tensor into a number of vectors. By regarding the vectors as points and their mutual inner products 
as quantities featuring distance relations among points, one can obtain a space with
topological and geometric properties extracted from a tensor. 
Here, topological properties are extracted through the second technique from topological data analysis called 
persistent homology \cite{Carlsson09}.\footnote{Some examples of physical applications include~\cite{Cirafici:2015pky, Cirafici:2015sdg, Cole:2017kve, Spreemann:2018}.} Geometric structure is extracted by virtual diffusion processes over points which are also often used in data analysis~\cite{Nadler:2005:DMS:2976248.2976368, Coifman7426}.

After introducing some notions and ideas, we consider a homogeneous fuzzy circle and homogeneous fuzzy two- and 
three-dimensional spheres to demonstrate the method.
We study the time evolution of the tensors corresponding to these fuzzy spaces under the classical equation of 
motion of the CTM and interpret them as the evolution of spacetime by the extraction procedure mentioned above. 
We compare the results with the classical equation of motion of the general relativistic system derived in 
a formal continuum limit of the CTM in a former paper \cite{Chen:2016ate} and find good agreement.

This paper is organized as follows.  In Section~\ref{sec:cp}, we review some elementary facts about the 
tensor-rank decomposition, and interpret the vectors obtained from the decomposition as points.
In Section~\ref{sec:fuzzyspace}, we give a systematic method of constructing real symmetric three-index tensors 
of fuzzy spaces corresponding to ordinary continuous spaces with any dimensions and topologies.
In Section~\ref{sec:neighbor}, we introduce the notion of neighborhoods in terms of mutual inner products 
among vectors representing points in the sense of Section~\ref{sec:cp}.
In Section~\ref{sec:persistent}, we review persistent homology, a technique from topological data analysis, and 
demonstrate how one can apply it to the fuzzy spaces.
In Section~\ref{sec:CPcont}, we point out that the derivative expansion previously performed in the 
formal continuum limit mentioned above can be represented in the form of a continuous tensor-rank decomposition.
Here the vectors of the decomposition are expressed with the scalar and metric fields of the general relativistic
system corresponding to the CTM.
In Section~\ref{sec:diffusion}, we present a method of obtaining the values of scalar and metric fields 
by virtual diffusion processes over continuously existing points, based on the expressions obtained in Section~\ref{sec:CPcont}.
In Section~\ref{sec:distance},  
the method developed for continuous cases in the preceding sections is generalized to discrete cases, 
namely for finite $N$, and a method to characterize the local distance structures in fuzzy spaces is presented.
In Section~\ref{sec:timeevolution}, the classical equation of motion of the CTM is applied to the real symmetric 
three-index tensors describing 
fuzzy spaces, and time-evolution is roughly described as an increasing process of number of points
and mutual distances among points.
In Section~\ref{sec:relativistic}, by applying the extraction procedure mentioned above, detailed analysis
of the time-evolutions of homogeneous fuzzy $S^1,\ S^2$ and $S^3$ is performed 
and good agreement is obtained with the general relativistic system corresponding to the CTM.
In Section~\ref{sec:constP}, we explicitly construct the real symmetric three-index tensors for fuzzy spaces with
various dimensions and topologies to demonstrate the absence of limitations of the procedure, 
showing the generality of the CTM.  
The last section is devoted to a summary and future prospects. 
In Appendix~\ref{app:CP}, we show the algorithm of the C++ program we made 
and used for the tensor-rank decomposition.

\section{Tensor-rank decomposition and notion of point}
\label{sec:cp}
In this section we introduce the tensor-rank decomposition, also often called CP-decomposition
\cite{SAPM:SAPM192761164,Carroll1970,harshman70,comon:hal-00923279}, 
and use it to interpret a tensor as a collection of points which form a space. Throughout this paper, unless otherwise stated, we consider tensors which are real, symmetric, and of three-way\footnote{A tensor with three indices is often called a rank-three tensor in physics literature, but this may cause confusion since the rank of a tensor discussed in this section has 
nothing to do with the amount of indices. To avoid this confusion we call 
a tensor with three indices a three-{\it way} tensor, which is often used in computational science.}:
\[
P_{abc}=P_{\sigma_a\sigma_b\sigma_c},
\]  
where $\sigma$ denotes arbitrary permutations of $a,b,c$, and the indices run from $1$ to $N$.
This particular choice of tensors is considered because we are interested in applying methods developed here to the CTM, which has a similar setup. It is however straightforward to 
generalize the contents of this section to other types of tensors. We also assume that the underlying vector space admits an $O(N)$ symmetry, which is the natural symmetry for real inner product spaces and is the kinematical symmetry of the CTM.

One may define a point by the simplest possible tensor. 
In the present case of a real symmetric three-way tensor, the simplest possibility is given by
\[
P_{111}\neq 0, {\rm others}=0.
\]
Using the $O(N)$ symmetry in the underlying vector space, the general form for a single point is given by 
\[
P_{abc}=v_a v_b v_c,
\label{eq:rankone}
\]
where $v$ is an $N$-dimensional real vector. This implies that arbitrary single points are equivalent 
under the  $O(N)$ symmetry up to the sizes.
The tensor of the form \eq{eq:rankone} is also called a rank-one tensor.

A space may be described by a collection of such single points, leading to a tensor of the form,
\[
P_{abc}=\sum_{i=1}^R v_a^i v_b^i v_c^i.
\label{eq:CP}
\]
A tensor represented by a sum of $R$ rank-one tensors is called a rank-$R$ tensor, for the smallest possible $R$
with a given $P$. 
Representing a given tensor in such a sum has various names such as 
tensor-rank decomposition, rank-one tensor decomposition, CP-decomposition, 
etc. \cite{SAPM:SAPM192761164,Carroll1970,harshman70,comon:hal-00923279}, and essentially generalizes the single value decomposition for matrices.
The decomposition always exists with a finite $R$ for finite $N$.

An important fact about the decomposition of a tensor in our usage is that
the set of vectors in the decomposition of a tensor has sorts of uniqueness \cite{Landsberg2012,Bocci2014}, 
and therefore a space can be represented by points unambiguously\footnote{
In fact, in the case of our present applications considering homogeneous fuzzy spaces with some Lie-group symmetries,
there exist some ambiguities under the Lie-group transformations. However, these ambiguities are not relevant, because
the relevant quantities we discuss later are obtained from some inner products, which are invariant under these 
transformations.} in our actual applications, unless 
the rank is taken to be unnecessarily too large in the approximate tensor-rank decomposition \cite{doi:10.1002/cem.1236},
which appears below.
This is different from the matrix case, because the vectors in the single value decomposition of a matrix 
always have a large continuous ambiguity.
For example, the expression, $M_{ab}=\sum_{i=1}^R v_a^i v_b^i$,
can be transformed by arbitrary orthogonal transformations, $v_a^i\rightarrow L^i{}_j v_a^j$
with $L\in O(R)$, without changing $M_{ab}$. 

There are other differences and subtleties in the decomposition of a tensor in comparison with the matrix case.
A tensor may have other tensor-rank decompositions with different $R$ and $v$,
though there are some proven cases with uniqueness (or a partial one).
Here the least value of $R$ is called the rank of the tensor.
The rank of a tensor depends on the base field (namely, real or complex numbers for instance) and 
whether each rank-one tensor in the decomposition is restricted to be symmetric or not. 
Since each term in \eq{eq:CP} is a symmetric real rank-one tensor, 
$R$ should be more precisely referred to as real symmetric rank, and the decomposition \eq{eq:CP} 
as symmetric tensor-rank decomposition over the reals.
Unless otherwise stated, the tensor-rank decomposition in this paper is always assuming the form \eq{eq:CP}
with real values, and we simply ignore these specifications for brevity. 

A typical rank is defined by any rank such that the set of tensors having the rank has positive measure in the whole space of the tensors.
This means that a given tensor can be approximated as closely as one likes with a finite probability by a tensor with such a typical rank. 
It is known that there exists only a single typical rank for complex symmetric tensors with given $w,N$, where
$w$ denotes the amount of ways (the amount of indices) of a tensor.  
This rank is called generic symmetric rank,  which we here denote by $R_g$, and is 
explicitly given by 
\[
R_g(w,N)=\left\lceil \frac{1}{N} \left(\begin{matrix}N+w-1 \\ w\end{matrix} \right)\right\rceil
\label{eq:rgdn}
\]   
with the following exceptions: $R_g(2,N)=N$, and $R_g$ is 
given by increasing the above formula by one for $(w,N)=(3,5),(4,3),(4,4),(4,5)$.
Here $\lceil \cdot \rceil$ denotes the ceiling function.
This statement is from the Alexander-Hirschowitz theorem \cite{zbMATH00773851} (See \cite{doi:10.1137/060661569,Landsberg2012} for more details.). 
The number on the right-hand side of \eq{eq:rgdn} is called expected rank, 
because it can be obtained by the simple number counting
of the degrees of freedom.

In the real case, however, there exist a number of typical ranks for given $w,N$,
the least value of which agrees with the generic rank of the complex case 
(See \cite{Sakata:2016:ACA:3027777} for more details.). 
So, the space of real tensors is divided into a number of subregions each of which has a certain typical rank.
The formula for the typical ranks is not known for general $w,N$ except for some specific cases. 
For example, the typical ranks are 2 and 3 for $(w,N)=(3,2)$ (See \cite{TenBerge2011}
for a table for three-way real tensors.).

The notion of typical rank implies that a given tensor can be approximated as closely as one likes by 
the form \eq{eq:CP} with a typical rank. 
However, due to the lack of a general formula for typical rank and a practical systematic procedure, 
the tensor-rank decomposition is to optimize vectors $v^i$
to approximate a given tensor as much as possible with the form \eq{eq:CP} with a value of $R$.
So, practically, what we obtain is an approximate tensor-rank decomposition,
\[
P_{abc}=\sum_{i=1}^R v_a^i v_b^i v_c^i+\Delta P_{abc},
\label{eq:cperror}
\]
rather than an exact \eq{eq:CP}, where the error $\Delta P_{abc}$ should be made as small as possible.
The error $\Delta P_{abc}$ can be made (numerically) vanish if one takes $R$ large enough, but
$R$ cannot be taken unconditionally large in practical computations.
This is not only because the optimization process takes longer time for large $R$ with larger degrees of freedom,
but also because for larger $R$ it becomes more difficult to avoid rough decompositions which contain mutual
cancellations of the rank-one components (See \cite{doi:10.1002/cem.1236,hackbusch-2012} for more details).
Therefore there exist various uncertainties in the decomposition. 
Is $R$ taken large enough? Are the vectors optimized? How much of an error is reasonable to allow?

These uncertainties introduce uncertainties in results and are potentially very harmful when actually doing computations.  
In our applications, however, reasonable results are obtained by taking $R$ reasonably large to make errors 
sufficiently small and repeating the optimization procedure several times to choose the best set of vectors.
Here, for the optimization, we made a C++ program which implements the greedy algorithm 
described in \cite{hackbusch-2012} with an additional constraint. The program is described in some detail 
in Appendix~\ref{app:CP}.
It is worth noting that despite the possible numerical problems, the tensor-rank decomposition 
is well defined, so at least in principle we have a good notion of points corresponding to a tensor.

\section{Real symmetric three-way tensors corresponding to fuzzy spaces}
\label{sec:fuzzyspace}
Real symmetric three-way tensors may be used to describe spaces through the algebra of functions acting on these spaces \cite{Sasakura:2005js, Sasakura:2006pq, Sasakura:2011ma, Sasakura:2011nj, Sasakura:2011qg}.
In this section we describe a systematic method to construct such tensors from their corresponding algebra. 
This method is particularly useful in constructing such tensors corresponding to homogeneous spaces 
invariant under Lie-group symmetries. 
A requirement for such tensors is that they should be invariant under the symmetric properties of the corresponding homogeneous spaces.

The rough idea of fuzzy spaces is to specify a space in terms of the algebra of functions on it rather than
a coordinate system. 
This would be in accord with the fact that the relevant objects in physics are fields on a space rather than a space itself.
Let us consider first an ordinary continuous space $\mathbb{R}^D$. 
In this case functions can be used to label points, because a single point, say $\omega_0$, 
can be identified by providing a localized function\footnote{This intuitive discussion is a bit formal, as this is not a proper function. We can take a function which is arbitrarily close to this localized distribution.} $f_{\omega_0}(\omega)=\delta^D(\omega-\omega_0)$. 
Therefore considering all the independent functions, which are $f_{\omega_0}$  with $\omega_0\in \mathbb{R}^D$, 
gives the whole space. The algebra of functions, 
$f_{\omega_0}(\omega)f_{\omega_1}(\omega)=\delta^D(\omega-\omega_0)\delta^D(\omega-\omega_1)
=\delta^D(\omega_0-\omega_1)f_{\omega_0}(\omega)$,
reflects the pointwise structure of the continuous space.
To get to more interesting cases one can modify this structure in various ways. Well known are the non-commutative 
spaces in which the function algebras are taken to be non-commutative (and usually associative)
\cite{Madore:2000aq}.

We modify the algebra in a different way, by picking up only functions corresponding to lower frequency modes 
than a cut-off and ignore all the other higher frequency modes.
In this case, functions cannot represent single points $\omega_0$ anymore, and the space necessarily becomes ``fuzzy". 
This truncation gives a finite number of functions $f_a(\omega)\ (a=1,2,\ldots,N)$ for a compact space ${\cal M}$
with a coordinate $\omega$. 
The simplest way to obtain an algebra of functions is to truncate the products of functions 
by ignoring higher frequency modes. 
Such an algebra has the form $f_a(\omega) f_b(\omega)=P_{ab}{}^c f_c(\omega)$ with structure coefficients
$P_{ab}{}^c$ taking the original values in the full algebra of the continuum case, 
but the summation over the modes $c$ is truncated by $c\leq N$.
This procedure gives a commutative non-associative algebra. 
Now the structure coefficients can be extracted by considering
\[
P_{abc}=\int_{\cal M} d\omega\,  f_a(\omega) f_b(\omega) f_c(\omega).
\label{eq:pfff}
\]  
This procedure naturally defines a three-way tensor corresponding to a space with 
fuzziness.\footnote{Similarly, one could consider a matrix, $M_{ab}=\int_{\cal M} d\omega\, f_a(\omega)_b(\omega)$. 
Though this also contains a product of two functions, it is projected to the zero mode by the integration, and 
the matrix cannot represent the full structure of a fuzzy space.
In this sense, taking the three-way tensor above is the minimum (and enough) choice.}
By considering real functions for all $f_a$, \eq{eq:pfff} gives a real symmetric three-way tensor representing a fuzzy space.
For a homogeneous space, $P_{abc}$ should be invariant under its symmetry, and therefore the function set 
must be taken so that it forms a certain representation of the symmetry and $P_{abc}$ is an invariant tensor.

A comment is in order. Comparing with the decomposition \eq{eq:CP},
one notices that \eq{eq:pfff} is nothing but a tensor-rank decomposition of $P_{abc}$, where
the index $i$ is replaced by a continuum one $\omega$.
This does not mean that a fuzzy space defined by \eq{eq:pfff} requires an infinite $R$ with a continuous index. 
In fact, for a finite $N$, the tensor-rank decomposition of $P_{abc}$ can always be performed with a finite $R$.
Therefore a compact fuzzy space defined by \eq{eq:pfff} is always represented by a finite number of points.
If one takes a limit back to the continuum (i.e. $N \rightarrow \infty$), $R$ should indeed become infinite, which 
is considered in a formal continuum limit of the CTM in Section~\ref{sec:CPcont}.

Another comment concerns the ordinary continuum space. Let us take a basis of real functions by 
the delta functions mentioned above. Then \eq{eq:pfff} is given by $P_{\omega_0\omega_1\omega_2}
=\delta^D(\omega_0-\omega_1)\delta^D(\omega_0-\omega_2)$.  
Thus the ordinary continuum space is described by a continuous fully diagonal three-way tensor.

As a concrete example, let us consider a homogeneous fuzzy two-sphere.
As described above, we take a cut-off $L$ and take the spherical harmonics with angular momenta $\leq L$ as a function set.
Then the  real symmetric three-way tensor corresponding to a homogeneous fuzzy two-sphere is given by
\[
P_{(l_1,m_1)\,(l_2,m_2)\,(l_3,m_3)}=
\int_{S^2} d\Omega\ \tilde Y_{l_1 m_1}(\Omega)\tilde Y_{l_2 m_2}(\Omega)\tilde Y_{l_3 m_3}(\Omega).
\label{eq:poftwosphere}
\]   
Here $(l,m)$ takes $l=0,1,\ldots, L,\ m=-l,-l+1,\ldots,l$, and $\tilde Y_{lm} $ are the real functions defined by
\[
\tilde Y_{lm}=
\left\{
\begin{matrix}
\frac{1}{\sqrt{2}} \left(Y_{lm}+Y_{lm}^*\right)e^{- l^2/L^2} ,& m>0,\\
Y_{l0}\,e^{-l^2/L^2} , & m=0 ,\\
\frac{1}{\sqrt{2}i} \left(Y_{lm}-Y_{lm}^*\right)e^{-l^2/L^2} ,  & m<0,
\end{matrix}
\right.
\label{eq:tildey}
\]
where $Y_{lm}$ are the spherical harmonics and the star represents taking a complex conjugation.\footnote{
More explicitly, we use a formula,
\[
\int_{S^2} d\Omega\ Y_{l_1 m_1}Y_{l_2 m_2}Y_{l_3 m_3}
=\prod_{i=1}^3 \sqrt{2l_i+1} \left(\begin{matrix}l_1&l_2&l_3 \\ 0&0&0\end{matrix}\right)
 \left(\begin{matrix}l_1&l_2&l_3 \\ m_1&m_2&m_3\end{matrix}\right) \nonumber
\]
with $3j$-symbols.
}
The exponential damping factor is to make the high frequency cut-off smoother,
which turns out to result in better behaved systems, as is explained in Section~\ref{sec:neighbor}.  
In the section we apply the tensor-rank decomposition to the $P$
in \eq{eq:poftwosphere}, and obtain the geometric picture in Figure~\ref{fig:sphere}, which
clearly represents a spherical object.

The above procedure is general enough to construct various real symmetric three-way tensors corresponding to 
fuzzy spaces.
Another simple example is a homogeneous fuzzy $S^1$, which can be obtained by considering a real basis for 
%the plane waves $e^{i l \theta-l^2/L^2}\ (l=0,\pm1,\ldots,\pm L)$ on a circle $0\leq \theta < 2\pi$. 
%these are not plane waves, plane waves have a plane as their wave fronts and that's by definition of S^1 not possible.
functions on a circle.
In Section~\ref{sec:constP}, we explicitly construct real symmetric three-way tensors corresponding to 
spaces with various dimensions and topologies, and also non-orientability.  
There we find that the tensor-rank decomposition leads to topological and geometric 
interpretations in agreement with the corresponding continuum spaces. 
This demonstration proves the generality of our construction and that real symmetric three-way tensors
can in principle represent any kinds of spaces. 
This last fact is particularly important for the generality of the CTM, in which the 
tensors are real symmetric three-way. This is in contrast with the other Euclidean tensor models 
\cite{Ambjorn:1990ge,Sasakura:1990fs,Godfrey:1990dt,Gurau:2009tw}, in which 
the number of ways (i.e., the amount of indices) of tensors
is supposed to be in accord with the dimension of building simplicial spaces one considers.   

\section{Notion of neighborhoods in fuzzy spaces}
\label{sec:neighbor}
In this section, we introduce the notion of neighborhoods around points in fuzzy spaces. 
Let us assume that a tensor-rank decomposition \eq{eq:CP} is obtained for a given real symmetric three-way tensor.  
A point $i$ is represented by a vector $v^i$, as discussed in Section~\ref{sec:cp}. 
Here we introduce the notion quite naively from the inner product,
but in Section~\ref{sec:diffusion} this form is justified in the CTM via a virtual diffusion process.

Let us define the neighborhood of a point $i$ by the following set of points:\footnote{This does not include all neighborhoods in a topological sense, which may be obtained by taking unions of these neighborhoods.}
\[
{\cal N}_c(i)=\left\{j \left|  v^i_a v^j_a >c \right. \right\},
\label{eq:neighbor}
\]
where the repeated index $a$ is assumed to be summed over. 
Hereafter this standard convention is implicitly assumed for the indices originated from the tensor indices, e.g., $a$ but 
not $i,j$ in \eq{eq:neighbor}.         
The paramater $c$ in \eq{eq:neighbor} determines the size of the neighborhood: 
For larger $c$, the neighborhood becomes smaller, and vice versa. 

\begin{figure}
\begin{center}
\includegraphics[scale=.4]{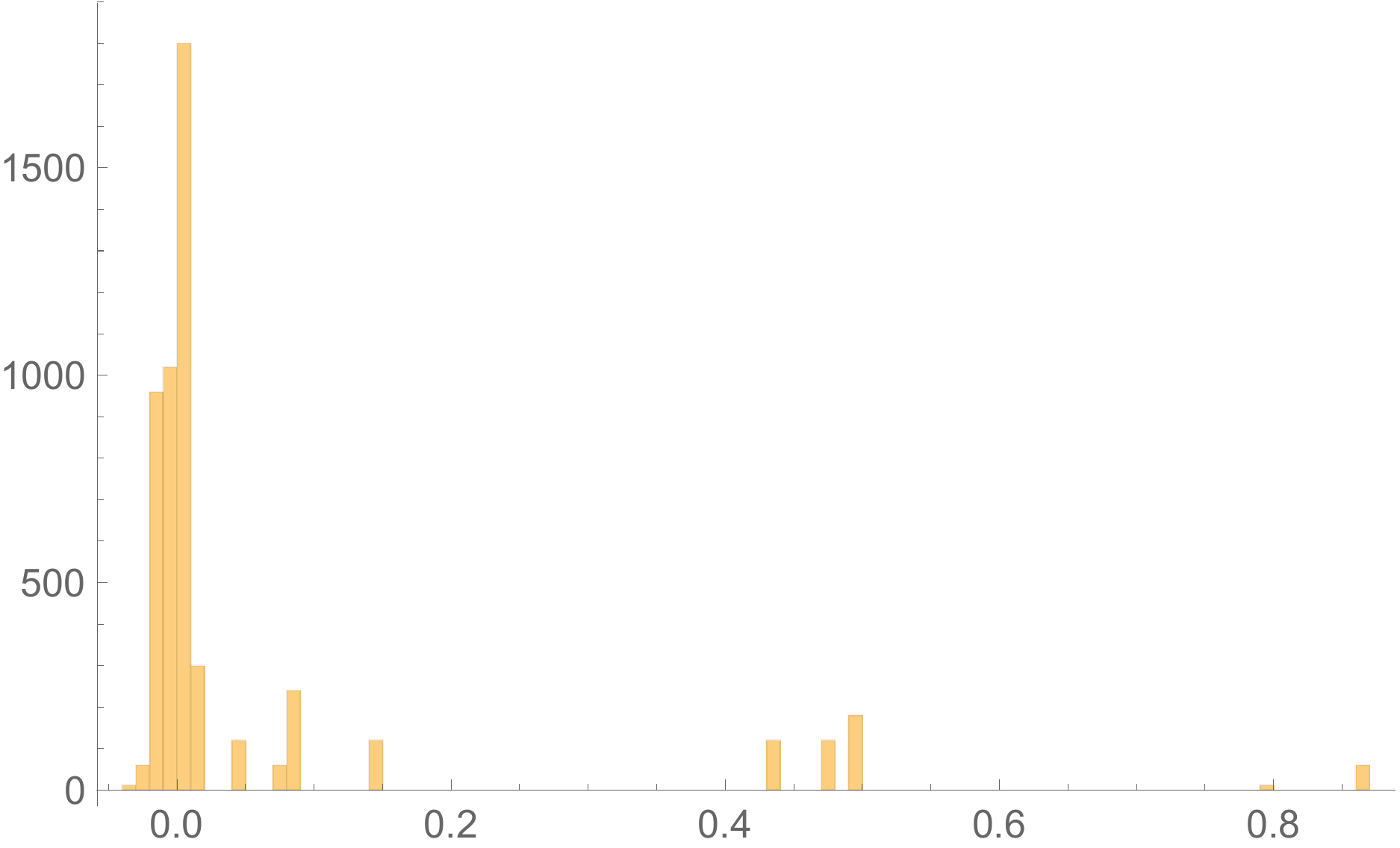}
\hfil
\includegraphics[scale=.6]{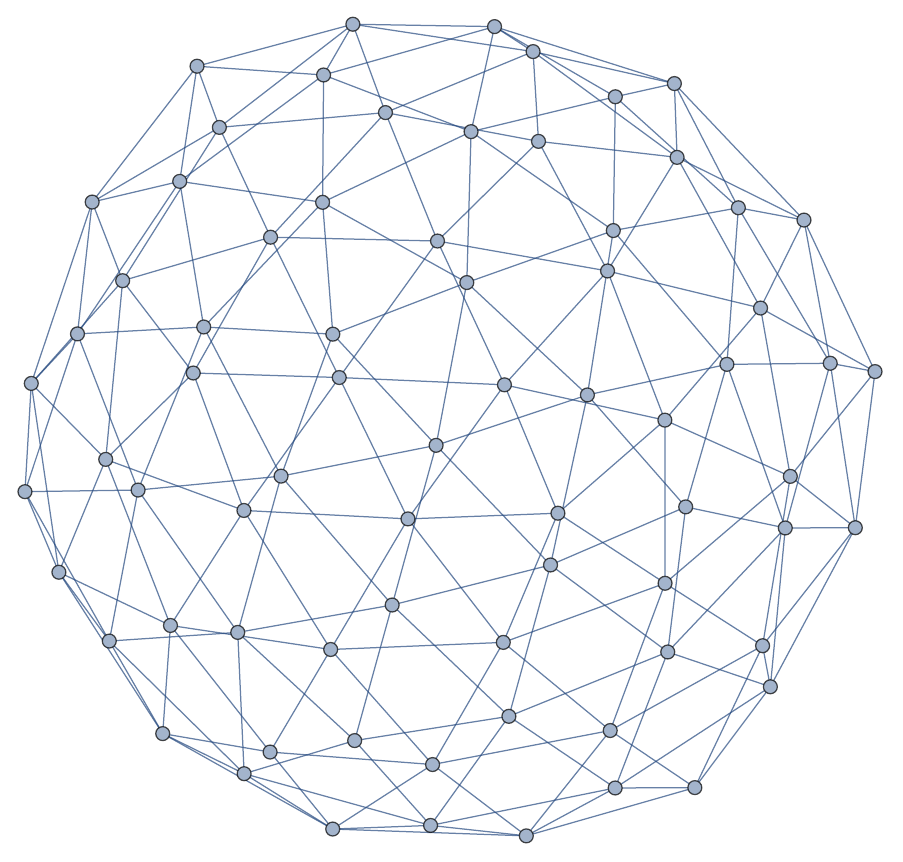}
\end{center}
\caption{Left: The histogram of the values of the inner products $v_a^i v_a^j\ (i,j=1,2,\ldots, R)$ for the fuzzy two-sphere with $L=5$.
Right: The diagram of connections of points obtained from the tensor-rank decomposition of 
$P$ of the fuzzy two-sphere.
Points are connected if two points $i$ and $j$ satisfy $v_a^iv_a^j>0.2$.}
\label{fig:sphere}
\end{figure}

\begin{figure}
\begin{center}
\includegraphics[scale=.4]{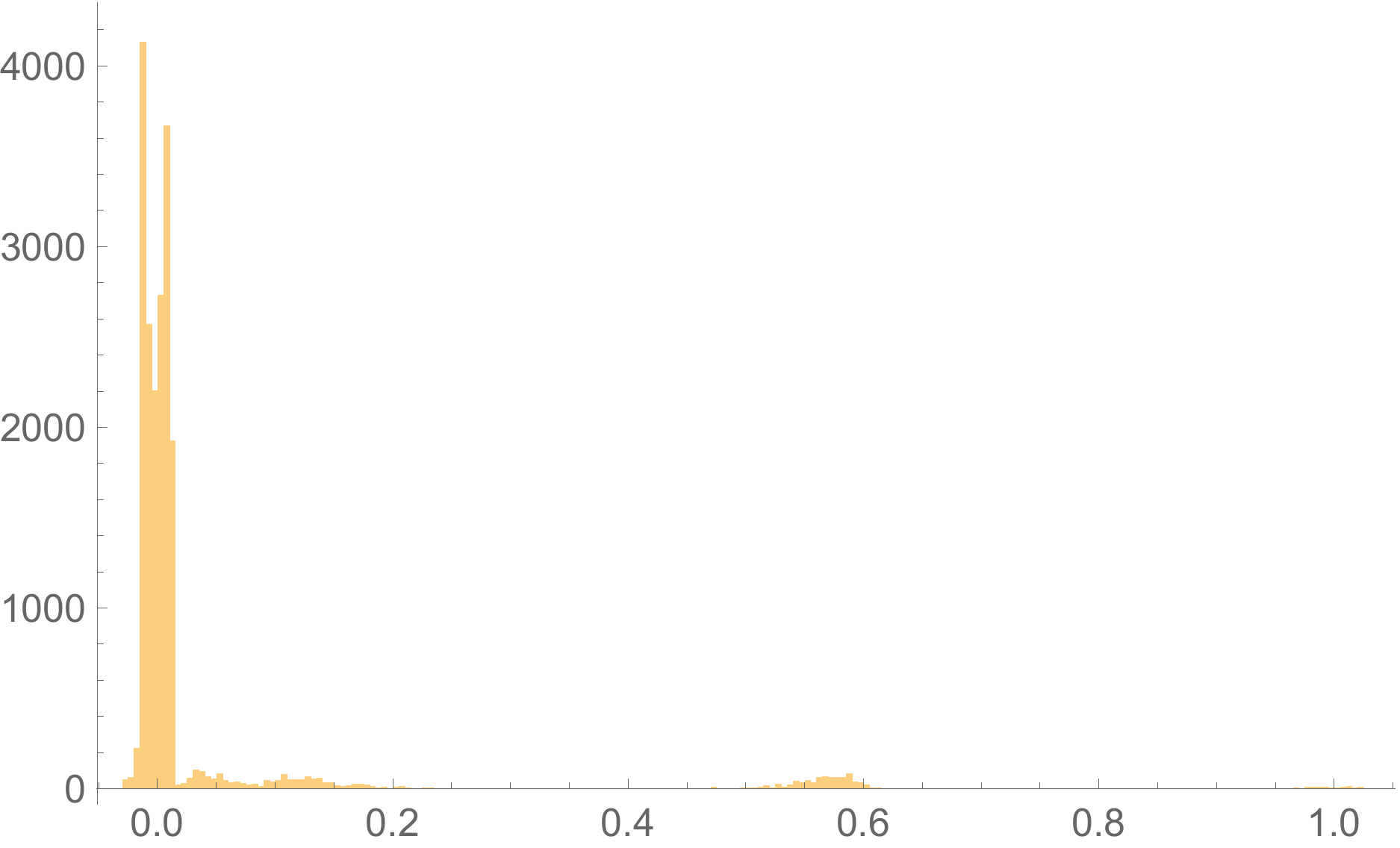}
\end{center}
\caption{The histogram of the inner products for the fuzzy two-sphere with $L=7$ and $R=142$.}
\label{fig:SphereLarge}
\end{figure}

As an example, let us consider a fuzzy two-sphere with $L=5$ defined in Section~\ref{sec:fuzzyspace}. 
The dimension of the vector space of $P$ is $N=(L+1)^2=36$. Taking the rank to be $R=72$, one can obtain a tensor-rank decomposition of $P$ within a 2 percent 
error\footnote{The percentage of error is defined from the ratio $\sqrt{\Delta P^2/P^2}$ for \eq{eq:cperror}, 
where $T^2\equiv T_{abc}T_{abc}$ for a 3-way tensor $T$.}.
The left of Figure~\ref{fig:sphere} is a histogram of the values of the inner products $v^i_a v^j_a$ for $i,j=1,2,\ldots,R$.
The rightmost bins around 0.8 are composed of the self-inner products $v_a^iv_a^i$. 
The middle ones around 0.4 are composed of the inner products between the nearest neighbor points. 
Most of the inner products concentrate in a small region around the origin, which means that most of the points are not in 
their mutual neighborhoods for $c>0$. The physical meaning of this concentration is that the fuzzy space respects locality, which is indeed what we hope for if we make $N$ sufficiently large. 
As can be seen in Figure~\ref{fig:SphereLarge}, this concentration 
around the origin becomes larger when the size of the fuzzy space is bigger, as more points are 
not in their mutual neighborhoods. This aspect can also be quickly understood by the fact that the probability 
for two independent $N$-dimensional vectors to have a relative angle $\theta$ is proportional to
$\sin^{N-1}\theta$, which is the surface volume on a unit sphere at an angle $\theta$ from a vector.
This phenomenon is called the concentration of measure in mathematical literature~\cite{matousek2002lectures}.
The right of Figure~\ref{fig:sphere} 
shows the neighborhood relations among the fuzzy space points, which are connected if two points $i$ and $j$ 
satisfy $v_a^iv_a^j>0.2$.
Here the cutoff value is chosen so that the middle bunch of bins around $0.4$ representing the nearest neighbor 
connections is well included. 
The figure clearly shows that the $P$ defined in Section~\ref{sec:fuzzyspace} 
represents a discrete analogue of a continuous two-sphere through our procedure.
The topological aspect is discussed more precisely in terms of persistent homology in Section~\ref{sec:persistent}.

\begin{figure}
\begin{center}
\includegraphics[scale=.4]{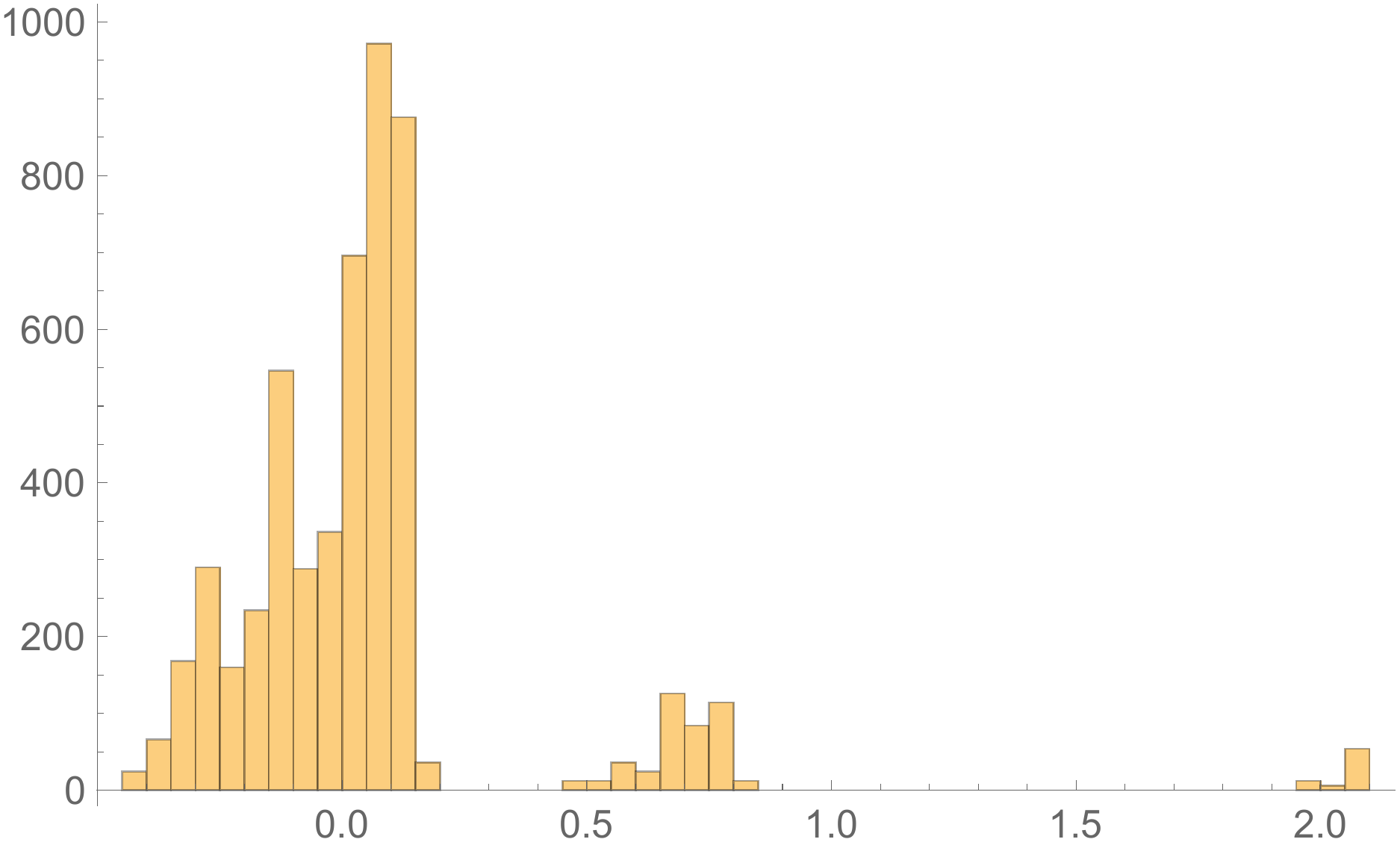}
\end{center}
\caption{The histogram of the inner products for the fuzzy two-sphere with $L=5$ without
the damping factor. 
The peak around the origin is broadened compared with the left of Figure~\ref{fig:sphere}. }
\label{fig:spherenoreg}
\end{figure}

Let us comment on the importance of the damping factor $e^{-l^2/L^2}$ in \eq{eq:tildey}, which smoothens the cut-off.
Figure~\ref{fig:spherenoreg} shows the 
histogram of the inner products obtained from the fuzzy two-sphere without the damping factor for $L=5$. 
Comparing with the left of Figure~\ref{fig:sphere}, 
one can see that the peak around the origin is broadened into the negative values. This situation can be
illustrated very roughly by comparing the following two elementary integrations:
\begin{align}
\begin{split}
\int_{-L}^L dp\, e^{ipx}&\propto \frac{\sin(L x)}{x},\\
\int_{-L}^L dp\, e^{-\frac{p^2}{L^2}+ipx}&\sim \int_{-\infty}^\infty dp\, e^{-p^2/L^2+ipx}
\propto  e^{-\frac{L^2}{4}x^2}.
\end{split}
\end{align}
While the former, corresponding to the sharp cut-off case, has a long-range oscillatory behavior with both
positive and negative values, the latter, corresponding to the case with the damping factor, has the fast 
exponential damping behavior with positive values only.  
The former aspects lead to two major difficulties.  
As is discussed in Section~\ref{sec:diffusion}, we use a virtual diffusion process to extract
geometrical information. However, the diffusion equation with negative coefficients can have very 
unusual behavior. 
For example, diffusion is usually a process of easing concentrations, but with negative coefficients
it can have diverging behavior, which contradicts the assumptions in Section~\ref{sec:diffusion}.
Another difficulty is that the broad distribution around the origin implies that
there exist a substantial number of pairs of points having non-negligible inner products. 
This means that the fuzzy space has strong non-local features violating locality of an ordinary continuous space.
These problems may also be solved by taking $L$ sufficiently large, but for doing actual numerical computations we have to restrict ourselves to finite tensors and the damping factor is very useful.

\def\VR{{\rm VR}}
\def\Betti{{\rm Betti}}

\section{Persistent homology}
\label{sec:persistent}
In Section~\ref{sec:neighbor}, we have introduced the notion of neighborhoods. 
This characterizes local topological structure of fuzzy spaces.
Global topological structure is also of much interest.
In this section, we introduce the notion called persistent homology \cite{Carlsson09}
as a method to extract the homological structure of the fuzzy spaces.

Let us first assume that a distance $d(\cdot,\cdot)$ between any pair of points on a fuzzy space (after a 
tensor-rank decomposition) is given.
How to construct such a distance is discussed in due course. 
Let us denote the set of points which represent the fuzzy space by $V$.

Let us introduce a family 
of abstract simplicial complexes, parameterized by $u$, associated to a fuzzy space, 
which is called a Vietoris-Rips stream\footnote{There also exist some other streams which can be more useful in some situations. See \cite{Carlsson09} and references therein for more details.}
and is denoted by $\VR(V,u)$. 
The complex, $\VR(V,u)$, is defined as follows:
\begin{itemize}
\item The vertex set is given by $V$.
\item For vertices $i$ and $j$, the edge $[ij]$ is included in $\VR (V,u)$ if and only if $d(i,j)\leq u$.
\item A higher dimensional simplex is included in $\VR (V,u)$ if and only if all of its edges are.
\end{itemize}
Since the Vietoris-Rips stream has the obvious property that $\VR (V,u) \subset \VR (V,u')$ for $u\leq u'$, 
it is called a filtered simplicial complex.
A filtered simplicial complex has the following functorial property: For $u\leq u'$, the inclusion 
$i: \VR (V,u) \rightarrow \VR (V,u')$ of
simplicial complexes induces a map $i_*: H_k(\VR (V,u))\rightarrow H_k(\VR (V,u'))$ between homology groups.

Given such a stream of simplicial complexes with the above functorial property, 
one can follow the creation and annihilation of the elements in the homology group of $\VR(V,u)$
while changing $u$. Here the filtration parameter $u$ roughly corresponds to the resolution of distances. 
When $u$ is smaller than any of the distances between points, all the points are independent; there is no non-trivial topological structure. 
When $u$ is increased, points get connected to one another, and there appear edges and higher-dimensional simplices, leading to some non-trivial topological structure.
An element of the $k$-homology group $H_k(\VR (V,u))$ corresponds to a $k$-cycle which is not the boundary of a $k+1$-cycle, and the dimension of this group (the Betti number) corresponds roughly to 
the amount of holes with $k$-dimensional boundaries. When $u$ is changed to become larger than the size of 
such a hole, the hole is filled by simplices and is not visible in the homology group.
Therefore such a hole can be represented by an interval $[u_{start},u_{end})$, which represents its creation 
and annihilation and is called a Betti interval.
For a given point set with mutual distances, there exist Betti intervals of various lengths. Each of them in principle is 
directly associated with the data, but the ones with long lengths
are considered to be the intrinsic feature of a fuzzy space.
On the other hand, the shorter ones are not stable against 
small perturbations, depending much on details, and are rather regarded as noises.
This summarizes the idea of persistent homology, which extracts a topological structure from a discrete set of points
with distances. 

We now describe a simple way to construct a distance function $d(\cdot, \cdot)$, also called a metric, 
which we can use in the analysis of persistent homology below. 
We define 
\[
d(i, j) := 1 \text{ if } j \in {\cal N}_c(i) \label{eq:persistent:d_ij}
\]
for fixed $c$, where the neighborhood ${\cal N}_c(i)$ is defined in \eq{eq:neighbor}.
A path between $i \in V$ and $j \in V$ is defined as an ordered sequence of points $p(i,j) = (p_1,\ldots,p_n)|_{p_1 = i, p_n = j}$, where the points in a pair $(p_k, p_{k+1})$ are always in each other's neighborhood. The length of a path is given by the sum of the distances of individual links, $L_p := \sum_{k=1}^{n-1} d(p_k,p_{k+1})$. The distance of two points is defined as the length of the shortest path between them, $d(i,j) := \min \{L_p | p(i,j)\}$. If two points $i, j$ are not connected by a union of neighborhoods, we say $d(i, j) = \infty$.
This distance function is relatively simple, whereas a more sophisticated notion of distance is introduced in Section~\ref{sec:diffusion}. This simple distance function however is easy
to calculate, and seems to be applicable to extract the intrinsic topological structure of a fuzzy space.
This is because long-lived stable Betti intervals are not affected by detailed choices of distances, while 
noisy short-lived Betti-intervals may be changed. 

\begin{figure}
\begin{center}
\includegraphics[scale=.4]{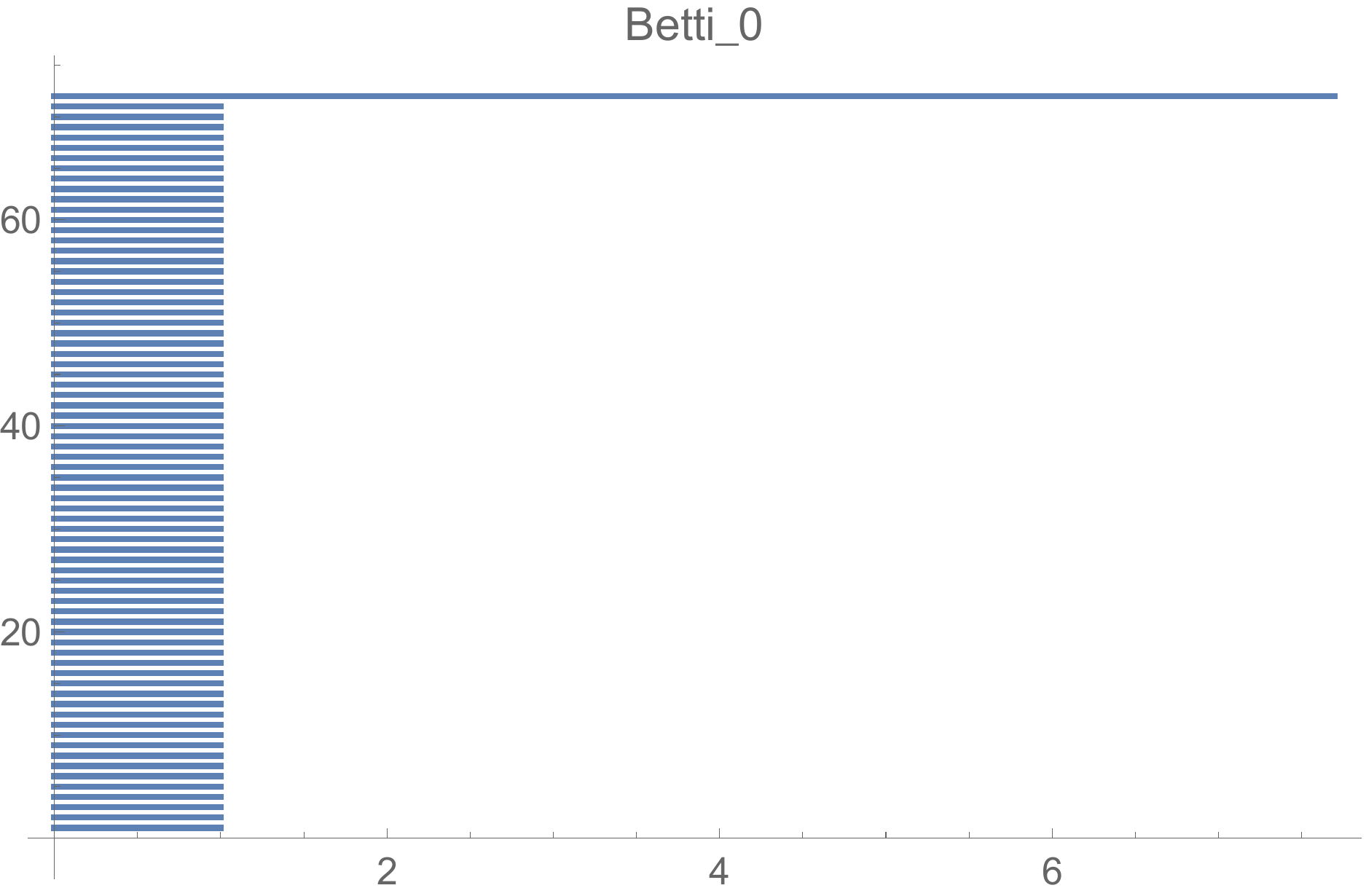}
\hfil
\includegraphics[scale=.4]{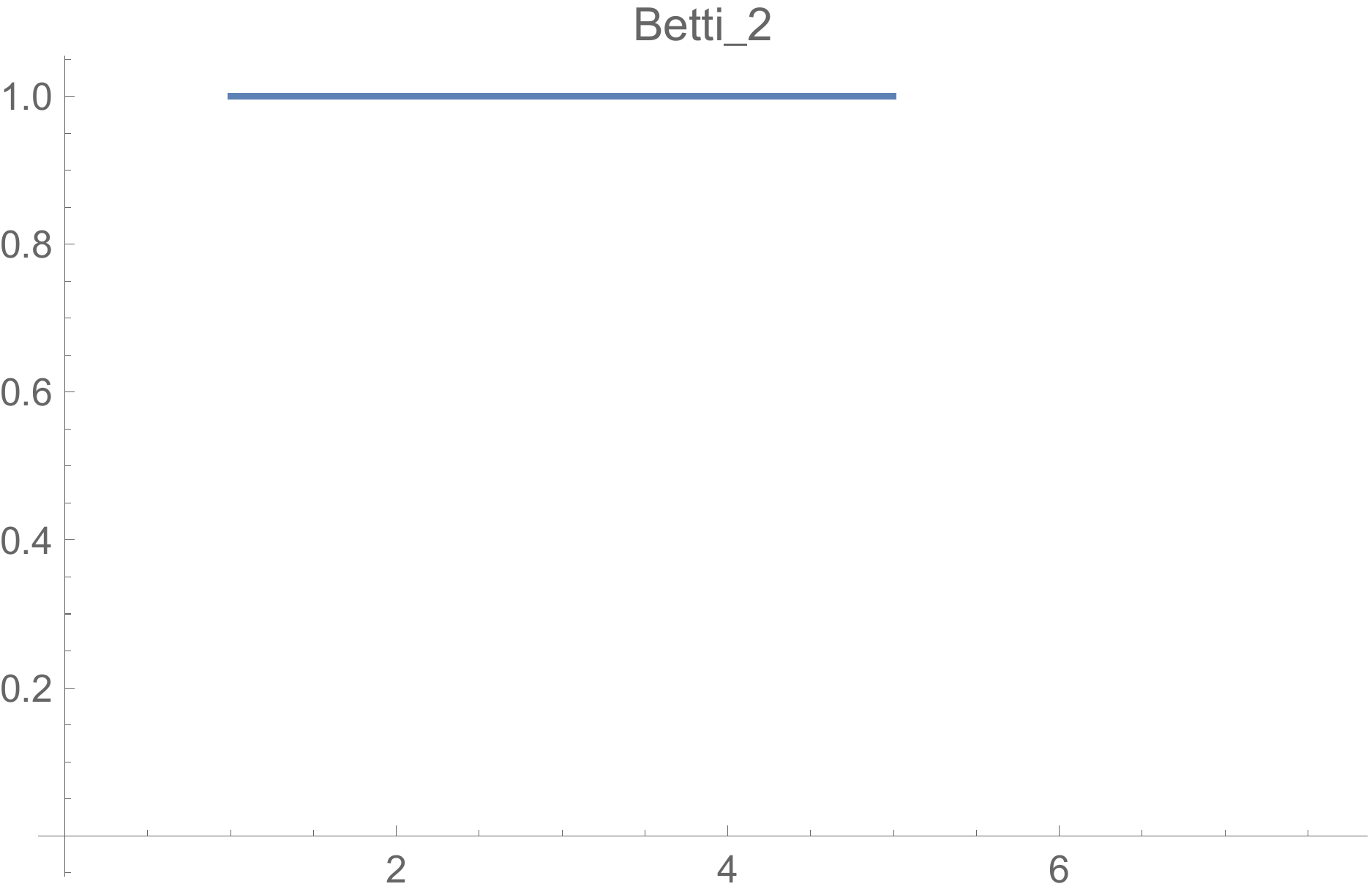}
\end{center}
\caption{The $\mathbb{Z}_2$-coefficient Betti intervals of the fuzzy two-sphere with $L=5,\ R=72$. 
We used the distance function defined around~\eqref{eq:persistent:d_ij} and chose $c=0.2$.
${\rm dim}(H_1)=0$ throughout.}
\label{fig:spherebetti}
\end{figure}

Figure~\ref{fig:spherebetti} shows the Betti intervals for the fuzzy two-sphere with $L=5$ and $R=72$, using the distance function defined above.
We used a program named ``Ripser"\footnote{The (open source) software can freely be downloaded from https://github.com/Ripser/ripser
in the GitHub repository.
This program computes the Betti intervals for the $\mathbb{Z}_n$-coefficient homology groups 
with the free choice of $n$ as an input.} 
to compute the so-called persistence barcodes of the Vietoris-Rips stream and imported the data in Mathematica to construct the images. 
Here the graph of $\Betti_k$ represents the Betti intervals for the homology group $H_k(\VR (V,u))$, which is the aforementioned persistence barcode.
For $0\leq u<1$, $u$ is smaller than any of the distances between the points by construction, and 
there are no edges.
Since $H_0$ represents the homology class of connected components,
the number of $\Betti_0$ intervals equals that of points or the rank $R$
in this region of $u$.
For $1\leq u$, the points are all connected to form one component such that there exists one long interval for $\Betti_0$
until the maximum distance on a fuzzy space, in this case $7$.
For $1\leq u < 4$, there exists one interval for $\Betti_2$, which represents the existence of 
a hole with a two-dimensional boundary.
It vanishes at $u=4$, when the hole is filled by simplices.   
$\Betti_1$ is vanishing throughout the range of $u$.
Thus, the long-life structure is observed to be ${\rm dim}(H_0)={\rm dim}(H_2)=1, \ {\rm dim}(H_1)=0$, 
topologically agreeing with an ordinary two-sphere.
 
\begin{figure}
\begin{center}
\includegraphics[scale=.4]{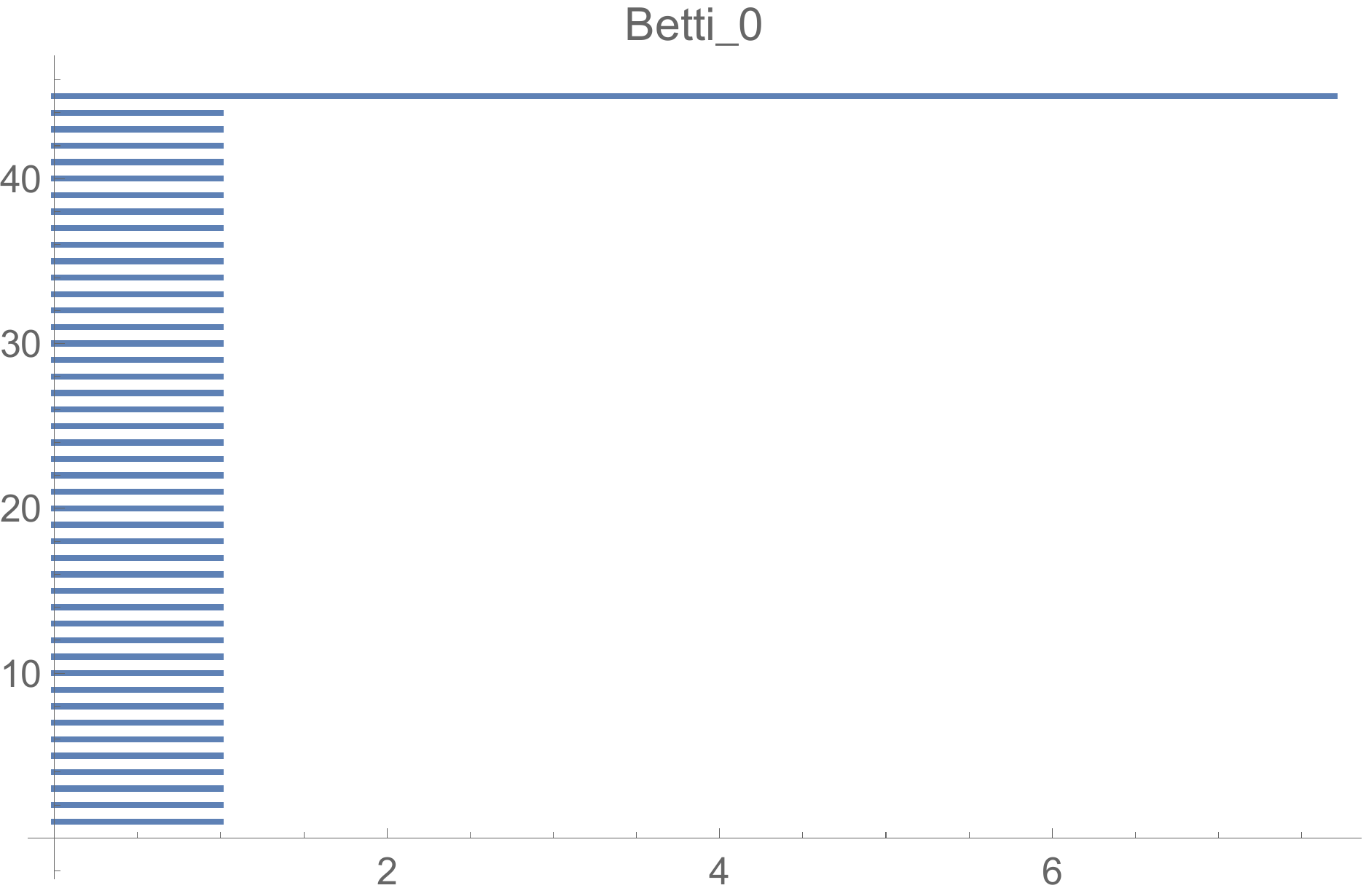}
\hfil
\includegraphics[scale=.4]{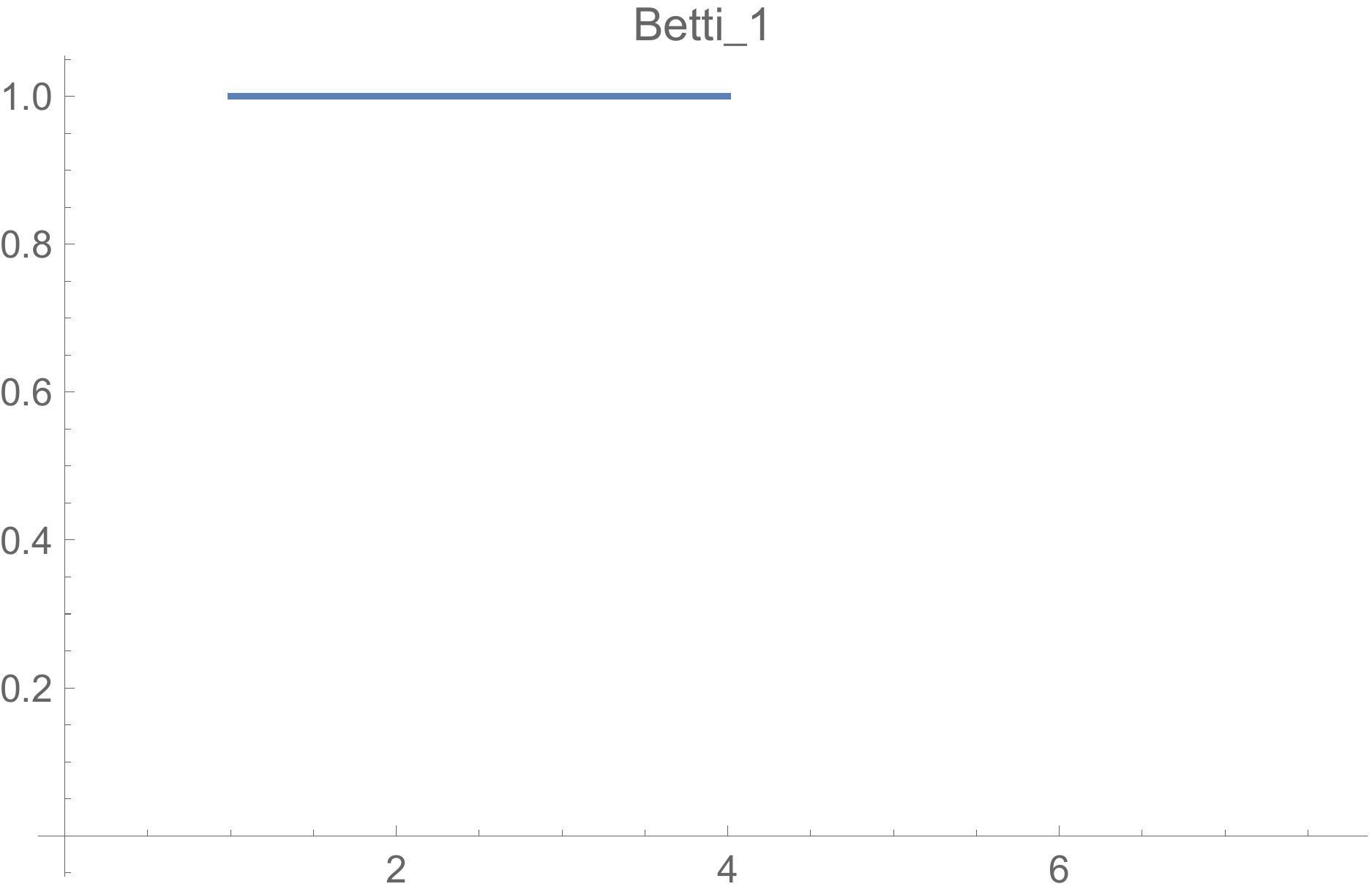}
\end{center}
\caption{The $\mathbb{Z}_2$-coefficient Betti intervals for the collection of points with the distances 
$3\leq d(\cdot,p) \leq 5$ from a reference point $p$
in the fuzzy space with $L=5,\ R=72$, again using $c=0.2$.
The collection has the persistent homology consistent with a circle implying that the topological dimension of the fuzzy space is two. }
\label{fig:strip}
\end{figure}

Another application of persistent homology is to determine the topological dimension of a fuzzy space.
If the analysis of persistent homology implies ${\rm dim}(H_k)\neq0$ with a certain $k$, 
one can know that the topological dimension of the fuzzy space should not be less than $k$.  
However, this method is not so useful, because, for example, a ball has a finite topological dimension 
but vanishing homologies except $H_0$. 
A more useful way is to consider a reference point, say $p$, and a collection of points within 
a certain range of distance from it, $d_{min}\leq d(\cdot,p) \leq d_{max}$, and study its persistent homology. 
One would expect that the collection of points forms a sphere of dimension being $D-1$, where $D$ is the dimension of the fuzzy space.
This yields a local definition of the topological dimension around a reference point,
%\footnote{This obviously only works if the space is ``nice''. The topological dimension of a space is certainly not a local 
%property.}, 
and if this is the same for any choice of reference point except special points such as those
on boundaries, the fuzzy space can be considered to have a well-defined topological dimension.   
In Figure~\ref{fig:strip}, an illustrative example is shown for the same fuzzy two-sphere as the previous one. 

\section{Tensor-rank decomposition in a formal continuum limit}\label{sec:CPcont}
In~\cite{Chen:2016ate} the authors discussed the correspondence between the CTM and a general relativistic system
by performing a derivative expansion of $P$ in a formal continuum limit of the CTM.
In the paper the authors considered derivatives up to the fourth order to analyze the equation of motion of 
the metric and a scalar field up to the second order of their derivatives, 
while additional higher spin fields must be taken into account in higher orders.  
In the current discussion, however, we are interested in the metric and the scalar field themselves with no necessity 
for their derivatives, and it is sufficient to consider a derivative expansion up to the second order.

In the formal continuum limit the indices of the tensor are assumed to become continuous coordinates in $\mathbb{R}^D$:
\[
P_{abc} \rightarrow P_{xyz},\ \  x,y,z \in \mathbb{R}^D.
\label{eq:pcont}
\]
Furthermore, a locality condition is imposed, which says $P_{xyz} \neq 0$ only if  $x \sim y \sim z$. This rough locality condition was mathematically translated to the tensor becoming a distribution and may be given by a derivative expansion:
\begin{equation}
 P_{xyz} = \int_{\mathbb{R}^D} d^D \omega\, \beta(\omega) \delta^D(x-\omega) \delta^D(y-\omega) \delta^D(z-\omega) + \text{\ derivatives of delta functions} .\label{eq:cont:P_der_exp}
\end{equation}
Distributions are defined by their action on test functions under integration. The authors showed that up to second order the expansion can be written as
\begin{equation}
 P f^3 := \int d^D x d^D y d^D z P_{xyz} f(x) f(y) f(z) :\approx \int d^D x \left(\beta(x) f(x)^3 + \beta^{\mu\nu}(x) 
 f(x)^2 f_{,\mu\nu}(x) \right),\label{eq:cont:P_testf}
\end{equation}
where the $\beta(x)$ and $\beta^{\mu\nu}(x)$, which is symmetric, are the expansion coefficients, and $f(x)$ is an arbitrary test function.
%note: test functions are already defined as functions with compact support right?
The $\beta$ and $\beta^{\mu\nu}$ in the expansion are fields on the space $\mathbb{R}^D$. These fields contain the degrees of freedom of the CTM in the formal continuum limit, which corresponds to a general relativistic system. 
The relation between the $\beta$ fields and the scalar field $\phi$ and the metric field $g_{\mu\nu}$ 
of the relativistic system has been found by analyzing the equations of motion of the CTM and is given by \cite{Chen:2016ate} 
%\[
%\beta(x) \beta^{\mu\nu}(x)=\frac{g^{\mu\nu}(x)}{\sqrt{g(x)}}
%\]
%and the $\beta$ fields were obtained from the metric and the scalar field
\s[
\beta(x)&=g(x)^{-1/4} e^{\phi(x)},  \\
\beta^{\mu\nu}(x)&=g(x)^{-1/4} e^{-\phi(x)}g^{\mu\nu}(x),
\label{eq:betagrel}
\s]
where $g={\rm Det}(g_{\mu\nu})$.
The test functions and $\beta$'s are not usual scalar functions but have non-vanishing density weights, 
and they were fixed from their transformation properties under spatial
diffeomorphisms which are part of the continuum limit of the $SO(N)$ symmetry transformation of the CTM.
The weights are $[f]=-[\beta]=-[\beta^{\mu\nu}]= [g^{1/4}]$ as in \eq{eq:betagrel}.
In particular, these weights are taken so that the weights associated to each index of $P_{xyz}$ are
$[g^{1/4}]$ and the integral for an index contraction $\int d^Dx\, P_{xab} P_{xcd}$ is invariant under diffeomorphisms.

\def\tomega{{\tilde \omega}}
Let us now consider a continuous analogue of the tensor-rank decomposition. For this we assume the form
\begin{equation}
    P_{xyz} = \int d^D \omega \beta(\omega)^{-2} w_x(\omega) w_y(\omega) w_z(\omega).\label{eq:cont:decomp}
\end{equation}
The integration form $d^D \omega \beta^{-2}(\omega)$ is chosen such that the weight of the integration form is $0$ so the $w_x(\omega)$ are of weight $0$ in $\omega$.
The $w_x(\omega)$ still has a weight in $x$ of $[g^{1/4}(x)]$, because each index of $P_{xyz}$
must have this weight as explained above. 
From \eqref{eq:cont:P_der_exp} one can see that the $w_x(\omega)$ can also be given by a derivative expansion of the form
\begin{equation}
 w_x(\omega) := w(\omega) \delta^D(x-\omega) + w(\omega)w^{\mu}(x) \delta_{\mu}^{(x)}(x-\omega) + w(\omega)w^{\mu\nu}(x) \delta_{\mu\nu}^{(x)}(x-\omega) + \text{h.o.},\label{eq:cont:v_exp_1}
\end{equation}
where h.o.~means higher orders, $\delta_{\mu}^{(x)}(x-\omega) := \nabla_\mu^{(x)} \delta^D(x-\omega)$, and $\nabla_\mu^{(x)}$ is 
the covariant derivative acting on $x$ (similarly $\delta_{\mu\nu}^{(x)} := \nabla_\mu^{(x)}\nabla_\nu^{(x)} \delta$). 
Here, as explained above, $w_x(\omega)$ must have the weight of $[g^{1/4}(x)]$.
Let us use the convention that the density weight of the delta function, say $\delta^D(x-y)$ 
having a total weight of $[g^{1/2}]$, 
is equally distributed over both arguments $x,y$. 
Then the weights of the fields in \eq{eq:cont:v_exp_1} are determined to be 
$[w]=[g^{-1/4}]$ and $[w^{\mu}]=[w^{\mu\nu}]=0$. 
By putting \eqref{eq:cont:v_exp_1} into \eqref{eq:cont:decomp}, 
multiplying test functions, integrating over their arguments,
and comparing the result to \eqref{eq:cont:P_testf}, one finds
\[
w_x(\omega) = \beta(\omega) \delta^D(x-\omega) + \frac{1}{3} \beta(\omega)\beta^{-1}(x) \beta^{\mu\nu}(x) \delta_{\mu\nu}^{(x)}(x-\omega) + \text{h.o.}\label{eq:cont:v_exp}
\]
According to the interpretation given in Section~\ref{sec:cp}, 
the vector $w_x(\omega)$ represents a single point labeled by $\omega$. 
With these $w_x(\omega)$ one can define a quantity similar to the Euclidean inner product between two points. Since the weight of $w_x(\omega)$ is $[g^{1/4}(x)]$ an invariant quantity can be obtained by
\[
    K(\omega, \tomega) &:= \int d^D x\ w_x(\omega) w_x(\tomega),
    \label{eq:K}  \\
    &= A(\omega,\tomega) \delta^{D}(\omega-\tilde{\omega}) + A^{\mu\nu}(\omega, \tilde{\omega}) \delta_{\mu\nu}^{(\tilde{\omega})}(\tilde \omega-\omega)
 +A^{\mu\nu}(\tilde \omega, \omega) \delta_{\mu\nu}^{(\omega)}(\omega-\tilde \omega)+\hbox{h.o.} ,
 \label{eq:expK}
\]
where
\[
\begin{split}
 A(\omega,\tomega) &:= \beta(\omega)\beta(\tomega)=g(\omega)^{-1/4}g(\tomega)^{-1/4} 
 e^{\phi(\omega)+\phi(\tomega)},\\
 A^{\mu\nu}(\omega,\tomega) &:= \frac{1}{3}\beta(\omega)\beta^{\mu\nu}(\tomega)=
 \frac{1}{3}g(\omega)^{-1/4}g(\tomega)^{-1/4} g^{\mu\nu}(\tomega).
 \end{split}
 \label{eq:abeta}
\]
Here we have used \eq{eq:betagrel} to obtain the last field theoretical expressions, and it is apparent that 
the density weights provided by $g^{-1/4}$ cancel the weights from the delta functions to make $K(\omega,\tomega)$
a scalar in $\omega$ and $\tomega$.

As above, the quantity $K(\omega, \tomega)$ transforms as a scalar under diffeomorphisms on $\omega,\tomega$. 
However, this invariant feature should be particular to the continuum case, 
because the (almost) uniqueness of the tensor-rank decomposition for finite $N$, mentioned in Section~\ref{sec:cp},
does not allow such degeneracies of expressions in the discrete case.
On the other hand, it would be important to use a corresponding similar form even in the discrete case, because
it can be expected to converge to this physically meaningful invariant form in a continuum limit with $N\rightarrow \infty$.
Therefore we follow similar steps taking care of invariant forms as in this section, 
when we discuss a discrete analogue in Section~\ref{sec:distance}.

The expression \eq{eq:expK} of $K(\omega, \tilde{\omega})$ is given by an expansion in terms of the derivatives of 
delta functions, inheriting the locality imposed for $P_{xyz}$ below \eq{eq:pcont}. 
The physical meaning of this fact is that locality is respected by the mutual relations among points.
Similar inner products, $v_a^iv_a^j$, can be considered for the discrete case and
characterize the local distance structures of a fuzzy space.
This aspect of the inner products has already been used to define local neighborhoods around points
in Section~\ref{sec:neighbor}.
In the following sections, this aspect is further pursued in more detail.

\section{Distances by a virtual diffusion process}\label{sec:diffusion}
The main purpose of the present and the following sections is to find a notion of distances
on fuzzy spaces in terms of a diffusion process by using the knowledge of the preceding section.
For this purpose we need to relate the distances given by the metric field in the continuous theory to the discrete case. This is done through $K(\omega,\tomega)$ defined by the inner product in \eqref{eq:K}, which can also be interpreted as a 
second order differential operator as shown in \eq{eq:expK}. This operator defines a virtual diffusion process 
on a continuum space, 
which can be easily replicated on a discrete space to extract corresponding continuum quantities.

Using virtual diffusion processes to interpret discrete systems similarly to continuous ones is common in the literature of data analysis and quantum gravity as they can be interpreted in a similar way for both the discrete and continuous cases. For instance in data analysis diffusion processes are often used in order to define distance 
functions in data sets \cite{Nadler:2005:DMS:2976248.2976368,Coifman7426}. 
In quantum gravity the use of diffusion processes is also appreciated~\cite{Calcagni:2013vsa}, as they can be defined similarly for continuous and discrete spaces and allow one to construct well-defined observables such as the spectral dimension~\cite{Ambjorn:2005db}. Our strategy is to
consider a diffusion process defined by $K(\omega, \tomega)$ in \eq{eq:expK} and extract the geometric 
and scalar field data from it. 
As defined in \eq{eq:K}, $K(\omega, \tomega)$ is the continuous analogue of the inner product $v_a^iv_a^j$, 
so we can easily relate it to the discrete model and find a notion of distances there,
which is done in the following section.

The diffusion equation we consider in the continuum case is given by
\[
\frac{d}{ds}\rho(\omega,s)=\int d^D \tomega\,\beta(\tomega)^{-2} \ K(\omega, \tilde \omega)\,  \rho (\tilde \omega,s).
\label{eq:diffusion}
\]
where $\rho$ is a scalar field representing the density of a virtual diffusing material, 
$K(\omega, \tilde \omega)$ is given in \eq{eq:expK},
and $\beta(\tomega)^{-2}$ makes the volume element to have weight zero.
By using \eqref{eq:expK} and \eqref{eq:abeta} and performing partial integrations, 
we obtain
\[
\frac{d}{ds} \rho(\omega,s)=B(\omega) \rho(\omega,s)+B^\mu(\omega) \nabla_\mu\rho(\omega,s)+B^{\mu\nu}(\omega)\nabla_{\mu}\nabla_{\nu}\rho(\omega,s),
\label{eq:diffusionrho}
\]
where
\s[
B(\omega)&=1+\frac{1}{3}\left(\beta^{\mu\nu}(\omega)\left(\frac{1}{\beta(\omega)}\right)_{,\mu\nu} + \beta(\omega) \left(\frac{\beta^{\mu\nu}(\omega)}{\beta^2(\omega)} \right)_{,\mu\nu}  \right),\\
B^\mu(\omega)&= \frac{2}{3}  \left( \beta^{\mu\nu}(\omega)\left(\frac{1}{\beta(\omega)}\right)_{,\nu}+\beta(\omega) \left(\frac{\beta^{\mu\nu}(\omega)}{\beta^2(\omega)} \right)_{,\nu}\right) ,\\
B^{\mu\nu}(\omega)&=\frac{2}{3} \frac{\beta^{\mu\nu}(\omega)}{\beta(\omega)}.
\label{eq:B}
\s]
In general the diffusion equation \eq{eq:diffusionrho} cannot be solved analytically and one would have to rely on numerics.
Rather than doing so, let us restrict ourselves to extracting only short distances by the diffusion equation. 
For this purpose, it is enough to consider a localized initial condition like $\rho(\omega,0)=\delta^D(\omega-\omega_0)$ for arbitrary location $\omega_0$ and a short time period of evolution $0\leq s \ll 1$. 
Since $\rho$ is non-vanishing only in a small distance region around $\omega_0$ under such a short period of time,
one can regard $B$'s as constants, assuming $B$'s are smooth enough in $\omega$. 
Then we can solve \eq{eq:diffusionrho} and obtain
\[
\rho(\omega,s)\simeq \rho_0 \, s^{-\frac{D}{2}}\exp \left [\left( B-\frac{1}{4}B^\mu B^{-1}_{\mu\nu}B^\nu \right)s
-\frac{1}{2}B^\mu B^{-1}_{\mu\nu}\delta \omega^\nu
-\frac{1}{4s}\delta \omega^\mu B^{-1}_{\mu\nu} \delta \omega^\nu
\right],
\label{eq:rhoomegas}
\]
where $\rho_0$ is an overall constant factor, and $\delta \omega=\omega-\omega_0$. 
%This seems like a very complicated expression to actually work with. However if we only look at small distances $\delta \omega = \omega - \omega_0$ we can simplify the expression given in \eqref{eq:B} up to lowest order 
The expressions are still complicated to actually work with, but we can further assume the covariant 
derivatives of $\beta$'s to vanish 
in the homogenous case and obtain 
\s[
B(\omega)&=1, \\
B^\mu(\omega)&=0,\\
B^{\mu\nu}(\omega)&=\frac{2}{3}\frac{\beta^{\mu\nu}(\omega)}{\beta(\omega)}=\frac{2}{3} 
e^{-2 \phi(\omega)}g^{\mu\nu}(\omega).
\label{eq:Bs}
\s] 
Thus the diffusion process can determine a conformally rescaled metric 
$e^{2 \phi(\omega)}g_{\mu\nu}(\omega)$.\footnote{It is curious to note that the metric which appears naturally
in string theory is also the one which is given by a conformal rescaling of the metric in the Einstein frame
with the dilaton field. 
This comes from the fact that the gravitational coupling constant depends on the dilaton field
in string theory \cite{polchinski_1998}.}

A comment is in order. We could have used $d^D\tomega \sqrt{g(\tomega)}$ as the volume element in \eq{eq:diffusion}.
In this case, \eq{eq:Bs} is changed to 
\s[
B(\omega)&=e^{2\phi (\omega)} , \\
B^\mu(\omega)&=0,\\
B^{\mu\nu}(\omega)&=\frac{2}{3}\frac{\beta^{\mu\nu}(\omega)}{\beta(\omega)}e^{2\phi(\omega)}
=\frac{2}{3}g^{\mu\nu}(\omega).
\label{eq:Bsd}
\s] 
So in this case the $\phi$ and $g^{\mu\nu}$ fields completely decouple in the diffusion process.
Namely, this choice gives more direct meaning to the coefficients of the diffusion equation from the point of 
view of the identification of the fields obtained in \cite{Chen:2016ate}.
However, this choice is practically difficult to implement for the discrete case, 
since there is no natural way to know $\sqrt{g}$
while defining the kernel. 
%That is; While the $\beta$ fields are part of the fundamental variables in the CTM,
%so one can expect in advance already that adding anoter variable to the fundamental treatment
%will not translate easily to the discrete case.
Moreover, in the presence of a scalar field, there is no canonical way to take a particular
choice of the metric from the ambiguity of the conformal transformation with the scalar field. 
This is what is called frame dependence, and various choices are possible depending on usages such as 
the Einstein frame normalizing the Einstein term.    
Therefore, we rather use $\beta(\tomega)^{-2}$ for the volume element as above, which 
is much easier to implement in the discrete case, as is done in Section~\ref{sec:distance}. 
Though the fields are not separated in the coefficients in this method, 
it gives a way to extract $\beta$ and $\beta^{\mu\nu}/\beta$
in a straightforward manner and can equivalently determine $\phi$ and $g^{\mu\nu}$ through the relation \eq{eq:betagrel}.

\section{Distances on fuzzy spaces}
\label{sec:distance}
In this section, we discuss the actual process of determining distances between points on fuzzy spaces
by considering discrete analogue of the method developed in Section~\ref{sec:diffusion}.
In fact, due to the difference between continuum and discrete spaces, 
we find an issue that there exist some offsets in the distances determined
by the procedure for the discrete case. We propose a provisional solution to this issue, 
and get acceptable results in the actual application in Section~\ref{sec:relativistic}.
However, a more satisfactory resolution is desirable.

Let us start with the continuum case.  
One can determine distances between nearby\footnote{This is because we are assuming the constancy of the parameters
in Section~\ref{sec:diffusion}.} points by measuring the third term in \eq{eq:rhoomegas}. 
This third term is a damping function in $s$ and generally tiny compared to the first term linear in $s$.
Therefore it is practically (or numerically) difficult to measure, if the first term exists. 
To circumvent the situation, it is more convenient to replace the kernel in \eq{eq:diffusion} by
\[
K(\omega,\tilde \omega)\rightarrow K(\omega,\tilde \omega)-
\beta(\tomega)^2 \delta^D(\omega-\tilde \omega).
\label{eq:tildeK}
\]
Then, assuming the homogenous case \eq{eq:Bs}, 
the problematic first term disappears from \eq{eq:rhoomegas}, as well as the second term. 
Thus, we obtain
\[
\tilde \rho(\omega,s)=\rho_0 \, s^{-\frac{D}{2}}\exp \left (
-\frac{1}{4s}\delta \omega^\mu B^{-1}_{\mu\nu} \delta \omega^\nu
\right),
\label{eq:rhotilde}
\]
where $\tilde \rho$ is the density function after the replacement \eq{eq:tildeK}.
The maximum of $\tilde \rho$ is located at $s=s_{max}$ satisfying 
\[
\delta \omega^\mu B^{-1}_{\mu\nu} \delta \omega^\nu=2 D s_{max}.
\label{eq:maximumofs}
\]
Thus, by measuring $s_{max}$ of diffusion processes, 
one can determine the distance squares between arbitrary nearby points.
Here note that the distances are defined with respect to the conformally rescaled metric 
$e^{2\phi(\omega)}g_{\mu\nu}(\omega)$, as noted below \eq{eq:Bs}.

Now let us discuss the discrete case. 
The discrete analogue of the tensor-rank decomposition to the continuum \eq{eq:cont:decomp} is given by
\[
P_{abc}=\sum_{i=1}^R w_a^iw_b^iw_c^i\, \beta^{-2}(i),
\label{eq:decompmeasure}
\] 
and the diffusion kernel corresponding to \eq{eq:K} is given by
\[
K(i,j)= w_a^iw_a^j.
\]
A non-trivial part is how to determine $\beta(i)$ from the tensor-rank decomposition. We 
consider a self-consistency condition given by 
\[
\sum_{j=1}^R w_a^iw_a^j\,\beta(j)^{-2}=1.
\label{eq:condmeasure}
\]
This is derived from the following continuum counterpart, $\int d^D\tomega\, \beta(\tomega)^{-2} K(\omega,\tomega)=1$,
which can be proven for the homogeneous case due to the vanishing of the derivatives of $\beta$'s. 
Since the relation between the two tensor-rank decompositions,  \eq{eq:CP} and \eq{eq:decompmeasure}, 
is given by
\[
v_a^i=w_a^i \, \beta(i)^{-2/3},
\label{eq:vwrel}
\]
the condition \eq{eq:condmeasure} can be rewritten as
\[
\sum_{j=1}^R v_a^iv_a^j\, \beta(j)^{-4/3}=\beta(i)^{-2/3}.
\label{eq:obtainbeta}
\]
This condition determines $\beta(i)$ from the decomposition \eq{eq:cperror}, and hence $w_a^i$ by \eq{eq:vwrel}.
This process is used in our numerical analysis.

Next, let us consider the discrete version of the diffusion equation \eq{eq:diffusion} with the substitution 
\eq{eq:tildeK}.
This is given by
\[
\begin{split}
\tilde K(i,j)&=\beta(i)^{-1}\beta(j)^{-1} w^i_a w^j_a-\delta_{ij} \sum_{k=1}^R w^i_a w^k_a \beta(k)^{-2}, \\
\frac{d}{ds} \tilde \rho(i,s;i_0)&=\sum_{j=1}^R \tilde K(i,j)\tilde \rho(j,s;i_0), \\
\tilde \rho(i,0;i_0)&=\delta_{i,i_0}. 
\end{split}
\label{eq:tildediffusion}
\]
Here note that the last term in the first line is actually $\delta_{ij}$ by using \eq{eq:condmeasure}.
Note also that $\tilde K(i,j)$ has been defined so that the weight associated to $i,j$ is 1/2 and the same for
 $\tilde \rho$. This is 
to make $\tilde K(i,j)$ symmetric to simplify the following discussions. The other assignments of weights 
would be possible,
like considering the diffusion equation with simpler assignments of weights, 
$d\tilde \rho(i)/ds=\sum_j \tilde K(i,j)\beta(j)^{-2} \tilde \rho(j)$, where $\tilde \rho$ and $\tilde K$ have no weights. 
This weight assignment was actually used in the continuum discussions.
However, this requires us to treat $\tilde K(i,j)\beta(j)^{-2}$,
which is asymmetric and makes things 
non-obvious about eigenvalue problems and the symmetry of distances under mutual permutations of points.
Therefore we employ the above symmetric assignment, which is indeed
equivalent to any asymmetric assignment, because they
are related by a similarity transformation $\beta(i)^{w} \tilde K(i,j) \beta(j)^{-w}$.    

The distance square between arbitrary nearby points, $i$ and $j$, can be determined from $s_{max}(i,j)$ at 
which $\tilde \rho(i,s;j)$ takes the maximum value in $s$. 
Because $\tilde K$ is symmetric, $\tilde \rho(i,s;j)$ and $\tilde \rho(j,s;i)$ 
give the same distance, which guarantees the symmetry of the distances.\footnote{
However, it does not seem guaranteed in general that the distances determined by this procedure 
satisfy the triangle inequality. If the violation occurs in a macroscopic scale, it is a problem, because 
the classical spacetime picture cannot be applied. 
On the other hand, the violation would be allowable on the order of the fundamental scale, where 
the classical spacetime picture is not required to hold.} 

To study time evolutions of fuzzy spaces, we are interested in the time dependence of their sizes. 
For a given fuzzy space, one can perform the tensor-rank decomposition and solve the diffusion equation 
\eq{eq:tildediffusion} to obtain $s_{max}(i,j)$ for arbitrary nearby points. 
This method cannot directly be used for long distances, because of the assumptions made in the derivation 
in Section~\ref{sec:diffusion}. 
One would also think that the distance between two arbitrary points could be determined, even if it is large, by 
considering the shortest path connecting the two points, where its length
is the sum of short lengths along the path.
However, this turns out not to be justified because of the existence of the offsets explained below:
The contribution of the offsets becomes considerable by being multiplied by the number of the short length 
portions along the path.
Instead, since we are only considering homogeneous fuzzy spaces in this paper,
we characterize the local distance structures of the fuzzy spaces and assume them to be proportional to their whole sizes.

To do so, let us first recall the distance introduced in Section~\ref{sec:persistent}, where we have 
discussed topological structures of fuzzy spaces.
The distance is defined to take $d(i,j)=1$, if two points $i$ and $j$ are in their local neighborhoods. 
Let us call this a topological distance and denote it by $d_t(i,j)$, 
because this is determined by topological relations of neighborhoods. 
Here, in the example of the fuzzy space in Figure~\ref{fig:sphere}, the two points $i,j$ with $d_t(i,j)=1$ are those 
which have the inner products $v_a^iv_a^j$ in the range between $0.4$ and $0.6$.  
Then topological distances $d_t(i,j)$ between any points $i,j$ are defined by taking the shortest paths 
as given below \eq{eq:persistent:d_ij}.

\begin{figure}
\begin{center}
\includegraphics[width=.4 \textwidth]{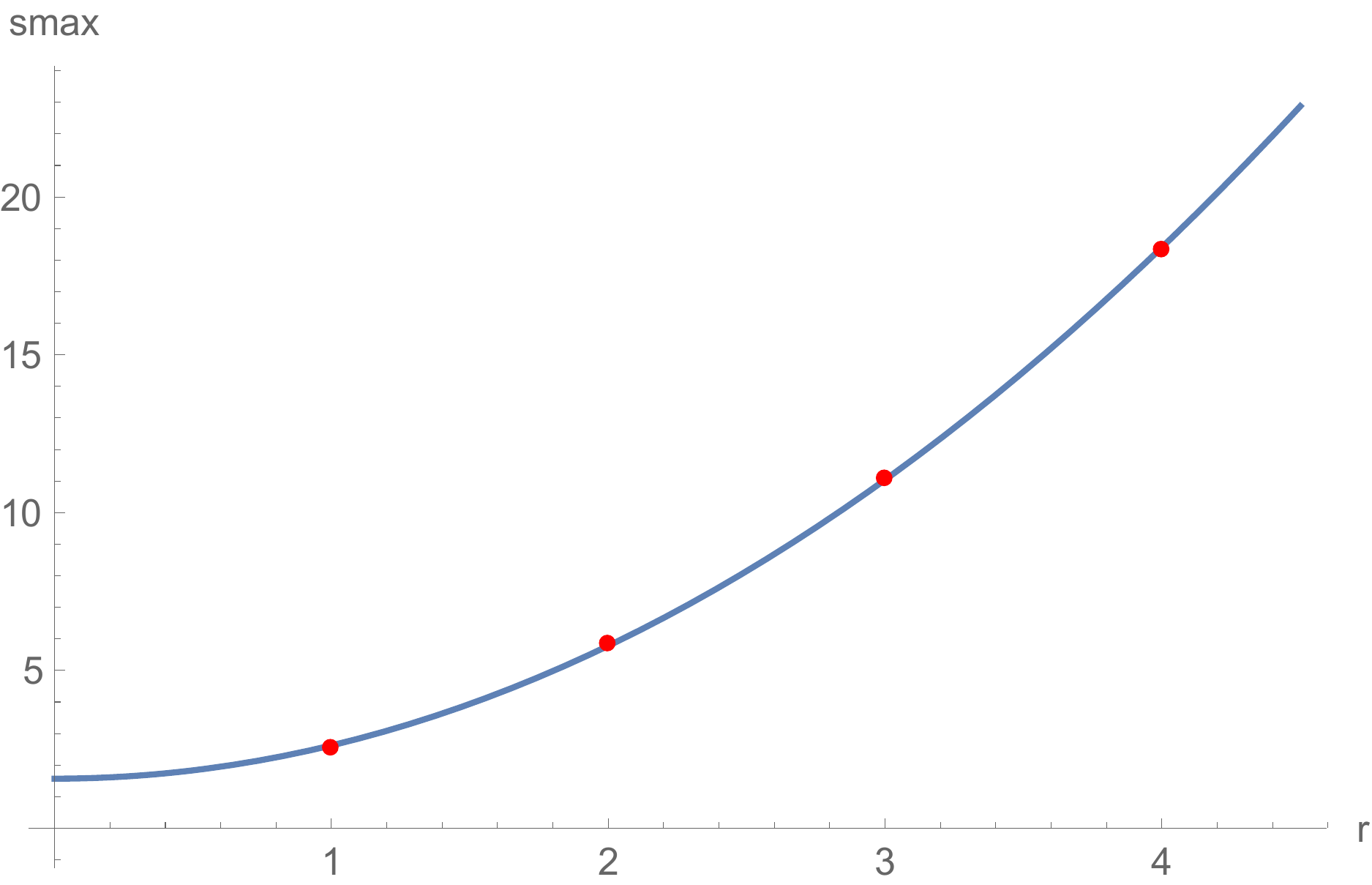}
\hfil
\includegraphics[width=.4 \textwidth]{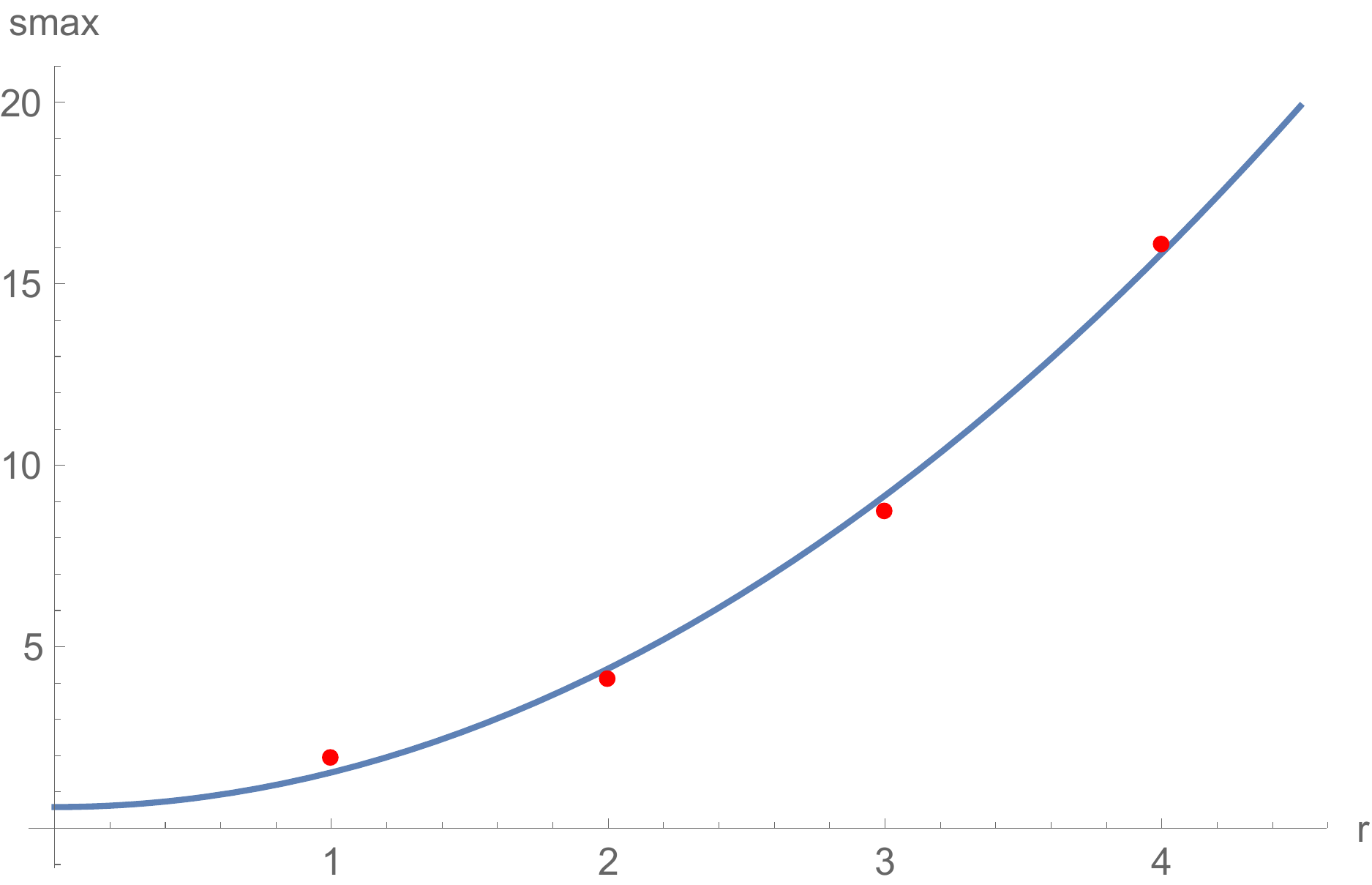}
\end{center}
\caption{
Left: The mean value $s_{max}(r)$ plotted against the topological distance $r$ 
for a fuzzy $S^1$ with $N=31$ and $R=46$. 
Right: The same for a fuzzy $S^2$ with $N=64$ and $R=146$. 
The data are fitted with a quadratic function. While the fitting for $S^1$ is really good, that for $S^2$ seems to 
have small deviations. The deviations probably represent an effect from the curvature on $S^2$, 
but this remains as an open question.
}
\label{fig:distance}
\end{figure}

Now characteristic local distances of a homogeneous fuzzy space can be obtained by considering $s_{max}(r)$, 
which is an average value of $s_{max}(i,j)$ over all the pairs of $i,j$ with $d_t(i,j)=r$. 
This is plotted against $r$ for the examples of fuzzy $S^1$ and $S^2$ in Figure~\ref{fig:distance}.
The data points in the figure are fitted with a quadratic function $a_0+a_1 r^2$ with coefficients $a_0,a_1$. 
In the continuum case, $a_0=0$ and the distance is strictly proportional to $\sqrt{s_{max}}$, but in the present case, 
the offset is non-vanishing, $a_0\neq 0$. This would be understandable because the diffusion process 
is from points to points at small $s$ and it becomes continuous only after larger $s$.  
Therefore the discreteness is apparent in small $s$, and may generate such a difference from the continuous case.
To regard $a_0$ negligible, we have to consider larger $r$ such that $s_{max}(r)\gg a_0$. 
On the other hand, we cannot take $r$ too large because this violates the assumptions made 
in Section~\ref{sec:diffusion}.  In Section~\ref{sec:relativistic}, we take $r=4$ and obtain some acceptable results.

\section{Time evolutions of fuzzy spaces in the CTM}
\label{sec:timeevolution}
As is explained, 
the equation of motion of the CTM gives a first-order differential equation in time for a real symmetric three-way tensor. 
In this section, we regard the solutions to the differential equation as time-evolutions of the tensors corresponding to 
fuzzy spaces, and study some of their elementary properties. One observation is that the time-evolutions 
increase the number of points forming fuzzy spaces starting from one,
in the sense which will be described more precisely. 

The degrees of freedom of the CTM are a pair of canonically conjugate real symmetric three-way tensors, which 
satisfy the fundamental Poisson brackets,
\[
\begin{split}
\{Q_{abc},P_{def}\}&=\sum_{\sigma} \delta_{a\sigma_d} \delta_{b\sigma_e} \delta_{c\sigma_f},\\
\{Q_{abc},Q_{def}\}&=\{P_{abc},P_{def}\}=0,
\end{split}
\] 
where the summation is over all the possible permutations of $d,e,f$ for the consistency with the 
permutation symmetry of the tensors. The classical equation of motion of the CTM is given by
\[
\frac{d X_{abc}}{dt}=\{ X_{abc}, H \},
\label{eq:EOMP}
\]
where $X$ is $Q$ or $P$, and the Hamiltonian $H$ is given by a linear combination of the first-class constraints, 
${\cal H}_a$ and ${\cal J}_{ab}$, as
\[
H={\cal N}_a {\cal H}_a +{\cal  N}_{ab} {\cal J}_{ab}.
\label{eq:CTMham}
\]
Here ${\cal N}_a$ and ${\cal N}_{ab}$ are freely choosable generally time-dependent variables corresponding to
the lapse function and the shift vector in the ADM formalism of general relativity.
The explicit expressions of the constraints are given by
\[
\begin{split}
{\cal H}_a&=\frac{1}{2} P_{abc} P_{bde}Q_{cde}, \\
{\cal J}_{ab}&=\frac{1}{4}\left( Q_{acd}P_{bcd}-Q_{bcd}P_{acd} \right).
\end{split}
\] 
In this paper, we put ${\cal N}_{ab}=0$, since the corresponding term in \eq{eq:CTMham} 
is just a generator of time-dependent $SO(N)$ transformations, which are irrelevant if we 
are only interested in $SO(N)$ invariant quantities like the inner products $v^i_a v^j_a$. 
Then the equation of motion of $P_{abc}$ is given by 
\[
\frac{dP_{abc}}{dt} =-{\cal N}_d \left(P_{dae}P_{ebc} +P_{dbe}P_{eca} +P_{dce}P_{eab} \right). 
\label{eq:eqofP}
\]
In this paper, we do not consider the equation of motion of $Q_{abc}$, 
because the interpretation of the equation of motion in the continuum language (namely, general relativity)
is only known for $P_{abc}$ \cite{Chen:2016ate}.
We also do not consider a term, $\lambda\, Q_{abb}$, which can be added to ${\cal H}_a$, 
because it causes an issue concerning locality in the classical equation of motion of $P_{abc}$ \cite{Sasakura:2015pxa}.

Let us consider the time evolution of the homogeneous fuzzy two-sphere defined in Section~\ref{sec:fuzzyspace}. 
The index set is given by $a=(l,m)$, where 
$l$ and $m$ are integers satisfying $0\leq l \leq L,\ -l\leq m \leq l$ with a cut-off $L$. Let us use $0$ to 
represent the index $(0,0)$ for notational simplicity.
We take the fuzzy two-sphere in \eq{eq:poftwosphere} as the input of $P_{abc}$ at $t=0$.
For the lapse function, we take ${\cal N}_0=\frac{1}{3}$ for convenience, where we also have to take ${\cal N}_a=0$ for
$a\neq 0$ to keep the $SO(3)$ symmetry of the homogeneous fuzzy two-sphere under
the time-evolution.\footnote{Physically, this is to consider the lapse which is uniform on the fuzzy two-sphere, 
or taking the space-like slices in which the evolution is described in a spatially uniform manner.
An important thing here is that, 
because of the first-class nature of the constraints, 
this is just a gauge choice, but not a particular choice of a time-evolution.}

First of all, let us point out an important property general for homogenous cases. 
This has a similarity to the property first pointed out for $N=1$ \cite{Sasakura:2014gia}.  
Due to the $SO(3)$ symmetry, $P_{00a}=0$ for $a\neq 0$.
Then, by putting $a=b=c=0$ in \eq{eq:eqofP},
one can find that the equation of motion of $P_{000}$ in \eq{eq:eqofP} decouples from the others, 
and obtain
\[
\frac{dP_{000}}{dt} =-P_{000} ^2.
\]
The solution is 
\[
P_{000}(t)=\frac{1}{1+t},
\label{eq:psol}
\]
where $P_{000}(0)=1$ has been assumed as the normalization of the initial condition. 
As can be seen in \eq{eq:psol}, the solution diverges at $t=-1$ and monotonically decreases as $t$ increases. 
The time $t=-1$ can be considered to be the time of birth of the fuzzy two-sphere as is explained below.
The time-dependence of the other components can be computed numerically, and we used the Runge-Kutta method
for the purpose.

Once a solution is obtained, one can perform the tensor-rank decomposition of $P_{abc}(t)$ at each $t$ by 
the program described in Appendix~\ref{app:CP}. 
Figure~\ref{fig:Rdepofdelp} plots the error ratio $\sqrt{\Delta P^2/P^2}$ of the decomposition,
where $\Delta P$ is the error of the approximate tensor-rank decomposition in \eq{eq:cperror}.
In Figure~\ref{fig:Rdepofdelp}, $P_{abc}(0)$ is taken to be the fuzzy two-sphere in \eq{eq:poftwosphere} with $L=5$.
The plot shows that the rank $R$ must be increased for larger $t$, if one wants to keep the error ratio 
being suppressed under a certain value.
In the $t\rightarrow -1$ limit, one can numerically find that $P_{000}$ dominates over all the other components, 
meaning that $P_{abc}(t)$ approaches a rank-one tensor. This explains the rapid decaying behavior of the error ratio
in the small $t$ region. By regarding the rank to be equivalent to the number of points forming a space,
the time evolution of $P_{abc}(t)$ can be regarded as that starting from one point at $t=-1$ and gradually increasing 
the number of points.\footnote{This should not be considered as a mathematically rigid statement
and is merely an approximate one, which would be relevant in practical physical applications 
containing errors or quantum fluctuations. 
One can easily see that, if an exact decomposition of $P_{abc}(t)$ is given at one time, the differential equation 
\eq{eq:eqofP} can be rewritten in a closed form with the vectors $v_a^i(t)$ only. 
This means that the rank of $P_{abc}(t)$ is constant in the course of changing $t$. 
Changing rank is needed only if a tensor-rank decomposition is approximate with an error
and one wants to keep the error ratio under a certain value throughout a time-evolution. 
A more mathematically rigid statement is left for future study.}
\begin{figure}
\centerline{
\includegraphics[width=.6\textwidth]{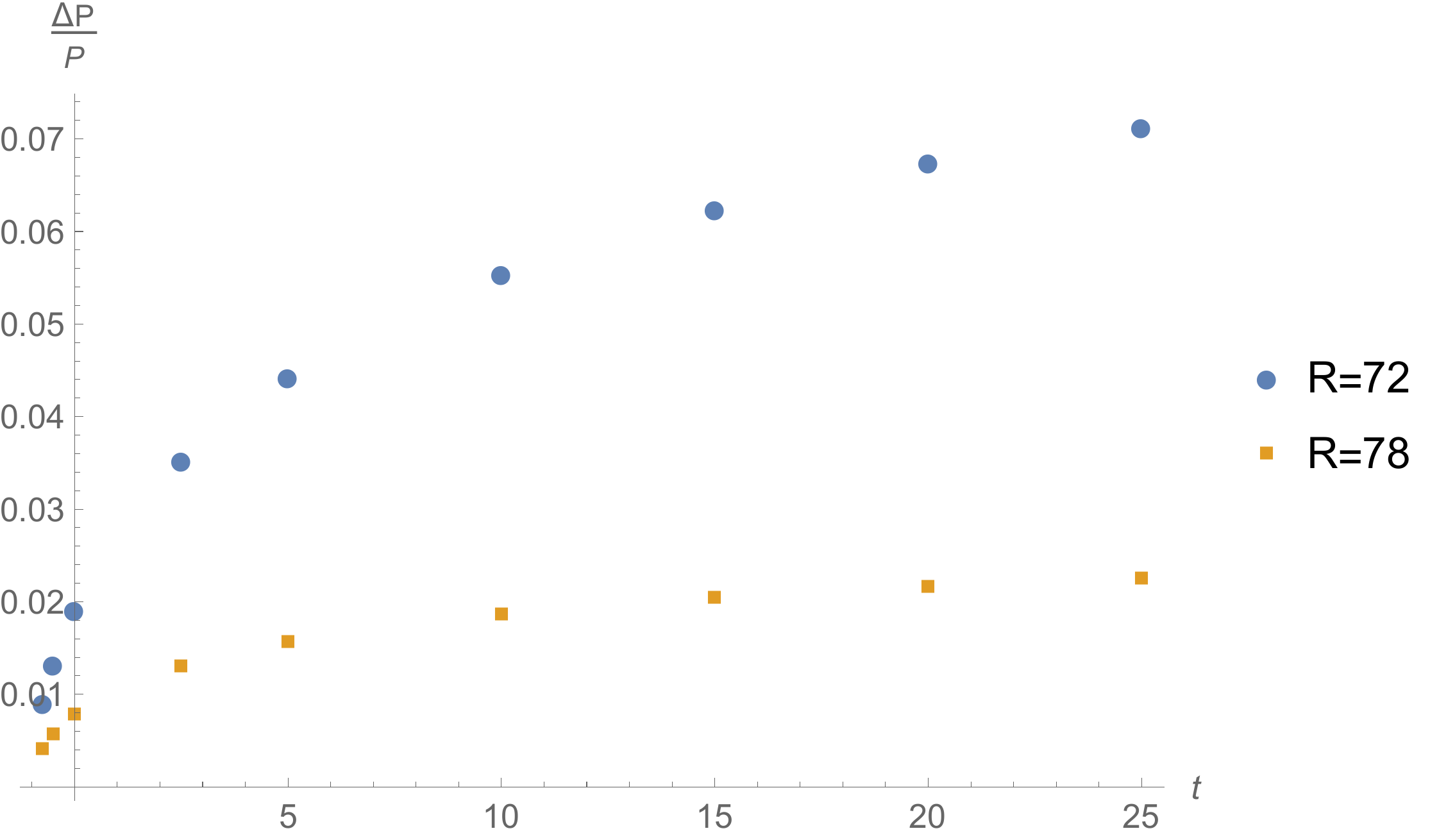}
}
\caption{
The error ratio $\sqrt{\Delta P^2/P^2}$ of the tensor-rank decomposition
of $P_{abc}(t)$ for a fuzzy two-sphere with $L=5$ taken for $P_{abc}(0)$. 
$P_{abc}(t)$ with larger $t$ requires higher ranks to keep the preciseness of the 
approximate tensor-rank decomposition.  
For $R=82$, the error is numerically consistent with zero, suggesting that the actual rank of the tensor is 82.}
\label{fig:Rdepofdelp}
\end{figure}

\begin{figure}
\centerline{
\includegraphics[width=.4\textwidth]{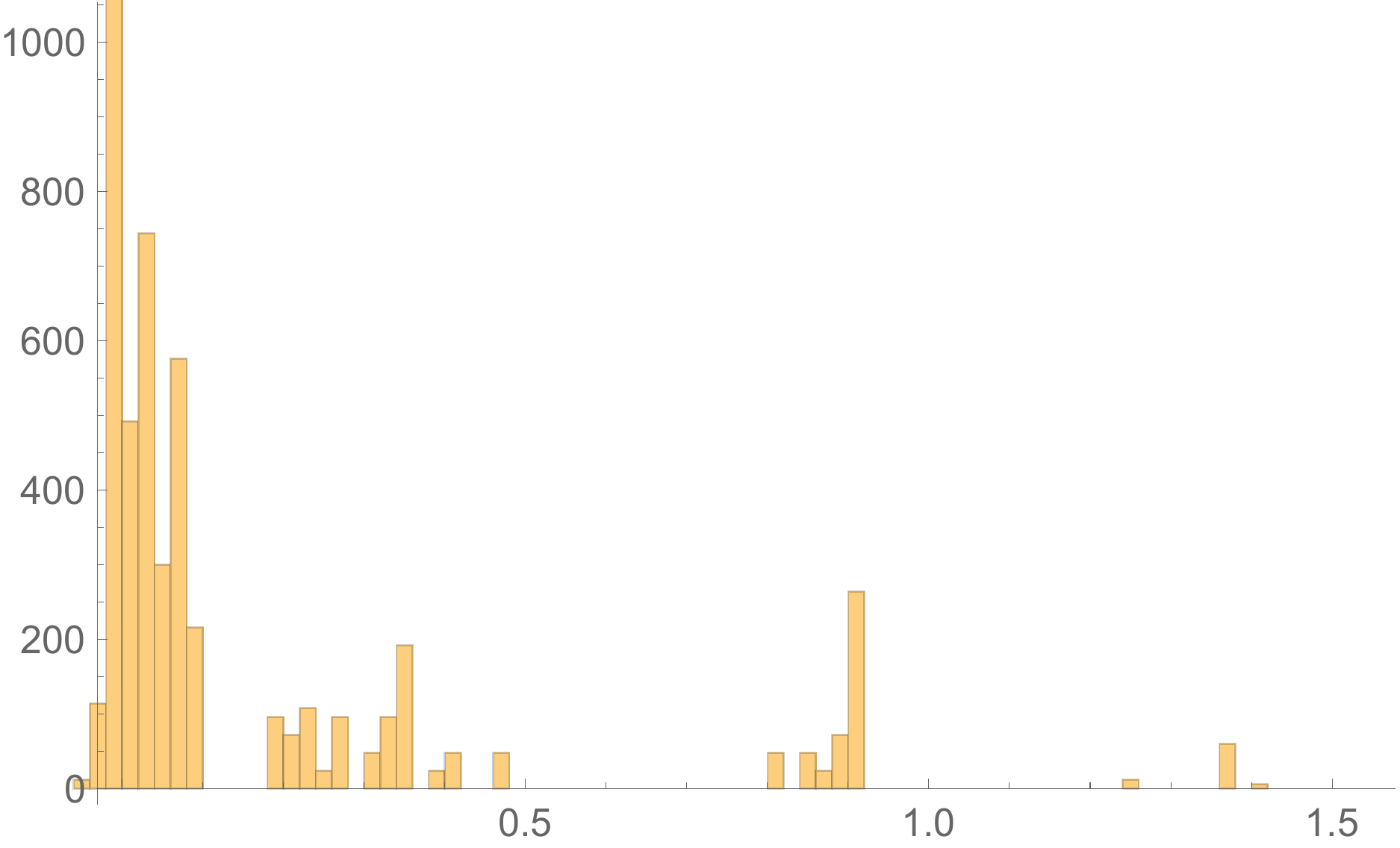}
\hfil
\includegraphics[width=.4\textwidth]{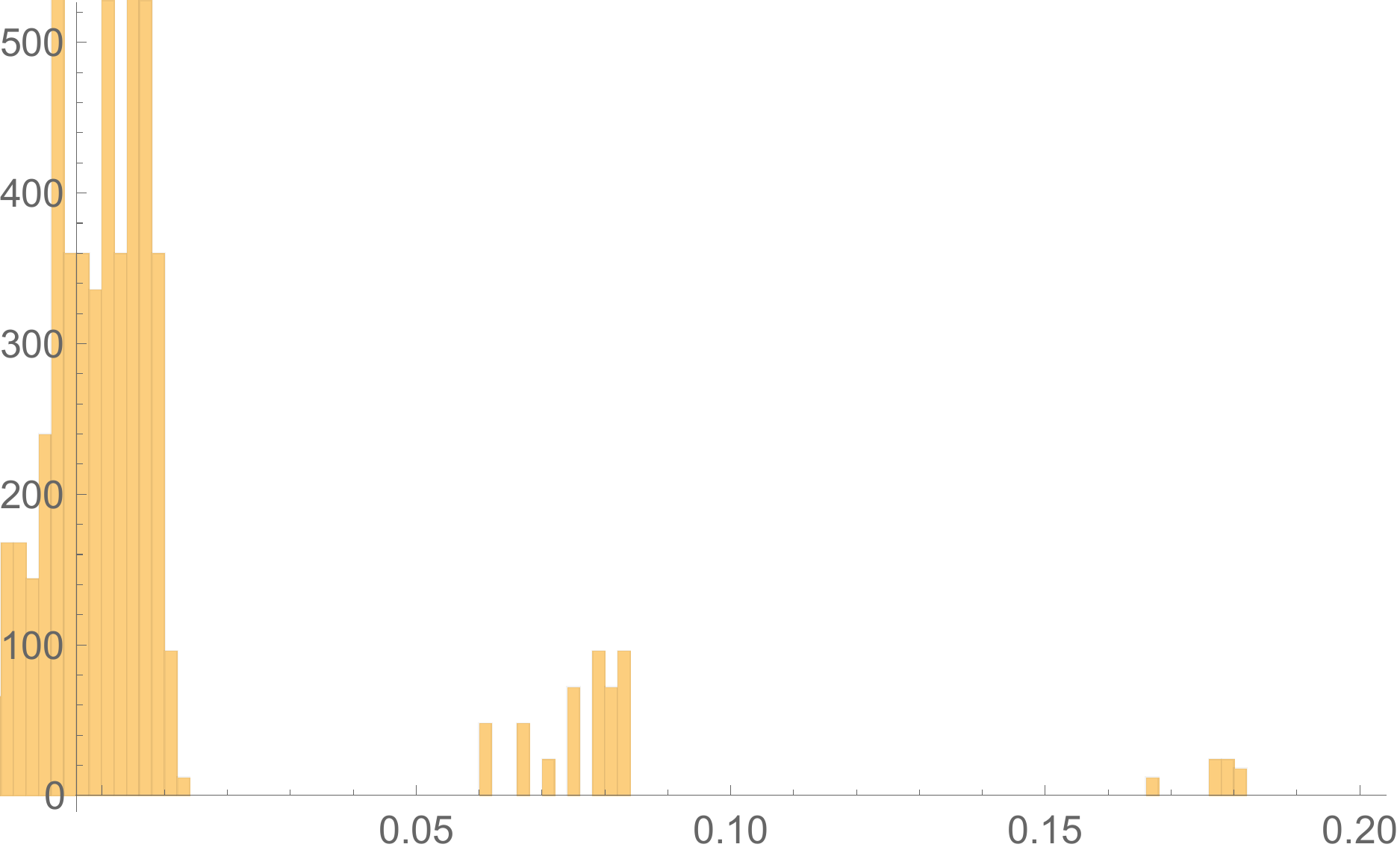}
}
\caption{
Histograms of $v^i_av_a^j$ for a fuzzy two-sphere with $N=36$ and $R=78$.
The left is for $t=-0.75$ and the right for $t=25$. 
}
\label{fig:histn36}
\end{figure}

Another aspect related to the time evolution appears in the inner products $v_a^iv_a^j$.
Figure~\ref{fig:histn36} shows the histograms of the inner products $v_a^iv_a^j$  at $t=-0.75$ and $t=25$ 
for a homogeneous fuzzy two-sphere with $N=36$ and $R=78$. 
The inner products are shifted to the positive values for smaller $t$, meaning that
the distances between points become shorter, and vice versa.  
This is consistent with the rough picture that the two-sphere becomes larger in time, as discussed above
in the sense of the number of points.
In the following section, we perform
more detailed analysis with comparison with the general relativistic system derived in \cite{Chen:2016ate}.   

\section{Correspondence to a general relativistic system} 
\label{sec:relativistic}
A discrete theory of quantum gravity is expected to reproduce a general relativistic system in a certain classical continuum limit.  In \cite{Chen:2016ate}, a general relativistic system corresponding to the CTM has been 
obtained by taking a formal continuum limit, which is to formally replace the discrete values taken by the indices,
$a=1,2,\ldots,N$, to the continuum coordinates, $x \in {\mathbb R}^D$. 
This formal replacement is obviously unsatisfactory from the view point of quantum gravity, because no considerations 
of dynamics to lead to the replacement were made. 
There is not only this difficult problem of dynamics of emergent macroscopic spaces, 
but also another related issue in this
formal continuum limit that it is not given as any limit of $N\rightarrow \infty$ but is rather given as a sudden formal
replacement from the discrete to the continuum indices.  
The question we consider in this section is whether fuzzy spaces with large $N$ 
can well be described by the continuum general relativistic theory obtained previously in \cite{Chen:2016ate} or not. 
We perform some detailed studies of the time evolutions of the homogeneous fuzzy $S^1$, $S^2$ and $S^3$ 
by using the methods developed in the former sections 
and compare the results with the equation of motion of the general relativistic system.     
We obtain good agreement with the continuum theory at least for these homogeneous cases.

Let us first discuss the continuum side. The equations of motion of $\beta(t,x)$ and 
$\beta^{\mu\nu}(t,x)$ in the continuum general relativistic theory are given by \cite{Chen:2016ate}\footnote{
The equations of motion are taken from Section VII of \cite{Chen:2016ate} with  
the consideration of the gauge condition $\beta \beta^{\mu\nu}=g^{\mu\nu}/\sqrt{g}$ and 
the change of the allover minus sign for a convention.}
\[
\begin{split}
\frac{d}{dt} \beta&=-9 n \beta^2, \\
\frac{d}{dt} \beta^{\mu\nu} &=-15 n \beta \beta^{\mu\nu}+2n \beta^{\mu\mu'}
\beta^{\nu\nu'} R_{\mu'\nu'},
\end{split}
\label{eq:eomclassical}
\]
where $n(t,x)$ is the lapse function, $R_{\mu\nu}$ is the Ricci curvature of the space, 
and we have ignored a number of covariant derivative terms
in the original equations, since we are considering homogeneous spaces. 
In the case of homogenous spaces with uniform time evolutions $n(t,x)=n(t)$, 
one can assume that $\beta(t,x),\ \beta^{\mu\nu}(t,x)$ are given by 
products of functions separately depending on time or space.  
Namely, we can write $\beta^{\mu\nu}(t,x)=\beta_2(t) \tilde \beta^{\mu\nu}(x)$ and 
$\beta(t,x)=\beta(t)$, and the second equation in \eq{eq:eomclassical} can be rewritten as
\[
\frac{d}{dt} \beta_2 &=-15\, n \, \beta\, \beta_2 + c_1 \, n\, \beta_2^2,
\label{eq:timebeta2}
\]
where $c_1$ is a constant proportional to the curvature on the space.
The equation for $\beta$ remains the same as in \eq{eq:eomclassical}.
Then, by taking $n=1/9$ for convenience, one can obtain the solution to the above equations as
\[
\beta(t)&=\frac{1}{t-t_0},
\label{eq:betatime}\\
\beta_2(t)&=\frac{\beta_2^0}{(t-t_0)^{\frac{5}{3}}\left(1+c_R(t-t_0)^{-\frac{2}{3}}\right)},
\label{eq:beta2time}
\] 
where $t_0$ is supposed to be the time of birth of the space, $\beta_2^0$ is an integration constant, and $c_R=c_1 \beta_2^0/6$. As explained in Section~\ref{sec:timeevolution}, $t_0=-1$ in our case.
The solution leads to the following time dependence of $B^{-1}_{\mu\nu}$ by taking the inverse of 
the expression in \eq{eq:Bs}:
\[
B^{-1}_{\mu\nu}(t) & \propto \frac{\beta}{\beta_2}\propto c_R+(t+1)^{\frac{2}{3}}.
\label{eq:B2t}
\]

\begin{figure}
\begin{center}
\includegraphics[width=.7 \textwidth]{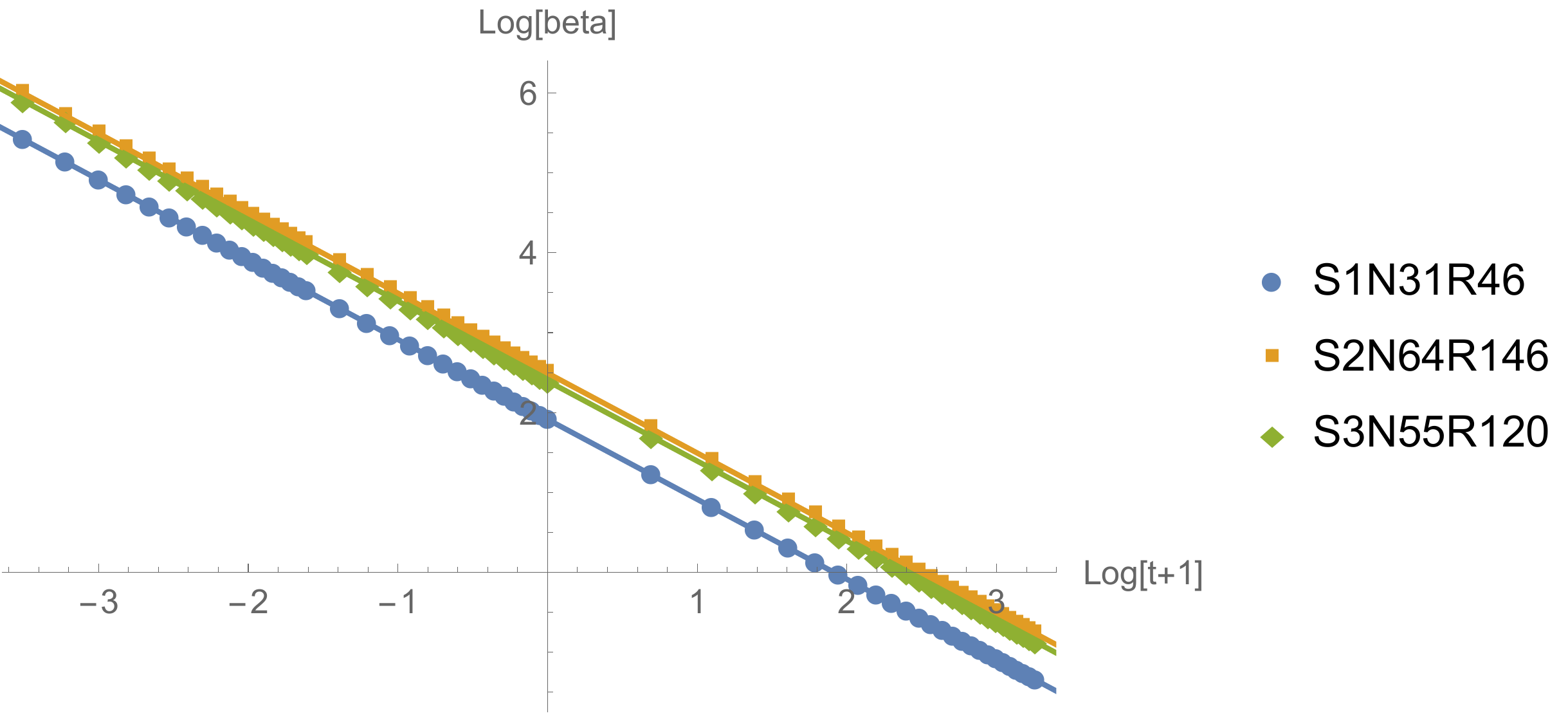}
\caption{Log-log plot of $\beta(t)$. The examples are fuzzy $S^1$ with $N=31,\ R=46$,
fuzzy $S^2$ with $N=64,\ R=146$, and fuzzy $S^3$ with $N=55,\ R=120$. 
The data points are plotted with intervals $0.01$ for $-0.97\leq t \leq -0.8$,  
$0.05$ for $-0.8\leq t \leq 0$, and 1 for $0\leq t \leq 25$,
to distribute the data points more or less evenly in log scale. 
The data are fitted with linear functions, where the gradients agree with $-1$ with the precisions down to 
the three decimal places.}
\label{fig:plotbeta}
\end{center}
\end{figure}

Let us compare the solution with the time evolutions of the fuzzy spaces.
We consider homogeneous fuzzy $S^1$ with $N=31,\ R=46$, $S^2$ with $N=64,\ R=146$,
and $S^3$ with $N=55,\ R=120$. 
We set the tensors corresponding to these fuzzy spaces explained in Section~\ref{sec:fuzzyspace} (see
Section~\ref{sec:constP} for the details of $S^3$) as the initial conditions at $t=0$ of the equation of motion of the CTM
shown in \eq{eq:eqofP}, and 
numerically obtained the solutions $P_{abc}(t)$ by the manner explained in Section~\ref{sec:timeevolution}.
Then we performed the tensor-rank decompositions of $P_{abc}(t)$
for a number of representative values of $t$.
Finally $\beta(i,t)$ were determined by solving \eq{eq:obtainbeta}. 
Figure~\ref{fig:plotbeta} plots the mean values, $\beta(t)\equiv \frac{1}{R} \sum_{i=1}^R \beta(i,t)$,
in log-log plot.
The gradients agree with $-1$ with the precisions down to the three decimal places, 
giving perfect agreement with \eq{eq:betatime}.

\begin{figure}
\begin{center}
\includegraphics[width=.45 \textwidth]{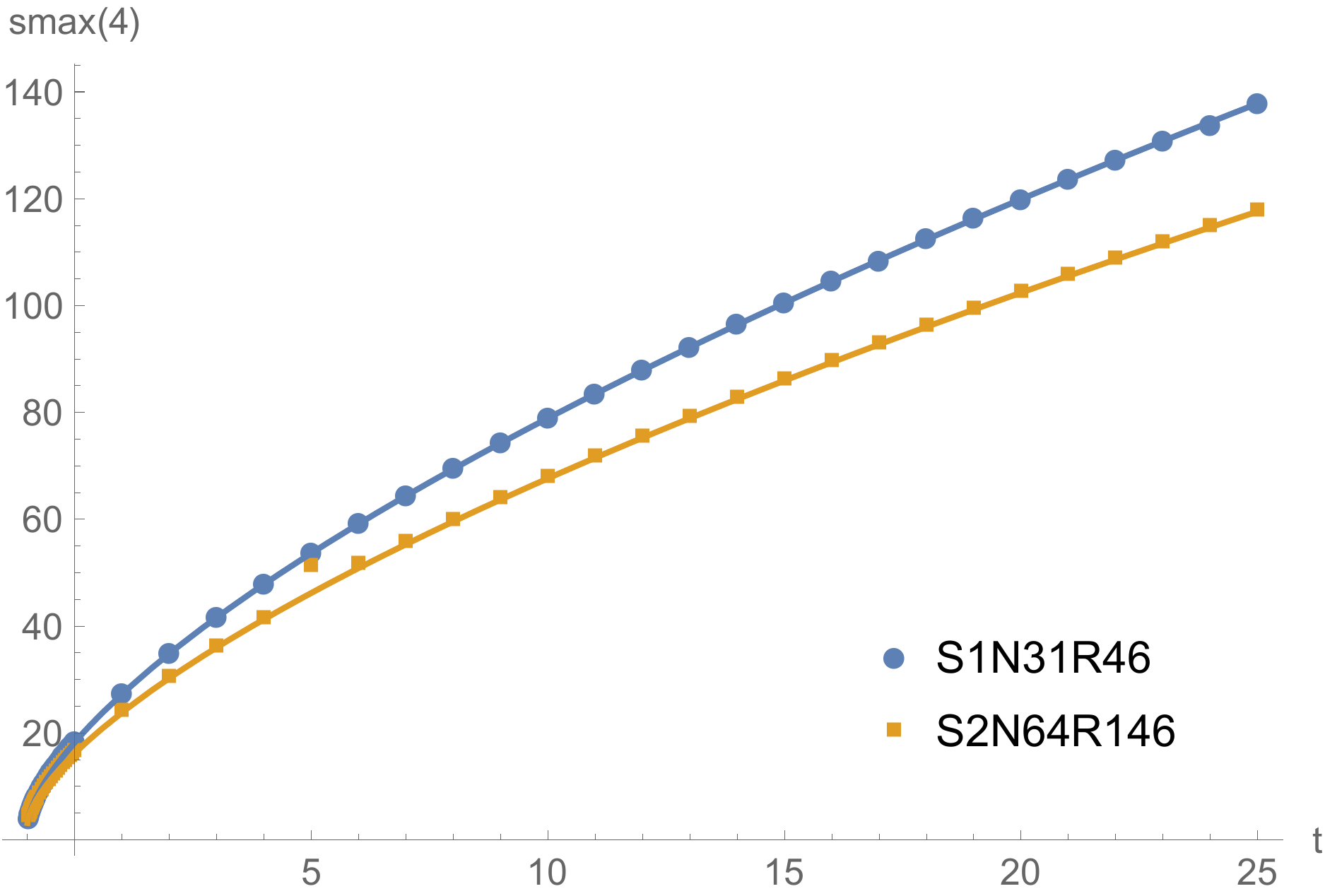}
\hfil
\includegraphics[width=.45 \textwidth]{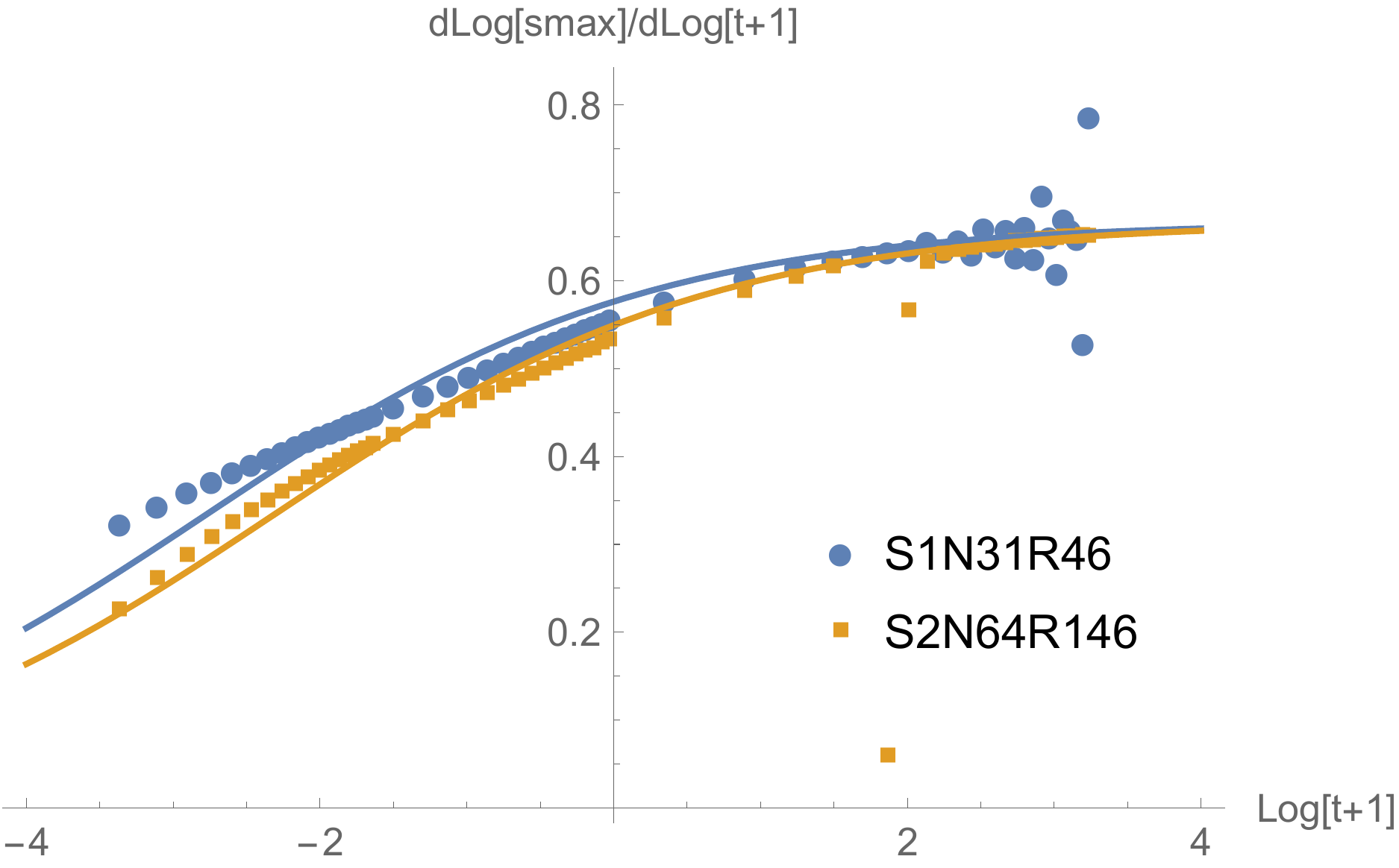}
\caption{Left: $s_{max}(4)$ is plotted against $t$ for fuzzy $S^1$ with $N=31,\ R=46$
and fuzzy $S^2$ with $N=64,\ R=146$.  The data are fitted with $a_0 (c_R+(1+t)^{2/3})$ (see \eq{eq:B2t})
with $c_R= 0.19,\ 0.24$, respectively. 
Right: The logarithmic derivatives of the left data obtained by subtracting the sequential data. 
They are fitted with the logarithmic derivative of \eq{eq:B2t}.
%The data of the fuzzy two-sphere show anomalous behavior around $t=5$, which seemed to come from 
%some singular behavior of the solution to \eq{eq:obtainbeta}. This aspect was not further explored.
This computation was not done for $S^3$ with $N=55,\ R=120$, 
since its size is too small to consider $s_{max}(4)$ as a reliable quantity (4 is well more than the half-size of $S^3$).
}
\label{fig:plotB2}
\end{center}
\end{figure}

As for $B^{-1}_{\mu\nu}(t)$, we have obtained the results shown in Figure~\ref{fig:plotB2}.
The left figure shows the time dependence of $s_{max}(4)$ defined in Section~\ref{sec:distance}.
This is expected to be proportional to $B^{-1}_{\mu\nu}$, and therefore the data are fitted with \eq{eq:B2t}.
The agreement is rather nice with non-zero values of $c_R$. 
This seems to contradict the supposed origin of $c_R$, 
since $S^1$ does not have a curvature.
The right figure shows the logarithmic derivative of the data obtained by subtracting the sequential data, and 
they are fitted with the corresponding derivative of \eq{eq:B2t}. 
The results of the fitting are not nice but only barely acceptable. 

The method above uses the short-time behavior of the virtual diffusion process, and is supposed to 
determine short distance structures of fuzzy spaces by measuring $s_{max}(i,j)$ between nearby points.
If the method is fully reliable, one should be able to determine the time-dependence of the whole size 
of a homogeneous fuzzy space up to an allover factor by the local distance structure, 
because they should be proportional.
On the other hand,  we want to believe the validity of the continuum theory, 
because a fuzzy space with large $N$ is made of many points and is expected to allow a continuum description.
We seem to have a tension between the measuring method and the continuum theory.

\begin{figure}
\begin{center}
\includegraphics[width=.3 \textwidth]{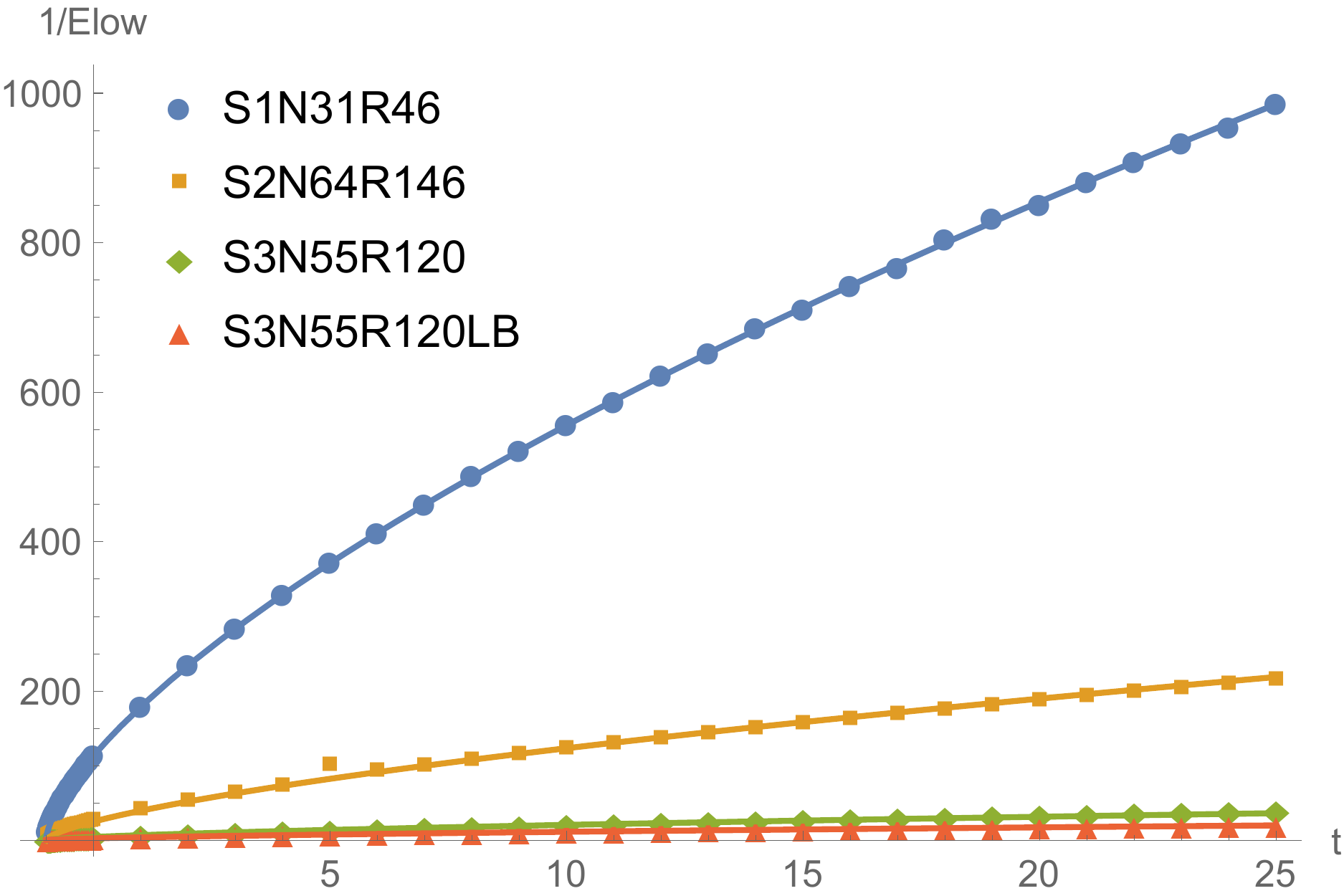}
\hfil
\includegraphics[width=.3 \textwidth]{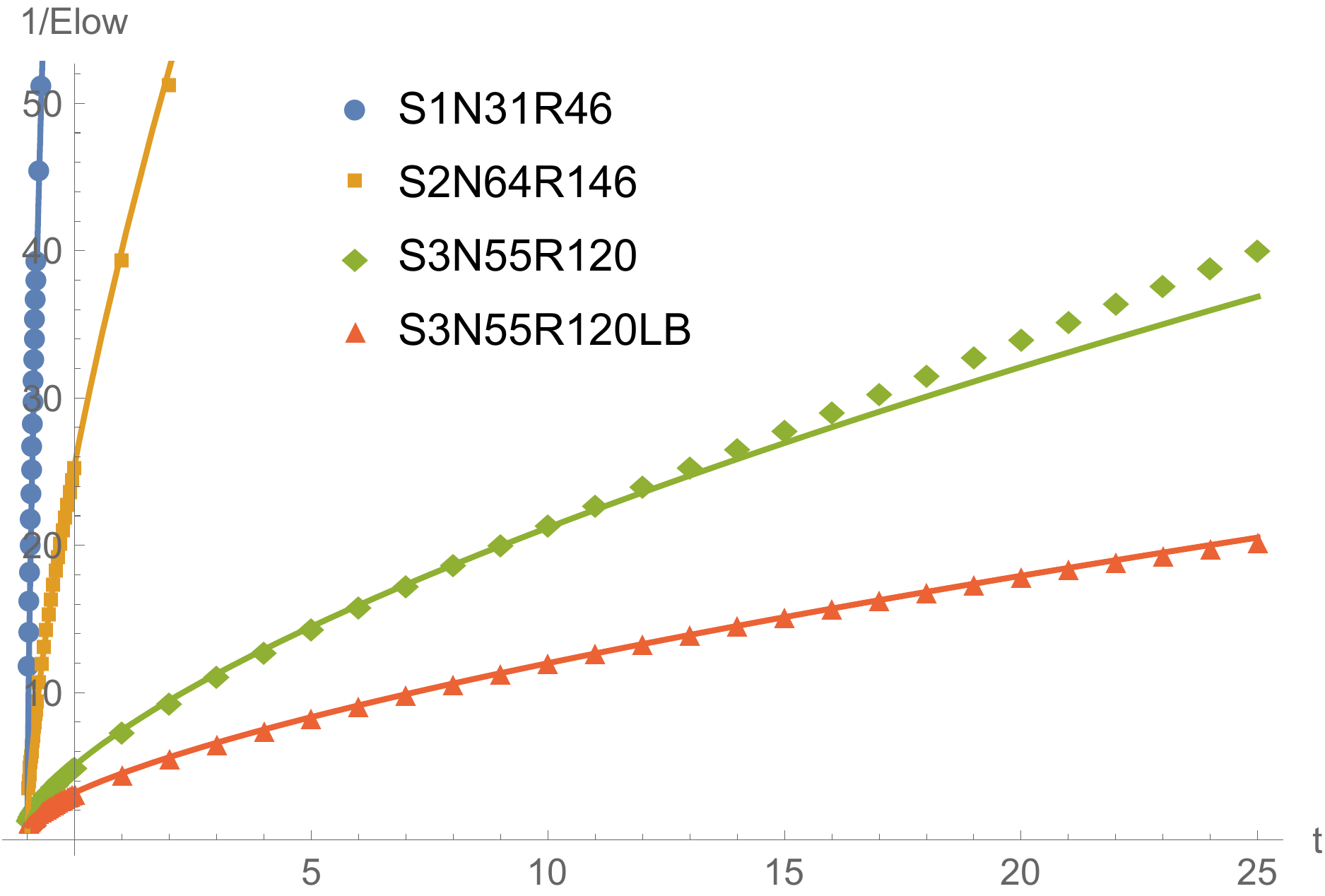}
\hfil
\includegraphics[width=.35 \textwidth]{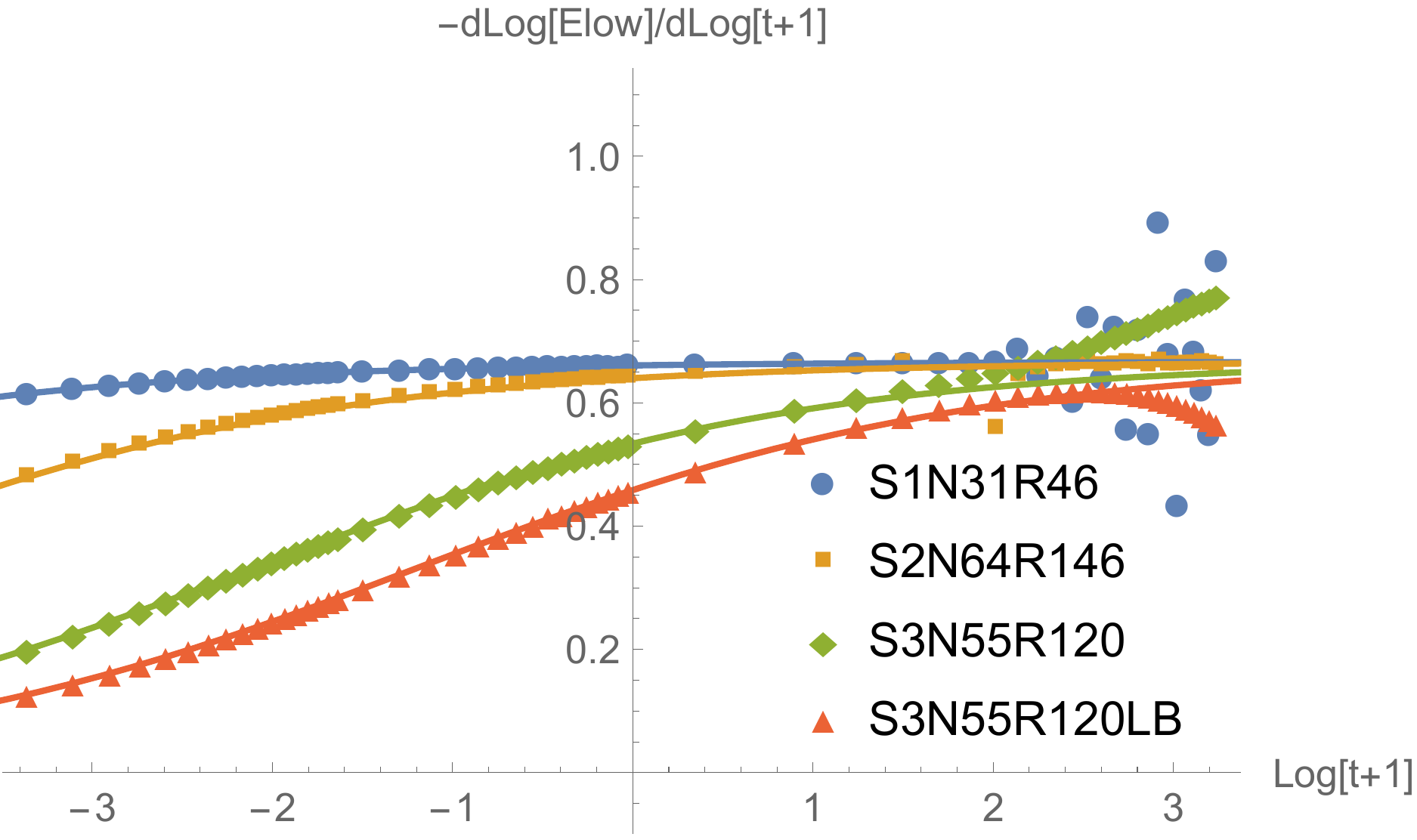}
\caption{Left: Time-dependence of the inverse of the lowest non-zero eigenvalue of $-\tilde K(i,j)$
in \eq{eq:tildediffusion} for homogeneous fuzzy $S^1$ with $N=31,\ R=46$, $S^2$ with $N=64,\ R=146$,
and two $S^3$'s with $N=55,\ R=120$ but with different damping factors.
Middle: The same graph with a different vertical axis scale to show more clearly the latter two cases. 
Right: Logarithmic derivatives of the left data obtained by subtracting the sequential data. 
The fitting lines are the logarithmic derivatives of \eq{eq:B2t} with 
$c_R=0.009,\ 0.04,\ 0.25,\ 0.45$, respectively.
 }
\label{fig:plotlaplacian}
\end{center}
\end{figure}

To study the issue from a different angle, let us measure the whole size in a different manner using
the lowest eigenvalue of a laplacian. The lowest non-zero eigenvalue of the minus of a laplacian is expected to
be proportional to the inverse of the square size of a space.
In our case, the inverse of the lowest non-zero eigenvalue of $-\tilde K(i,j)$ 
in \eq{eq:tildediffusion} is expected to behave in the same manner as $B^{-1}_{\mu\nu}$.
The left and the middle of Figure~\ref{fig:plotlaplacian} plot the data of the inverse, 
and they are fitted with \eq{eq:B2t}.
The right one shows the logarithmic derivatives of the data
obtained by subtracting the sequential data, and they are fitted with the logarithmic derivative of 
\eq{eq:B2t}. They are in much better agreement with the continuum theory than Figure~\ref{fig:plotB2}.

The two results above seem to conclude that the continuum theory is right, but our former method of 
measuring local distances for sizes is not fully reliable at least in our situation, while the latter method of using the lowest
eigenvalue is.
In fact, we have already pointed out the subtlety of the former method in Section~\ref{sec:distance}.
The former method would become more reliable and interesting for much larger fuzzy spaces with
points existing more densely enough to validate the continuum description.
There is also another possibility that the short-length dynamics is actually different from the global one.  
Note that the main contributors in the former method are the modes with large eigenvalues of $-\tilde K$ 
in \eq{eq:tildediffusion} because of short diffusion time, while the latter is the lowest one. 
   
Another interesting thing in Figure~\ref{fig:plotlaplacian} in comparison with Figure~\ref{fig:plotB2} 
is that the value of $c_R$ for $S^1$ has substantially decreased 
from the former method to the latter, while it keeps a more or less similar value for $S^2$.
This seems to suggest that a large portion of $c_R$ for $S^1$ comes from the small scale
rather than the global scale. 
This would explain the presence of $c_R$ even for $S^1$, which has no curvatures,
in the following sense.
As derived in \cite{Chen:2016ate}, the right-hand side of the second equation in
\eq{eq:eomclassical} actually contains the terms like $\beta^{\mu\mu'} \beta^{\nu\nu'}(\nabla_{\mu'} \beta)(\nabla_{\nu'} \beta)/\beta^2$ and $\beta^{\mu\nu} \beta^{\mu'\nu'}(\nabla_{\mu'}\nabla_{\nu'}\beta)/\beta$,
which can potentially contribute to $c_R$.
We have ignored these terms because of the homogeneity of the spaces,
but the discreteness locally violates this assumption in short distances.
Therefore, it does not seem obvious that these terms can really be ignored
in short distances.

Though the agreement of the fitting in Figure~\ref{fig:plotlaplacian} is really good
especially in the small $t$ region,
there exist some deviations in the large 
$t$ region, as can clearly be seen in the right figure. 
In fact, the continuum description cannot be expected to be right in this region, 
because the sizes of the fuzzy spaces are so large that the points exist sparsely on them, 
or, in other words, discreteness is macroscopic.  
To see this from a different viewpoint, 
we show the data from two fuzzy $S^3$'s with $N=55,\ R=120$ but with different damping factors. 
The damping factor used for the data S3N55R120 is $e^{-l^2/L^2}$ with $L=3$,
where $l$ denotes the angular momenta of the modes. 
On the other hand, for S3N55R120LB, it is $e^{-l(l+2)/L^2}$ with $L=4$,
where $l(l+2)$ comes from the eigenvalues of the Laplace-Beltrami operator on $S^3$.
It is clearly seen that the behaviors are qualitatively different from each other in the large $t$ region. 
This implies that,
while the small $t$ region can well be described by the continuum theory irrespective of the 
damping factor, the large $t$ region can be affected by small-distance details of fuzzy spaces.  
This seems to be consistent with the fact that the discrete structure is macroscopic in the large $t$ region. 
It would be an interesting future problem how such deviations from the continuum can be described.

\section{Generality of real symmetric three-way tensors}
\label{sec:constP}
In section \ref{sec:fuzzyspace}, we explained how to calculate the 3-way tensor $P_{abc}$ by using an example of 2-sphere $S^2$, and showed the realization of homogeneous fuzzy 2-sphere. In this section we show that this methodology can be applied to other fuzzy spaces by using some low dimensional manifolds with various topologies.
The main point of this section is to show the generality of real symmetric three-way tensors by these demonstrations 
and consequently the generality of the CTM. 
One can explicitly see that real symmetric three-way tensors can in principle represent any spaces with free choices of 
dimensions and topologies. 
So the way such tensors realize spaces is essentially distinct from that in the other Euclidean tensor 
models \cite{Ambjorn:1990ge,Sasakura:1990fs,Godfrey:1990dt,Gurau:2009tw},
in which the dimensions of building simplicial blocks are supposed to be equivalent to the numbers of ways 
(the amount of indices) of tensors.
 
First let us summarize the procedure to construct a fuzzy space corresponding to a compact manifold $\mathcal{M}$:
\begin{enumerate}
\item Take a coordinate $x^\mu$ and a positive-definite metric $g_{\mu\nu}$ on the considering manifold $\mathcal{M}$. 
\item Prepare a set of real basis functions $\left\{f_a(x)\right\}$ on the manifold. It is convenient to impose an orthonormalization condition: for all combination $a,b$,
\[
\int_{\mathcal{M}} \mathrm{d}^Dx\sqrt{g} f_a(x)f_b(x)=\delta_{ab},
\]
where $g=\det(g_{\mu\nu})$.
Since the dimension of the function space is infinite in general, we have to choose a finite subset from the complete 
basis suited for a practical purpose. There are no general procedures for that, but in each individual case, 
there is a proper one. 
For example, let us suppose that basis functions $f_a(x)$ are taken to satisfy the Helmholtz equation,
\[
(\Delta+m_a^2)f_a(x)=0,
\label{eq:KGlike}
\]
where $\Delta$ is Laplace-Beltrami operator on the manifold $\mathcal{M}$.
Here $m_a$ plays the role as a ``frequency'' associated to each value $a$ of the indices, and provides a natural way 
to choose a subset from the complete basis by $\left\{f_a(x)\,|\,m_a^2\leq \Lambda^2\right\}$ with some parameter $\Lambda$. 
This $\Lambda$ determines the part of the basis which is considered, and effectively determines the value of $N$.
The physical reason for considering such lower frequency modes than a cut-off is that  
we are interested in defining a space which is modified in a small scale but keeps its ordinary properties otherwise. 
\item Define ``regularized'' functions $\tilde{f}_a(x)$ from $f_a(x)$. There also exist a freedom in the way to regularize, 
but in the case $f_a(x)$ satisfies (\ref{eq:KGlike}), a natural definition of $\tilde{f}_a(x)$ is
\[
\tilde{f}_a(x)=f_a(x) e^{-m_a^2/L^2}=e^{\Delta/L^2}f_a(x)\label{eq:damping}
\] 
with some damping scale $L$. It is good to choose $L\lesssim \Lambda$ in general.
Here, the damping factor can be another function damping with $m_a^2$ or with a similar damping behavior.
As discussed in Section~\ref{sec:neighbor}, this regularization smoothens the cutoff and is important for locality of 
fuzzy spaces and good behavior of the virtual diffusion process.
\item Calculate $P_{abc}$ by using
\[
P_{abc}=\int_{\mathcal{M}} \mathrm{d}^Dx\sqrt{g} \tilde{f}_a(x)\tilde{f}_b(x)\tilde{f}_c(x).
\]
\end{enumerate}
Then $P_{abc}$ defines a fuzzy space. 
It is expected that the tensor-rank decomposition of the tensor and connecting neighboring points
will give a discretized counterpart of the manifold $\mathcal{M}$.
\subsection{Spheres}
The square integrable functions on an $n$-dimensional sphere can be represented by a linear combination of $n$-dimensional (generalized) spherical harmonics. 
Therefore, let us take the generalized spherical harmonics as the set of the orthonormal basis functions on $S^n$.

Let us start with some setups~\cite{doi:10.1063/1.527513}. Let us choose local coordinates such that the metric tensor on $S^n$ (with radius $r=1$) is given by
\[
(g_{\mu\nu})&=\mathrm{diag}\left[1,\sin^2\theta_1,\sin^2\theta_1\sin^2\theta_2,\ldots,\prod_{i=1}^{n-1}\sin^2\theta_i\right],\\
\theta_i&\in
\begin{cases}
[0,\pi]&i=1,\ldots,n-1\\
[0,2\pi]&i=n
\end{cases},
\]
where $\mu,\nu=1,\ldots,n$. One can obtain Laplace-Beltrami operator in local coordinates for any Riemannian manifold by
\[
\Delta=\frac1{\sqrt{g}}\partial_\mu\left(\sqrt{g} g^{\mu\nu}\partial_\nu\right),
\]
where $g=\det(g_{\mu\nu})$.
Then the $n$-dimensional spherical harmonics $Y_{l_1l_2\cdots l_n}(\theta_1,\ldots,\theta_n)$ are defined as 
the solutions of this equation:
\[
[\Delta+l_n(l_n+n-1)]Y_{l_1l_2\cdots l_n}(\theta_1,\ldots,\theta_n)=0,
\] 
where all $l_i$'s are integer and $|l_1|\leq l_2\leq \ldots \leq l_{n-1}\leq l_n$ is satisfied.
Some explicit formulas to represent $Y_{l_1l_2\ldots l_n}(\theta_1,\ldots,\theta_n)$ are known,
but we rather used a Mathematica package \cite{packharmonic}, which can produce 
the set of $n$-dimensional spherical harmonics automatically for any $n$ and $l_n$. 
By using this $n$-dimensional spherical harmonics, one can obtain the orthonormal basis functions in our previous notation by
\[
f_{(l_1,\ldots,l_n)}(\theta_1,\ldots,\theta_n)=Y_{l_1l_2\ldots l_n}(\theta_1,\ldots,\theta_n),
\]
and the ``regularized'' basis functions can be defined by
\[
\tilde{f}_{(l_1,\ldots,l_n)}(\theta_1,\ldots,\theta_n)=Y_{l_1l_2\ldots l_n}(\theta_1,\ldots,\theta_n)e^{-l_n^2/L^2}
\]
with a damping scale $L$. 
It is also possible to take $-l_n(l_n+n-1)/L^2$ as the exponent of the damping factor, faithfully following  \eq{eq:damping}.
%We note that this basis naturally takes in the orthogonality of spherical harmonics.
\subsubsection{Circle $S^1$}
We take for the coordinate on $S^1$, $\theta\in[0,2\pi]$ and $\sqrt{g}=1$. The set of basis functions is
\[
f_a(\theta)=\left(\frac{1}{\sqrt{2\pi}},\left\{\frac{1}{\sqrt{\pi}}\sin n\theta\right\},\left\{\frac{1}{\sqrt{\pi}}\cos n\theta\right\}\right). \label{eq:S1}
\]
The notation like $\left\{\sin n\theta\right\}$ is the abbreviation of $\left\{\sin n\theta\right|n\in\mathbb{N}^+\}$. Since these functions satisfy (\ref{eq:KGlike}), we can use the procedure (\ref{eq:damping}) to regularize the basis and the results are
\[
\tilde{f}_a(\theta)=\left(\frac{1}{\sqrt{2\pi}},\left\{\frac{1}{\sqrt{\pi}}\sin n\theta \,e^{-n^2/L^2}\right\},\left\{\frac{1}{\sqrt{\pi}}\cos n\theta\, e^{-n^2/L^2}\right\}\right),
\label{eq:S1damping}
\]
with $n\in\mathbb{N}^+$. $P_{abc}$ can be calculated by
\[
P_{abc}=\int_0^{2\pi}\mathrm{d}\theta \tilde{f}_a(\theta)\tilde{f}_b(\theta)\tilde{f}_c(\theta),
\]
and one can get homogeneous fuzzy circles, which look like polygons, from this $P_{abc}$.
\subsubsection{Three-dimensional sphere $S^3$}
\begin{figure}
\centering
\includegraphics[width=.4\textwidth]{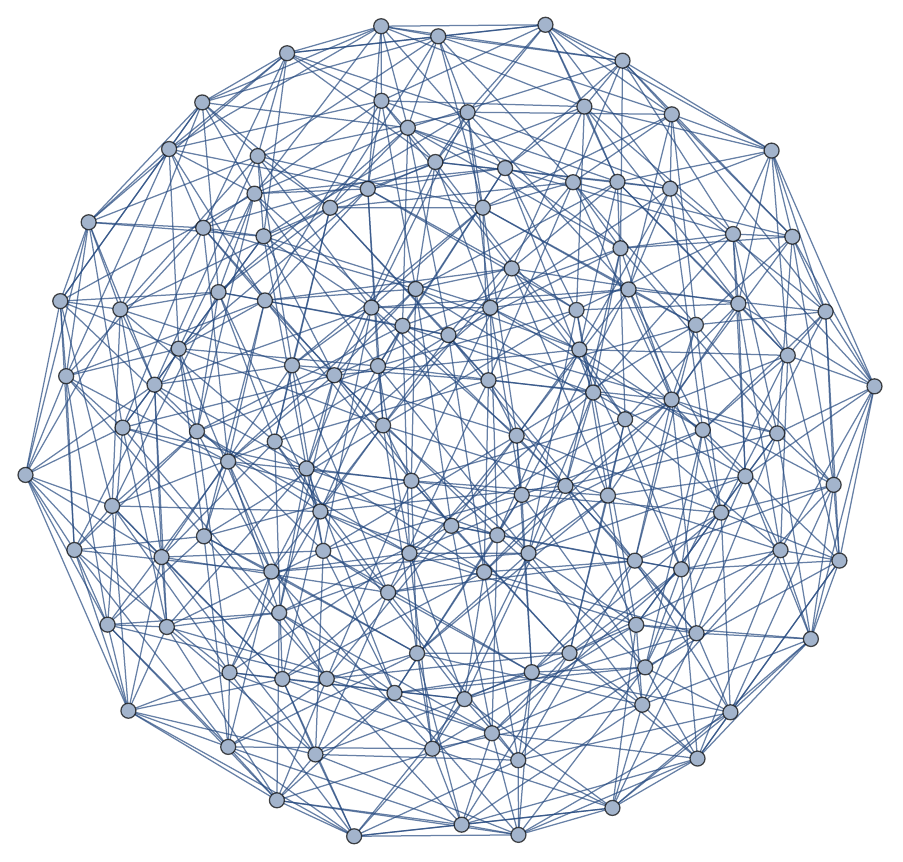}
\caption{The homogeneous fuzzy 3-sphere with $N=55$ and $R=120$. 
The edges are drawn between the neighboring points by the criterion $v_a^iv_a^j>0.2$, after
the tensor-rank decomposition.
}
\label{fig:S3}
\end{figure}
\begin{figure}
\centering
\includegraphics[width=.4\textwidth]{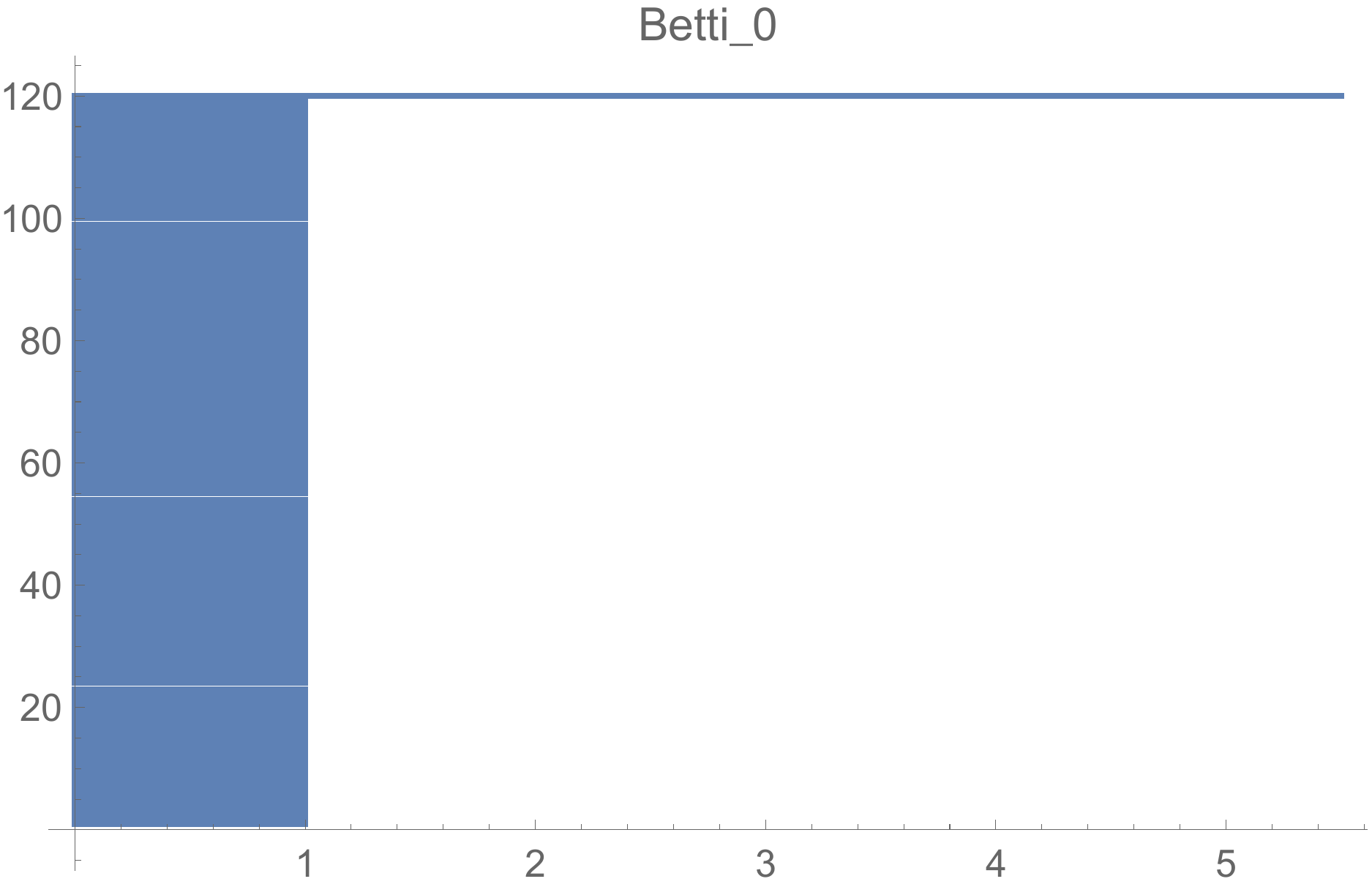}
\hfil
\includegraphics[width=.4\textwidth]{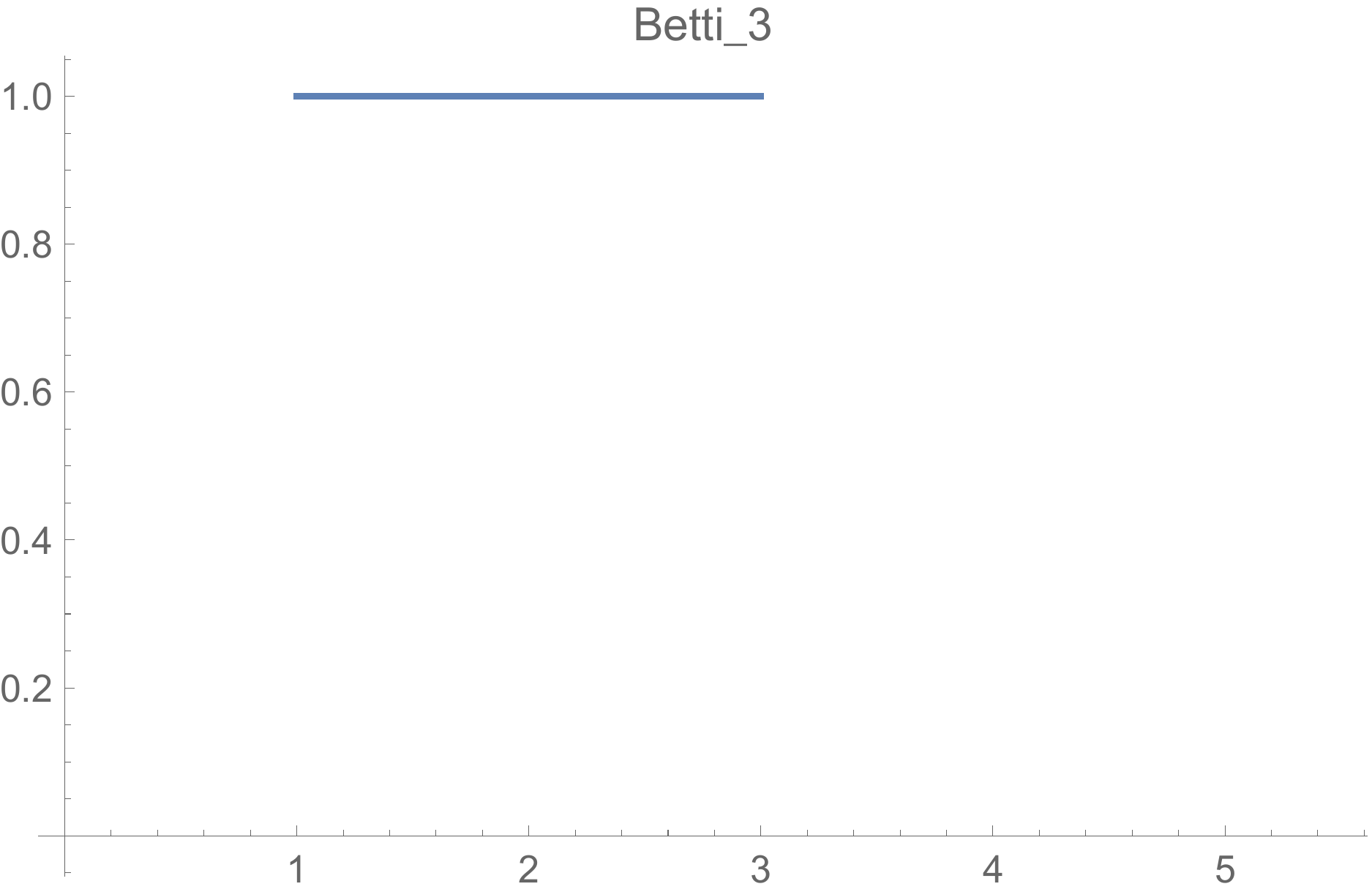}
\caption{The $\mathbb{Z}_2$-coefficient Betti intervals for the fuzzy 3-sphere. There are no intervals 
for $H_{1,2,4}$.}
\label{fig:S3interval}
\end{figure}

Figure \ref{fig:S3} shows a homogeneous fuzzy 3-sphere obtained from the three-way tensor constructed
from the above procedure. For this, we took $n=3$, and $l_3$ was taken up to $l_3\leq 4$,
which resulted in $N=55$. 
The tensor-rank decomposition was carried out with $R=120$, and the points have connections if $v_a^iv_a^j>0.2$
in Figure~\ref{fig:S3}.
Though it is really hard to recognize this object as $S^3$,
the topological data analysis method discussed in Section~\ref{sec:persistent}, namely, persistent homology, 
is quite helpful.
The analysis of Betti intervals is shown in Figure~\ref{fig:S3interval}. 
The result tells the homology groups to be
\s[
&H_0(S^3,\mathbb{Z}_2)=\mathbb{Z}_2,\\
&H_3(S^3,\mathbb{Z}_2)=\mathbb{Z}_2,
\s]
and ${\rm dim}(H_{1,2,4})=0$. The homology groups agree with those of the 3-sphere, 
supporting our construction procedure of the fuzzy 3-sphere.

\subsection{Line segments}
One can also consider manifolds with boundaries.
In the case of spaces with boundaries, one needs to set boundary conditions for its basis functions.
There exist various choices such as Dirichlet, Neumann, their mixtures, and so on. 
In the following analysis, we simply consider the standard Dirichlet and Neumann boundary conditions.

Let us consider line segments.
We take the coordinate of a line segment to be given by $x\in[-\pi,\pi]$ and $\sqrt{g}=1$.
In the case of Dirichlet boundary condition, one imposes  $f^{\mathrm{D}}_a(\pm\pi)=0$ for all $a$, and then one finds two types of functions: $\sin(nx)$ and $\cos((n-1/2)x)$ where $n\in\mathbb{N}^+$.
Thus the set of an orthonormal basis can be taken to be 
\[
f_a^{\mathrm{D}}(x)=\left(\left\{\frac{1}{\sqrt{\pi}}\cos (n-1/2)x\right\},\left\{\frac{1}{\sqrt{\pi}}\sin nx\right\}\right).
\]
In the case of Neumann boundary condition, 
$\left.\frac{\mathrm{d}}{\mathrm{d}x}f^{\mathrm{N}}_a(x)\right|_{x=\pm\pi}=0$,
there are three types of functions: the constant function, $\cos(nx)$ and $\sin((n-1/2)x)$ with $n\in\mathbb{N}^+$.
Then the set of an orthonormal basis can be taken to be 
\[
f_a^{\mathrm{N}}(x)=\left(\frac{1}{\sqrt{2\pi}},\left\{\frac{1}{\sqrt{\pi}}\sin (n-1/2)x\right\},\left\{\frac{1}{\sqrt{\pi}}\cos nx\right\}\right).
\]

The $P_{abc}$ can be computed from the regularized functions $\tilde{f}^{\mathrm{N,D}}_a(x)$. 
The regularization factor can be taken for example to be $\exp(-k^2/L^2)$ for a trigonometric function with frequency $k$.
From these regularized functions, the three-way tensors can be obtained by 
\[
P^{\mathrm{N,D}}_{abc}=\int_{-\pi}^\pi \mathrm{d}x\, \tilde f_a^{\mathrm{N,D}}(x) \tilde f_b^{\mathrm{N,D}}(x)
\tilde f_c^{\mathrm{N,D}}(x),
\]
where we are supposed to take a finite number of low-frequency modes. 
We have explicitly checked that fuzzy line segments are obtained from the tensor-rank decompositions
of these tensors for both Dirichlet and Neumann boundary conditions. 

\subsection{Fiber bundles}
In this subsection, let us construct more non-trivial fuzzy spaces, namely from fiber bundles. 

\subsubsection{Trivial bundles}
%Non-orientable manifolds, for example M\"obius strip or Klein bottle, can't construct from simple manifolds like above. So
%when we consider some unorientable space, we have to investigate basis functions by indivisual way.
Let us consider a manifold $\mathcal{M}$ which is isomorphic to the Cartesian product of two manifolds 
$\mathcal{M}_1$ and $\mathcal{M}_2$. There exist the following relations:
\[
\begin{array}{|c|c|c|c|}\hline
\mathrm{Manifold}&\mathcal{M}_1&\mathcal{M}_2&\mathcal{M}=\mathcal{M}_1\times\mathcal{M}_2\\ \hline
\mathrm{Coordinate}&x_1&x_2&x=(x_1,x_2)\\ \hline
\mathrm{Index}&a_1&a_2&a=(a_1,a_2)\\ \hline
\mathrm{Basis}&f_{a_1}(x_1)&f_{a_2}(x_2)&f_{a}(x)=f_{a_1}(x_1)f_{a_2}(x_2)\\ \hline
\mathrm{Tensor}&P_{a_1b_1c_1}&P_{a_2b_2c_2}&P_{abc}=P_{a_1b_1c_1}P_{a_2b_2c_2}\\ \hline
\end{array}\nonumber
\]
The basis functions $f_{a}(x)=f_{a_1}(x_1)f_{a_2}(x_2)$ on $\mathcal{M}$ are normalized 
properly by the normalizations on $\mathcal{M}_1$ and $\mathcal{M}_2$. 

As an example, let us consider the flat two-torus $T^2 \cong S^1\times S^1$. Using the basis on $S^1$ given in \eqref{eq:S1}, the set of the orthonormal basis functions on $T^2$ is given by
\s[
f_a(\theta_1,\theta_2)=&\biggl(\biggr.\frac{1}{2\pi},\left\{\frac{1}{\sqrt{2}\pi}\sin n\theta_1\right\},\left\{\frac{1}{\sqrt{2}\pi}\cos n\theta_1\right\},\\
&\left\{\frac{1}{\sqrt{2}\pi}\sin m\theta_2\right\},\left\{\frac{1}{\pi}\sin n\theta_1\sin m\theta_2\right\},\left\{\frac{1}{\pi}\cos n\theta_1\sin m\theta_2\right\}\\ 
&\left\{\frac{1}{\sqrt{2}\pi}\cos m\theta_2\right\},\left\{\frac{1}{\pi}\sin n\theta_1\cos m\theta_2\right\},\left\{\frac{1}{\pi}\cos n\theta_1\cos m\theta_2\right\}\biggl.\biggr),
\s]
where $\theta_1,\theta_2\in[0,2\pi]$ and $n,m\in\mathbb{N}^+$. The regularized basis functions 
$\tilde{f}_a(\theta_1,\theta_2)$ can be obtained by using (\ref{eq:damping}), but this can also be 
obtained by the product of the regularized basis functions (\ref{eq:S1damping}) of $S^1$:
$\tilde{f}_{a}(\theta_1,\theta_2)=\tilde{f}_{a_1}(\theta_1)\tilde{f}_{a_2}(\theta_2)$.
Then the three-way tensor $P_{abc}$ can be obtained by 
\[
P_{abc}=\int_{[0,2\pi]^2}\mathrm{d}\theta_1\mathrm{d}\theta_2 \tilde{f}_a(\theta_1,\theta_2)\tilde{f}_b(\theta_1,\theta_2)\tilde{f}_c(\theta_1,\theta_2),
\]
which defines a fuzzy two-torus.

\subsubsection{M\"obius strip}
The M\"obius strip is an example of a nontrivial bundle, which is a bundle of a line segment over a circle. It can practically be built by considering a square and gluing a pair of opposite edges with a twist. 
From this we can easily find the conditions on the functions $f(x,y)$ on a M\"obius strip. 
We assume $x,y\in[-\pi,\pi]$ and suppose that the edges at $y=\pm\pi$ are glued with a twist in the $x$ direction. 
Then the condition on these edges gives the periodic boundary condition,
\[
f(x,y)&=f(-x,y+2\pi).\label{eq:twist1}
\]
The boundary condition on $x=\pm\pi$ can be freely chosen for instance from the Dirichlet boundary condition,
\[
f(\pm\pi,y)&=0,\label{eq:Dirichlet}
\]
or the Neumann boundary condition,
\[
\left.\frac{\partial}{\partial x}f(x,y)\right|_{x=\pm\pi}&=0,\label{eq:Neumann}
\]
for all $y\in[-\pi,\pi]$.
Using the periodic condition (\ref{eq:twist1}) twice, one has
\[
f(x,y)=f(x,y+4\pi),
\]
so $f(x,y)$ can be expanded by a linear combination of $\left\{g_m(x)e^{imy/2}\right\}$ 
with integer $m$ and functions of $x$, $g_m(x)$. 

If one imposes Dirichlet boundary condition (\ref{eq:Dirichlet}), $g_m(x)$ is further restricted to be a linear combination
of $\left\{\cos(n-1/2)x,\sin nx\right\}$ with $n\in \mathbb{N}^+$. 
Therefore the set of the basis functions is a subset of the Cartesian product of $\left\{\cos(n-1/2)x,\sin nx\right\}$ 
and $\left\{1,\cos my/2,\sin my/2\right\}\ (n,m\in \mathbb{N}^+)$. 
Finally, by taking into account  (\ref{eq:twist1}), 5 types of orthonormal basis functions are obtained:
\s[
f^{\mathrm{D}}_a(x,y)=&\biggl(\biggr.\left\{\frac{1}{\sqrt{2}\pi}\cos(n-1/2)x\right\},\\
&\left\{\frac{1}{\pi}\cos(n-1/2)x \cos my\right\},\left\{\frac{1}{\pi}\sin nx\cos(m-1/2)y\right\},\\
&\left\{\frac{1}{\pi}\cos(n-1/2)x\sin my\right\},\left\{\frac{1}{\pi}\sin nx\sin(m-1/2)y\right\}\biggl.\biggr)
\s]
with $n,m\in\mathbb{N}^+$. 

In the case of imposing Neumann boundary condition (\ref{eq:Neumann}), the basis functions are given by 
a subset of the Cartesian product of $\left\{1, \sin(n-1/2)x,\cos nx\right\}$ and 
$\left\{1,\cos my/2,\sin my/2\right\}$ $(n,m\in \mathbb{N}^+)$ by a similar argument. 
After taking into account (\ref{eq:twist1}), 8 types of orthonormal basis functions are obtained:
\s[
f^{\mathrm{N}}_a(x,y)=&\biggl(\frac{1}{2\pi},\biggr.\left\{\frac{1}{\sqrt{2}\pi}\cos nx\right\},\left\{\frac{1}{\sqrt{2}\pi}\cos my\right\},\left\{\frac{1}{\sqrt{2}\pi}\sin my\right\},\\
&\left\{\frac{1}{\pi}\cos nx \cos my\right\},\left\{\frac{1}{\pi}\sin(n-1/2)x\cos(m-1/2)y\right\},\\
&\left\{\frac{1}{\pi}\cos nx\sin my\right\},\left\{\frac{1}{\pi}\sin(n-1/2)x\sin(m-1/2)y\right\}\biggl.\biggr)
\s]
with $n,m\in\mathbb{N}^+$. 
The regularized basis functions $\tilde{f}_a(x,y)$ can be obtained by the procedure (\ref{eq:damping}),
and one obtains the three-way tensors by
\[
P^{\mathrm{N,D}}_{abc}=\int_{[-\pi,\pi]^2} \mathrm{d}x\mathrm{d}y\, \tilde f_a^{\mathrm{N,D}}(x,y) \tilde f_b^{\mathrm{N,D}}(x,y)\tilde f_c^{\mathrm{N,D}}(x,y).
\]
We have checked that connecting neighboring points by the result of the tensor-rank decompositions of the $P$ produces discrete analogues of the M\"obius strip for both Dirichlet and Neumann boundary conditions.

\subsubsection{Klein bottle $K^2$}
The Klein bottle is another nontrivial bundle of a circle over a circle and can be constructed by considering a square, gluing one pair of opposite edges, and gluing the other pair with a twist. This procedure tells us how to obtain the basis functions on a Klein bottle. 
We again take the coordinates $(x,y)\in[-\pi,\pi]^2$ and suppose that the glued edges with twisting correspond 
to those at $y=\pm\pi$. Then the periodic boundary conditions are given by 
\[
f(x,y)&=f(x+2\pi,y),\\
f(x,y)&=f(-x,y+2\pi).\label{eq:twist2}
\]
Using the condition~\eqref{eq:twist2} twice, one obtains
\[
f(x,y)=f(x,y+4\pi).
\]
So we see that $f(x,y)$ can be expanded in a linear combination of $\exp(inx+imy/2)$ $(n,m\in\mathbb{Z})$. 
Taking a real basis and requiring \eqref{eq:twist2}, 8 types of orthonormal basis functions are obtained:
\s[
f_a(x,y)=&\biggl(\biggr.\frac{1}{2\pi},\left\{\frac{1}{\sqrt{2}\pi}\cos nx\right\},\left\{\frac{1}{\sqrt{2}\pi}\cos my\right\},\left\{\frac{1}{\sqrt{2}\pi}\sin my\right\},\\
&\left\{\frac{1}{\pi}\cos nx\cos my\right\},\left\{\frac{1}{\pi}\sin nx\cos (m-1/2)y\right\},\\
&\left\{\frac{1}{\pi}\cos nx\sin my\right\},\left\{\frac{1}{\pi}\sin nx\sin (m-1/2)y\right\}\biggl.\biggr)
\label{eq:KBbasis}
\s]
with $n,m\in\mathbb{N}^+$. 
By the procedure explained before at \eqref{eq:damping}, one can obtain regularized basis functions $\tilde{f}_a(x)$ and the tensor $P_{abc}$ from them.

\begin{figure}
\centering
\includegraphics[width=.4\textwidth]{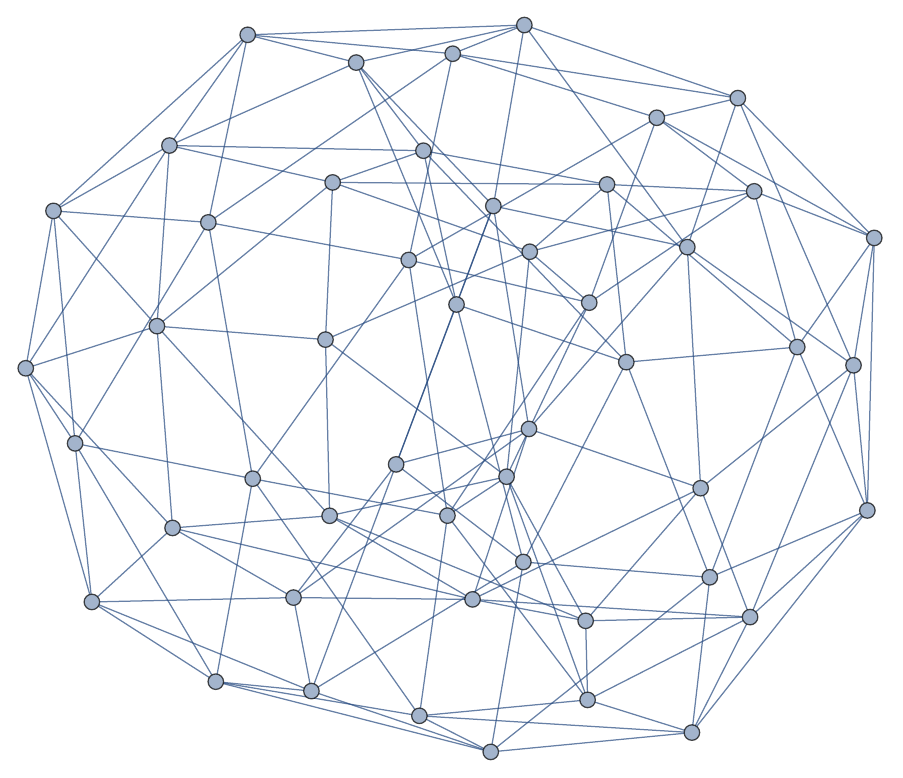}
\caption{The fuzzy Klein bottle with  $N=49$,  $R=49$.
The edges are drawn between the neighboring points by the criterion $v_a^iv_a^j>0.05$, after 
the tensor-rank decomposition.}
\label{fig:KL}
\end{figure}

Figure \ref{fig:KL} shows the fuzzy Klein bottle with $N=49$,  the damping scale $L=3$, and the tensor rank $R=49$.
Here $N=49$ comes from setting the parameter $\Lambda$ below \eq{eq:KGlike} by $\Lambda=4$. More explicitly, 
in the case $\Lambda=4$, 49 is the summation of the numbers of the modes as $N=1+4+4+4+8+10+8+10=49$,
where the summands are ordered in the same way as in the expression (\ref{eq:KBbasis}). Note that the numbers of 
the combinations $(n,m)\in \mathbb{N}^+\times\mathbb{N}^+$ which satisfy $n^2+m^2\leq4^2$ and $n^2+(m-1/2)^2\leq4^2$ are 8 and 10, respectively.
The object in Figure \ref{fig:KL}  
can be seen as a discretized two-dimensional closed surface with the structure of self-intersection,
which is the characteristics of Klein bottle. 
\begin{figure}
\centering
\includegraphics[width=.3\textwidth]{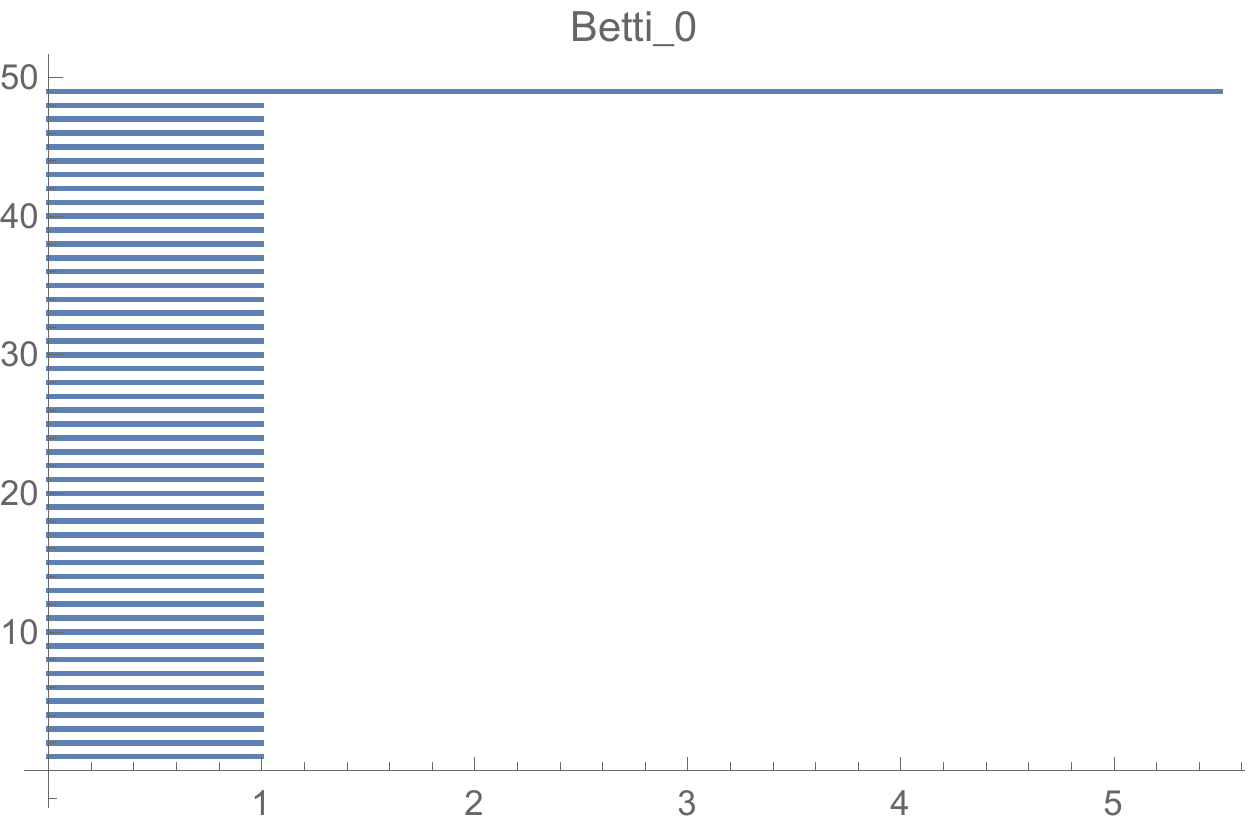}
\hfil
\includegraphics[width=.3\textwidth]{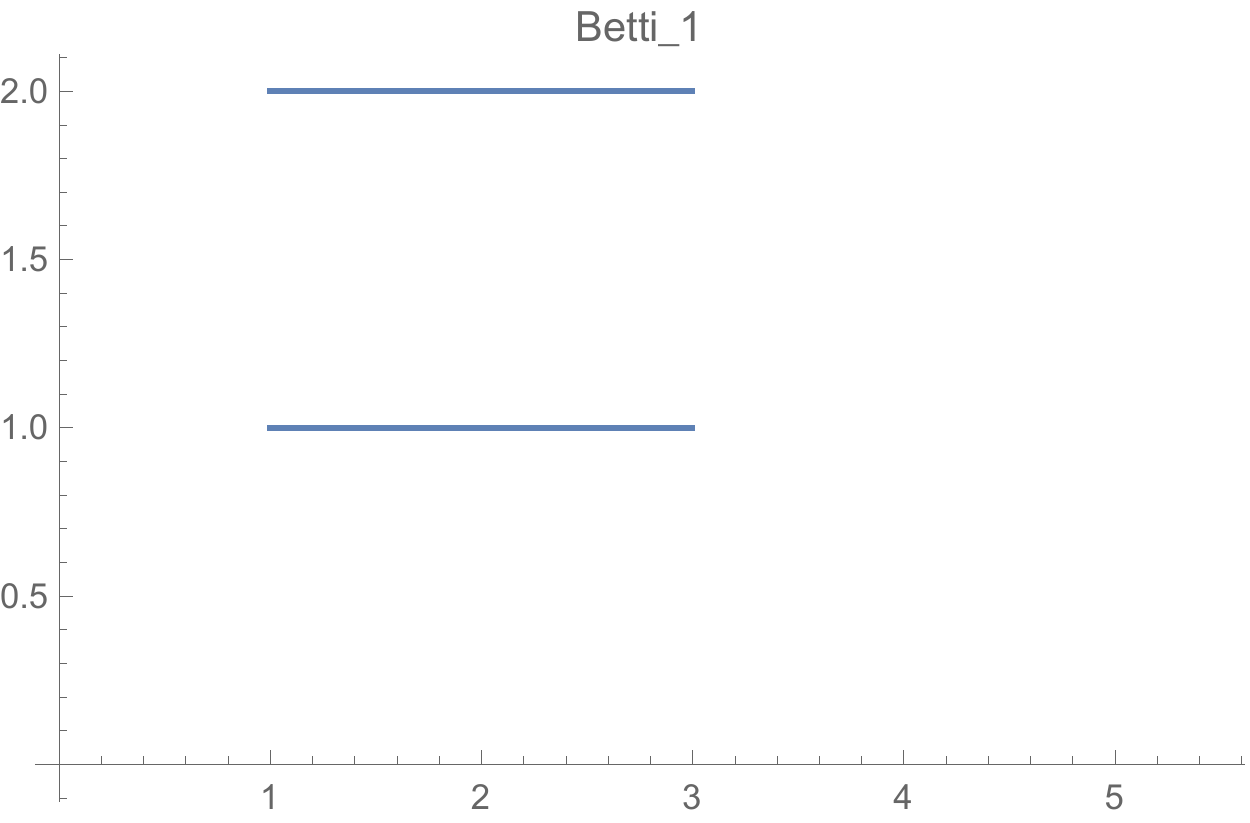}
\hfil
\includegraphics[width=.3\textwidth]{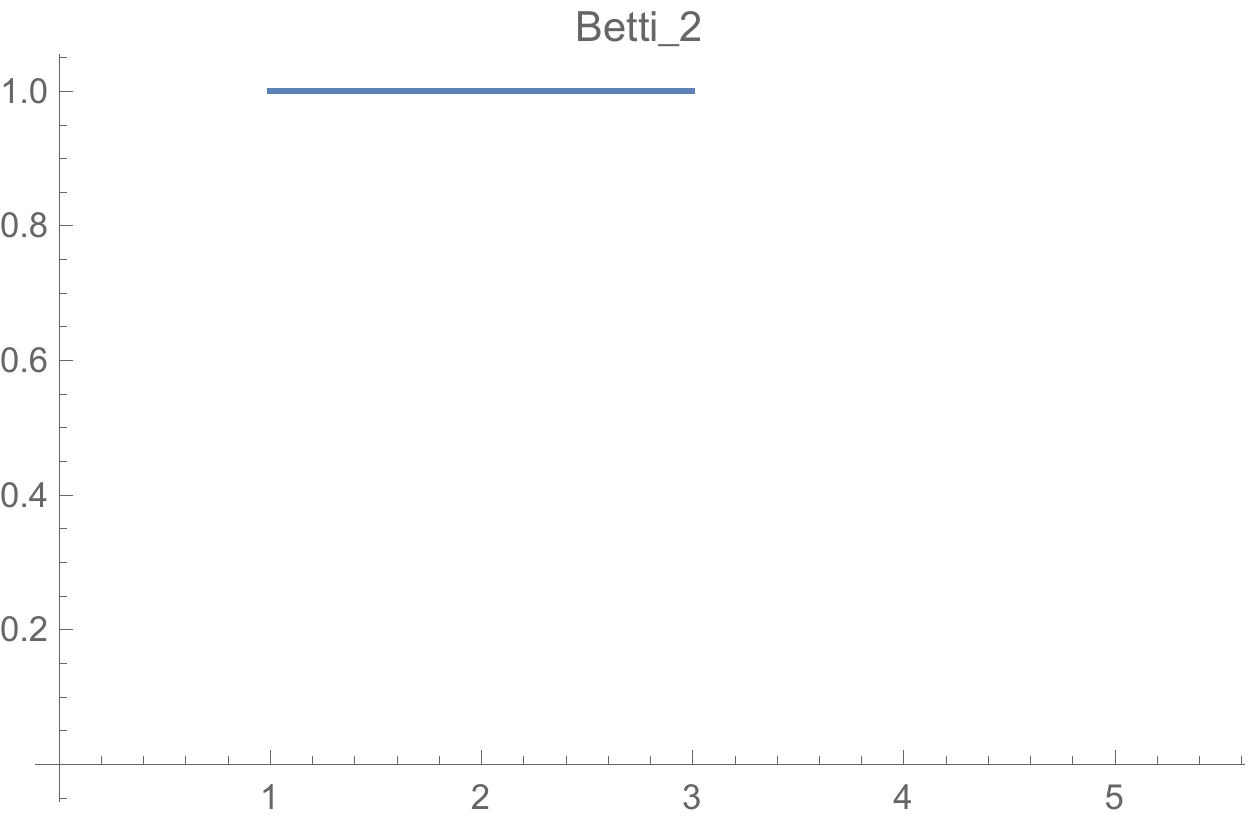}
\caption{The $\mathbb{Z}_2$-coefficient Betti intervals for the fuzzy Klein bottle.}
\label{fig:KLbettiZ2}
\end{figure}
Figure \ref{fig:KLbettiZ2} shows the $\mathbb{Z}_2$-coefficient Betti intervals for the fuzzy Klein bottle. This result shows that the homology groups of the fuzzy space are
\[
&H_0(K^2,\mathbb{Z}_2)=\mathbb{Z}_2,\\
&H_1(K^2,\mathbb{Z}_2)=\mathbb{Z}_2\oplus\mathbb{Z}_2,\\
&H_2(K^2,\mathbb{Z}_2)=\mathbb{Z}_2,
\]
and we have also checked $H_n(K^2,\mathbb{Z}_2)=0$ at least for $n=3,4,5$.
\begin{figure}
\centering
\includegraphics[width=.3\textwidth]{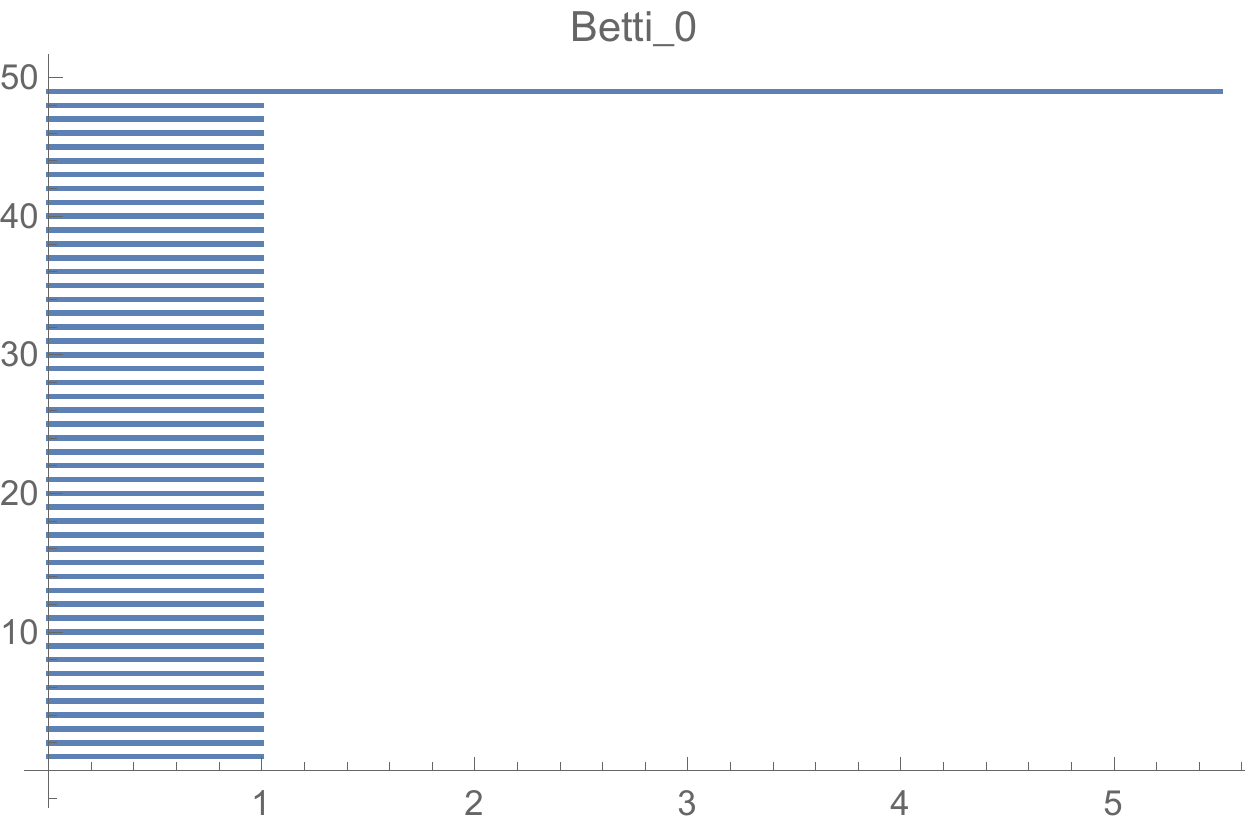}
\hfil
\includegraphics[width=.3\textwidth]{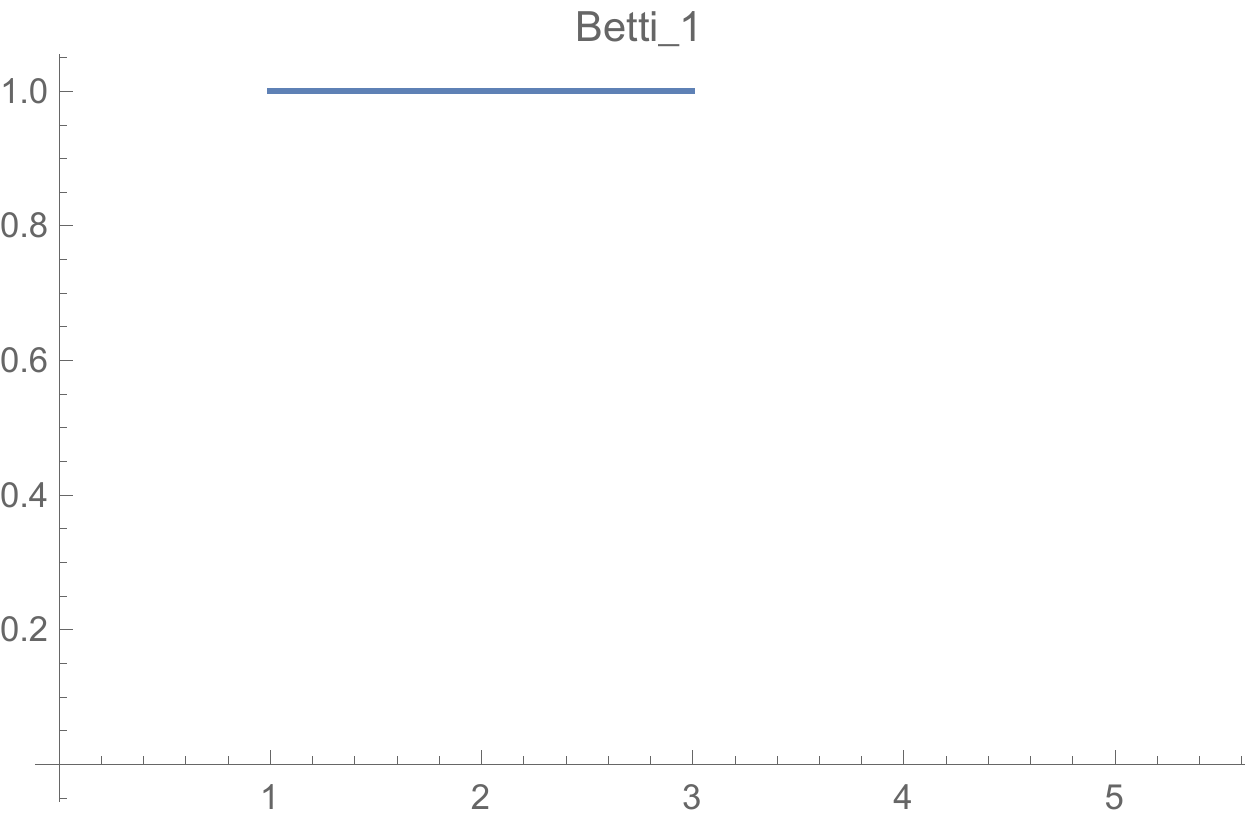}
\caption{The $\mathbb{Z}_3$-coefficient Betti intervals for the fuzzy Klein bottle. 
There are no intervals for $H_2$.}
\label{fig:KLbettiZ3}
\end{figure}
Figure \ref{fig:KLbettiZ3} shows the $\mathbb{Z}_3$-coefficient Betti intervals. 
This result also shows that the homology groups of the fuzzy space are
\[
&H_0(K^2,\mathbb{Z}_3)=\mathbb{Z}_3,\\
&H_1(K^2,\mathbb{Z}_3)=\mathbb{Z}_3,\\
&H_2(K^2,\mathbb{Z}_3)=0,
\]
and we have also checked  $H_n(K^2,\mathbb{Z}_2)=0$ at least for $n=3,4,5$. 
These homology groups agree with those of the ordinary continuous Klein bottle,
supporting the validity of our construction of the fuzzy Klein bottle.

The contents of this section are limited to homogeneous fuzzy spaces, and we are successful at least in these cases.
However, the construction procedure explained at the beginning of this section (also in Section~\ref{sec:fuzzyspace}) 
is not limited to homogeneous spaces, 
and it should be straightforward to construct inhomogeneous fuzzy spaces in a similar manner.

\section{Summary and future prospects}
The canonical tensor model (CTM) is a discrete model of gravity, which has a canonical conjugate pair of 
real symmetric three-way tensors as its dynamical variables. 
A question about the model was how to interpret the tensors as spacetimes.
We have solved this question by using two well-known techniques in data analysis, 
namely the tensor-rank decomposition and persistent homology, and
have formulated a mathematical procedure to extract topological and geometric properties 
from the real symmetric three-way tensors. 
We have also provided a systematic method to construct real symmetric three-way tensors corresponding to 
fuzzy spaces with any dimensions and topologies.
We demonstrated these techniques by considering 
the real symmetric three-way tensors corresponding to homogeneous fuzzy $S^1,\ S^2$, and $S^3$, 
solved the equations of motion of the CTM with these tensors as the initial conditions,
and interpreted the time-dependent solutions as time-evolutions of geometric spaces. 
We have found that the results coincide with the expectation from the general relativistic system derived previously 
in a formal continuum limit of the CTM \cite{Chen:2016ate}.
We have also explicitly constructed real symmetric three-way tensors for a variety of homogeneous fuzzy spaces
with various dimensions and topologies, demonstrating the generality of 
the construction and extraction procedures, and hence of the CTM.

It is now apparent that the CTM is not limited to a particular dimension: The real symmetric three-way tensors can 
represent spaces of any dimension. 
This is a strong advantage of the CTM in relation to quantum gravity, because now generic spacetimes, including their dimensions, should emerge from the dynamics in the macroscopic limit rather than input parameters. 
This is in sharp contrast with the other Euclidian types of tensor 
models \cite{Ambjorn:1990ge,Sasakura:1990fs,Godfrey:1990dt,Gurau:2009tw}, in which the numbers of 
ways (indices) of tensors are directly related to the dimensions of simplicial building blocks of spaces.

Another important implication of this paper is that the formal continuum limit of the CTM discussed 
previously in \cite{Sasakura:2015pxa,Chen:2016ate} can actually be realized in large $N$ limits. 
In these previous papers, the limit was formally put in by hand by performing an immediate replacement
of discrete values of 
indices to continuum ones. In other words, the arguments were valid after the emergence of 
continuous macroscopic spacetimes, but did not tell anything about how they emerged.
Though this paper is limited to the classical cases, 
we have explicitly shown that such limits can be realized by some large $N$ cases 
by choosing appropriate initial conditions representing fuzzy spaces 
for the classical equation of motion. 
An obvious remaining problem here is how such initial conditions and classical trajectories
are generated in the quantum framework, and we have a plausible hint for this: 
The physical wave function of the CTM has strong peaks at the tensor 
configurations invariant under Lie-groups with indefinite signatures \cite{Obster:2017dhx,Obster:2017pdq}. 
As in the constructions of homogeneous spaces, such Lie-group symmetries can play vital roles
in spacetime emergence.

Though this paper has introduced some interesting tools to interpret the dynamics of the tensors in the CTM 
as the dynamics of spacetimes,
the applications are largely immature. This paper only dealt with the dynamics of zero modes in spaces, 
but for real physical interests, one has to deal with local dynamics in
three dimensional spaces.  Though there are no theoretical difficulties, there is a technical issue: 
Such studies require much larger fuzzy spaces, but the present performance of the tensor-rank decomposition is 
too slow. We are aware of high interest in studies in this direction of computer science,
and hope that we are able to overcome this main technical difficulty in near future 
by incorporating recent developments. 
Other directly related future research directions would be trying to find and examine observables on the discrete geometry generated by the three-way tensors. Examples of such observables are already known from different discrete approaches, such as the spectral dimension~\cite{Ambjorn:2005db}, volume profile, Hausdorff dimension or the recently introduced quantum Ricci curvature~\cite{Klitgaard:2017ebu, Klitgaard:2018snm}.
It would also be interesting to see if this method can be extended to include matter fields to see the influence of matter on the dynamics of the fuzzy spaces. Furthermore, we have an interpretation for one of the tensors, but what would the geometric interpretation of the canonical conjugate be? It would also be interesting to see if the vectors from the tensor rank decomposition would have a quantum mechanical analogue.

The novel connection between gravity and data analysis shown in this paper stimulates some new kinds of, more speculative, questions. 
Can the Universe purely be described by data? How can one identify physically significant observables from random data? How do black holes appear in data? 
What is mass or energy in data?
%Are there anything like blackholes in data? Do the condition $P_{abc}v_av_bv_c>0$ set in the tensor-rank decomposition
%in Appendix~\ref{app:CP} correspond to anything in general relativity? How about the positive energy condition?
%What is mass?
Is the equation of motion of the CTM useful in data analysis? We hope mutual communications of ideas in
different fields stimulate new questions and studies to benefit them altogether.

\vspace{1cm}
\section*{Acknowledgements}
%%%%%%%%%%%%%%%%%%%%%%%%%%%%%%%%%%%%%%%%%%%%%%%
%\centerline{\bf Acknowledgements} 
The work of N.S. is supported in part by JSPS KAKENHI Grant No.15K05050. 
N.S. thanks H.~Fuji for some communications about recent developments in topological data analysis.
D.O. thanks R.~Loll for interesting discussions and advice.
 %%%%%%%%%%%%%%%%%%%%%%%%%%%%%%%%%%%%%%%%%%%%%%%%

\appendix

\section{The program for the tensor-rank decomposition}
\label{app:CP}
We made our own C++ program to obtain an approximate tensor-rank decomposition 
of a real symmetric three-way tensor as in \eq{eq:cperror} for an $R$ given as an input.
The program is roughly divided into two major parts. The former part is to set up an initial approximate 
tensor-rank decomposition of $P$, and 
the latter is to improve it as closely as possible. 

Let us begin with the former. 
Setting $P^0=P$, $v^i\ (i=1,2,\ldots,R)$ are iteratively determined by minimizing the size of 
\[
P^i_{abc}=P^{i-1}_{abc}-v^i_a v^i_b v^i_c,
\]
starting from a random value of $v^i$.
The actual minimization method is described in the end.
This iterative process sets the initial approximate tensor-rank decomposition as
\[
P_{abc}= \sum_{i=1}^R v^i_a v^i_b v^i_c+\Delta P_{abc}
\label{eq:primitive}
\] 
with an error $\Delta P_{abc}$.
In this iterative process, each $v^i$ has been optimized in the absence of the later vectors, 
$v^j\ (j=i+1,i+2,\ldots,R)$, and therefore they are not optimized as a whole. 
Further optimization is possible to reduce $\Delta P_{abc}$.

This is done in the latter part of the program. Each $v^i\ (i=1,2,\ldots, R)$ is iteratively improved with the presence of 
the other vectors by minimizing the size of 
\[
 \tilde P^{i}_{abc}-v^i_a v^i_b v^i_c,
\]
where
\[
%\tilde P^{i}_{abc}=P_{abc}-\sum_{j=1 \atop j\neq i}^R v^j_a v^j_b v^j_c,
\tilde P^{i}_{abc}=P_{abc}-\sum_{\genfrac{}{}{0pt}{3}{j=1}{j\neq i}}^R v^j_a v^j_b v^j_c,
\]
is kept fixed during the optimization of $v^i$. This iteratively goes through $i=1,2,\ldots,R$, forming one cycle.
After every cycle, it is checked whether $P_{abc} v_a^i v_b^i v_c^i>0$ is satisfied by every $i$ or not.
If not, the $v^i$ which does not satisfy the condition is discarded. 
This cycle is repeated many times until the error $\Delta P_{abc}$ cannot be reduced or becomes smaller than
a criterion.   
  
In our application, it is observed that the above condition $P_{abc} v_a^i v_b^i v_c^i>0$ tends to avoid
rough tensor-rank decompositions containing mutual cancellations among large $v^i$'s
\footnote{See for example \cite{doi:10.1002/cem.1236,hackbusch-2012} for more details about this numerically (and theoretically) 
serious problem.}, 
and gives a decomposition with $v^i$ of nearly equal sizes. This is useful in our application, 
because we are considering homogeneous fuzzy spaces, in which all the points should be more or less
uniformly weighted.

Finally let us explain the actual minimization method used as a subroutine in the above procedure.
The subroutine minimizes the size of 
\[
\tilde P_{abc}-v_a v_b v_c
\label{eq:pmvvv}
\]
for a given $\tilde P$ by optimizing $v$. By taking the square, 
$(P_{abc}-v_a v_b v_c)(P_{abc}-v_a v_b v_c)$, this is to find a minimum of 
\[
(v^2)^3 -2 \tilde P_{abc}v_a v_b v_c.
\label{eq:vsix}
\]
Though this is a well-defined problem, it is not easy to obtain a global minimum, and we 
restrict ourselves to finding a local one. This limitation is a disadvantage which cannot be underestimated in general, 
but, since this subroutine is called many times, it is also unlikely that $v$ stays in a bad local minimum 
through the whole process.  
The condition for a local minimum of \eq{eq:vsix} is given by the vanishing of its first derivative, 
\[
(v^2)^2 v_a=\tilde P_{abc}v_b v_c,
\label{eq:veqpvv}
\]
and the absence of negative eigenvalues of its second derivative matrix (Hessian) given by
\[
4 v^2 v_a v_b+(v^2)^2 \delta_{ab}-2 \tilde P_{abc} v_c.
\label{eq:twoderv}
\]

Rather than trying to directly solve the above problem which has strong non-linearity, 
let us consider a simpler form than \eq{eq:vsix}, 
\[
\frac{1}{4} (w^2)^2 -\frac{1}{3} \tilde P_{abc}w_a w_b w_c,
\label{eq:vsimple}
\]
The minimization of this is also well-defined.
The reason for considering \eq{eq:vsimple} rather than \eq{eq:vsix} becomes evident in due course.
In the same way as above, the conditions for a local minimum of \eq{eq:vsimple} are given by
\[
w^2 w_a=\tilde P_{abc}w_b w_c
\label{eq:weqpww}
\]
and the non-negativity (in the same meaning as above) of 
\[
2 w_a w_b+(w^2) \delta_{ab}-2 \tilde P_{abc} w_c.
\label{eq:twoderw}
\]
Comparing the two problems, one can see that a local minimum of the latter gives one of the former 
by doing a rescaling $v^2 v_a=w_a$.
Here it is important that the non-negativity of \eq{eq:twoderw} readily implies that of \eq{eq:twoderv}, 
because the difference $2 v^2 v_a v_b$ is non-negative.

A way to obtain a local minimum of \eq{eq:vsimple} is to apply the steepest descent method.
By taking the first derivative of \eq{eq:vsimple} and choosing a step size of $\gamma/w^2$ with $\gamma>0$, one obtains a sequence,
\[
w_a'=(1-\gamma)w_a +\gamma  \frac{P_{abc} w_b w_c}{w^2}.
\label{eq:gooditer}
\]
A convergent vector of the sequence gives a local minimum of \eq{eq:vsimple}.
 
An advantage of considering \eq{eq:gooditer} from \eq{eq:vsimple}
rather than what can be obtained from \eq{eq:vsix} is that
\eq{eq:gooditer} is more controllable. The second term of \eq{eq:gooditer} is bounded for any $w$, and one can easily prove that, 
for $0 < \gamma < 2$, the sequence does not diverge. 
Therefore, the worst behavior is a bounded non-convergent sequence, which would be changed to a convergent one 
by an appropriate choice of $\gamma$, typically by making it smaller. 
In our application, however, the simplest choice $\gamma=1$ suffices\footnote{
However, this is just an accidental fact. One can easily construct examples which have oscillatory behaviors for 
$\gamma=1$. This is given by $P_{111}=1,\ P_{122}<-0.5$ for $N=2$.}, 
where only the second term needs to be computed on the righthand side of \eq{eq:gooditer}.  
 
%\bibliographystyle{utphys}
%\bibliography{main} 

\begin{thebibliography}{10}

\bibitem{GOROFF1986709}
M.~H. Goroff and A.~Sagnotti, ``The ultraviolet behavior of einstein gravity,''
  \href{http://dx.doi.org/http://dx.doi.org/10.1016/0550-3213(86)90193-8}{{\em
  Nuclear Physics B} {\bfseries 266} no.~3, (1986) 709 -- 736}.
  \url{http://www.sciencedirect.com/science/article/pii/0550321386901938}.

\bibitem{Eichhorn:2017egq}
A.~Eichhorn, ``{Status of the asymptotic safety paradigm for quantum gravity
  and matter},'' in {\em {Black Holes, Gravitational Waves and Spacetime
  Singularities Rome, Italy, May 9-12, 2017}}.
\newblock 2017.
\newblock \href{http://arxiv.org/abs/1709.03696}{{\ttfamily arXiv:1709.03696
  [gr-qc]}}.
\newblock
\url{https://inspirehep.net/record/1623009/files/arXiv:1709.03696.pdf}.
\newblock
%%CITATION = ARXIV:1709.03696;%%.

\bibitem{Reuter:2012id}
M.~Reuter and F.~Saueressig, ``{Quantum Einstein Gravity},''
  \href{http://dx.doi.org/10.1088/1367-2630/14/5/055022}{{\em New J. Phys.}
  {\bfseries 14} (2012) 055022},
\href{http://arxiv.org/abs/1202.2274}{{\ttfamily arXiv:1202.2274 [hep-th]}}.
%%CITATION = ARXIV:1202.2274;%%.

\bibitem{Ambjorn:1990ge}
J.~Ambjorn, B.~Durhuus, and T.~Jonsson, ``{Three-dimensional simplicial quantum
  gravity and generalized matrix models},''
\href{http://dx.doi.org/10.1142/S0217732391001184}{{\em Mod. Phys. Lett.}
  {\bfseries A06} (1991) 1133--1146}.
%%CITATION = MPLAE,A6,1133;%%.

\bibitem{Sasakura:1990fs}
N.~Sasakura, ``{Tensor model for gravity and orientability of manifold},''
\href{http://dx.doi.org/10.1142/S0217732391003055}{{\em Mod. Phys. Lett.}
  {\bfseries A06} (1991) 2613--2624}.
%%CITATION = MPLAE,A6,2613;%%.

\bibitem{Godfrey:1990dt}
N.~Godfrey and M.~Gross, ``{Simplicial quantum gravity in more than
  two-dimensions},''
\href{http://dx.doi.org/10.1103/PhysRevD.43.1749}{{\em Phys. Rev.} {\bfseries
  D43} (1991) R1749--1753}.
%%CITATION = PHRVA,D43,1749;%%.

\bibitem{Klebanov:2017nlk}
I.~R. Klebanov and G.~Tarnopolsky, ``{On Large $N$ Limit of Symmetric Traceless
  Tensor Models},'' \href{http://dx.doi.org/10.1007/JHEP10(2017)037}{{\em JHEP}
  {\bfseries 10} (2017) 037},
\href{http://arxiv.org/abs/1706.00839}{{\ttfamily arXiv:1706.00839 [hep-th]}}.
%%CITATION = ARXIV:1706.00839;%%.

\bibitem{Gurau:2017qya}
R.~Gurau, ``{The $1/N$ expansion of tensor models with two symmetric
  tensors},'' \href{http://dx.doi.org/10.1007/s00220-017-3055-y}{{\em Commun.
  Math. Phys.} {\bfseries 360} no.~3, (2018) 985--1007},
\href{http://arxiv.org/abs/1706.05328}{{\ttfamily arXiv:1706.05328 [hep-th]}}.
%%CITATION = ARXIV:1706.05328;%%.

\bibitem{Gurau:2009tw}
R.~Gurau, ``{Colored Group Field Theory},''
  \href{http://dx.doi.org/10.1007/s00220-011-1226-9}{{\em Commun. Math. Phys.}
  {\bfseries 304} (2011) 69--93},
\href{http://arxiv.org/abs/0907.2582}{{\ttfamily arXiv:0907.2582 [hep-th]}}.
%%CITATION = ARXIV:0907.2582;%%.

\bibitem{Gurau:2013cbh}
R.~Gurau and J.~P. Ryan, ``{Melons are branched polymers},''
  \href{http://dx.doi.org/10.1007/s00023-013-0291-3}{{\em Annales Henri
  Poincare} {\bfseries 15} no.~11, (2014) 2085--2131},
\href{http://arxiv.org/abs/1302.4386}{{\ttfamily arXiv:1302.4386 [math-ph]}}.
%%CITATION = ARXIV:1302.4386;%%.

\bibitem{Bonzom:2011zz}
V.~Bonzom, R.~Gurau, A.~Riello, and V.~Rivasseau, ``{Critical behavior of
  colored tensor models in the large N limit},''
  \href{http://dx.doi.org/10.1016/j.nuclphysb.2011.07.022}{{\em Nucl. Phys.}
  {\bfseries B853} (2011) 174--195},
\href{http://arxiv.org/abs/1105.3122}{{\ttfamily arXiv:1105.3122 [hep-th]}}.
%%CITATION = ARXIV:1105.3122;%%.

\bibitem{Bonzom:2015axa}
V.~Bonzom, T.~Delepouve, and V.~Rivasseau, ``{Enhancing non-melonic
  triangulations: A tensor model mixing melonic and planar maps},''
  \href{http://dx.doi.org/10.1016/j.nuclphysb.2015.04.004}{{\em Nucl. Phys.}
  {\bfseries B895} (2015) 161--191},
\href{http://arxiv.org/abs/1502.01365}{{\ttfamily arXiv:1502.01365 [math-ph]}}.
%%CITATION = ARXIV:1502.01365;%%.

\bibitem{Lionni:2017xvn}
L.~Lionni and J.~Thurigen, ``{Multi-critical behaviour of 4-dimensional tensor
  models up to order 6},''
\href{http://arxiv.org/abs/1707.08931}{{\ttfamily arXiv:1707.08931 [hep-th]}}.
%%CITATION = ARXIV:1707.08931;%%.

\bibitem{Ambjorn:2004qm}
J.~Ambjorn, J.~Jurkiewicz, and R.~Loll, ``{Emergence of a 4-D world from causal
  quantum gravity},''
  \href{http://dx.doi.org/10.1103/PhysRevLett.93.131301}{{\em Phys. Rev. Lett.}
  {\bfseries 93} (2004) 131301},
\href{http://arxiv.org/abs/hep-th/0404156}{{\ttfamily arXiv:hep-th/0404156
  [hep-th]}}.
%%CITATION = HEP-TH/0404156;%%.

\bibitem{Ambjorn:2012jv}
J.~Ambjorn, A.~Goerlich, J.~Jurkiewicz, and R.~Loll, ``{Nonperturbative Quantum
  Gravity},'' \href{http://dx.doi.org/10.1016/j.physrep.2012.03.007}{{\em Phys.
  Rept.} {\bfseries 519} (2012) 127--210},
\href{http://arxiv.org/abs/1203.3591}{{\ttfamily arXiv:1203.3591 [hep-th]}}.
%%CITATION = ARXIV:1203.3591;%%.

\bibitem{Ambjorn:2013eha}
J.~Ambjorn, L.~Glaser, A.~Goerlich, and J.~Jurkiewicz, ``{Euclidian 4d quantum
  gravity with a non-trivial measure term},''
  \href{http://dx.doi.org/10.1007/JHEP10(2013)100}{{\em JHEP} {\bfseries 10}
  (2013) 100},
\href{http://arxiv.org/abs/1307.2270}{{\ttfamily arXiv:1307.2270 [hep-lat]}}.
%%CITATION = ARXIV:1307.2270;%%.

\bibitem{Coumbe:2014nea}
D.~Coumbe and J.~Laiho, ``{Exploring {Euclidean} Dynamical Triangulations with
  a Non-trivial Measure Term},''
  \href{http://dx.doi.org/10.1007/JHEP04(2015)028}{{\em JHEP} {\bfseries 04}
  (2015) 028},
\href{http://arxiv.org/abs/1401.3299}{{\ttfamily arXiv:1401.3299 [hep-th]}}.
%%CITATION = ARXIV:1401.3299;%%.

\bibitem{Sasakura:2011sq}
N.~Sasakura, ``{Canonical tensor models with local time},''
  \href{http://dx.doi.org/10.1142/S0217751X12500200}{{\em Int. J. Mod. Phys.}
  {\bfseries A27} (2012) 1250020},
\href{http://arxiv.org/abs/1111.2790}{{\ttfamily arXiv:1111.2790 [hep-th]}}.
%%CITATION = ARXIV:1111.2790;%%.

\bibitem{Sasakura:2012fb}
N.~Sasakura, ``{Uniqueness of canonical tensor model with local time},''
  \href{http://dx.doi.org/10.1142/S0217751X12500960}{{\em Int. J. Mod. Phys.}
  {\bfseries A27} (2012) 1250096},
\href{http://arxiv.org/abs/1203.0421}{{\ttfamily arXiv:1203.0421 [hep-th]}}.
%%CITATION = ARXIV:1203.0421;%%.

\bibitem{Obster:2017dhx}
D.~Obster and N.~Sasakura, ``{Emergent symmetries in the canonical tensor
  model},'' \href{http://dx.doi.org/10.1093/ptep/pty038}{{\em PTEP} {\bfseries
  2018} no.~4, (2018) 043A01},
\href{http://arxiv.org/abs/1710.07449}{{\ttfamily arXiv:1710.07449 [hep-th]}}.
%%CITATION = ARXIV:1710.07449;%%.

\bibitem{Sasakura:2015pxa}
N.~Sasakura and Y.~Sato, ``{Constraint algebra of general relativity from a
  formal continuum limit of canonical tensor model},''
  \href{http://dx.doi.org/10.1007/JHEP10(2015)109}{{\em JHEP} {\bfseries 10}
  (2015) 109},
\href{http://arxiv.org/abs/1506.04872}{{\ttfamily arXiv:1506.04872 [hep-th]}}.
%%CITATION = ARXIV:1506.04872;%%.

\bibitem{Sasakura:2014gia}
N.~Sasakura and Y.~Sato, ``{Interpreting canonical tensor model in
  minisuperspace},''
  \href{http://dx.doi.org/10.1016/j.physletb.2014.03.006}{{\em Phys. Lett.}
  {\bfseries B732} (2014) 32--35},
\href{http://arxiv.org/abs/1401.2062}{{\ttfamily arXiv:1401.2062 [hep-th]}}.
%%CITATION = ARXIV:1401.2062;%%.

\bibitem{Chen:2016ate}
H.~Chen, N.~Sasakura, and Y.~Sato, ``{Equation of motion of canonical tensor
  model and Hamilton-Jacobi equation of general relativity},''
  \href{http://dx.doi.org/10.1103/PhysRevD.95.066008}{{\em Phys. Rev.}
  {\bfseries D95} no.~6, (2017) 066008},
\href{http://arxiv.org/abs/1609.01946}{{\ttfamily arXiv:1609.01946 [hep-th]}}.
%%CITATION = ARXIV:1609.01946;%%.

\bibitem{SAPM:SAPM192761164}
F.~L. Hitchcock, ``The expression of a tensor or a polyadic as a sum of
  products,'' \href{http://dx.doi.org/10.1002/sapm192761164}{{\em Journal of
  Mathematics and Physics} {\bfseries 6} no.~1-4, (1927) 164--189}.
  \url{http://dx.doi.org/10.1002/sapm192761164}.

\bibitem{Carroll1970}
J.~D. Carroll and J.-J. Chang, ``Analysis of individual differences in
  multidimensional scaling via an n-way generalization of ``eckart-young''
  decomposition,'' \href{http://dx.doi.org/10.1007/BF02310791}{{\em
  Psychometrika} {\bfseries 35} no.~3, (Sep, 1970) 283--319}.
  \url{https://doi.org/10.1007/BF02310791}.

\bibitem{harshman70}
R.~Harshman, ``Foundations of the parafac procedure: models and conditions for
  an 'exploratory' multimodal factor analysis,'' in {\em UCLA Working Papers in
  Phonetics}, no.~16, pp.~1--84.
\newblock University Microfilms, Ann Arbor, Michigan, No. 10,085, 1970.
\newblock \url{http://psychology.uwo.ca/faculty/harshman/wpppfac0.pdf}.

\bibitem{comon:hal-00923279}
P.~Comon, ``{Tensors: a Brief Introduction},''
  \href{http://dx.doi.org/10.1109/MSP.2014.2298533}{{\em {IEEE Signal
  Processing Magazine}} {\bfseries 31} no.~3, (May, 2014) 44--53}.
  \url{https://hal.archives-ouvertes.fr/hal-00923279}.

\bibitem{Carlsson09}
G.~Carlsson, ``Topology and data,''
  \href{http://dx.doi.org/10.1090/S0273-0979-09-01249-X}{{\em Bull. Amer. Math.
  Soc.} {\bfseries 46} (2009) 225--308}.

\bibitem{Cirafici:2015pky}
M.~Cirafici, ``{Persistent Homology and String Vacua},''
  \href{http://dx.doi.org/10.1007/JHEP03(2016)045}{{\em JHEP} {\bfseries 03}
  (2016) 045},
\href{http://arxiv.org/abs/1512.01170}{{\ttfamily arXiv:1512.01170 [hep-th]}}.
%%CITATION = ARXIV:1512.01170;%%.

\bibitem{Cirafici:2015sdg}
M.~Cirafici, ``{BPS Spectra, Barcodes and Walls},''
\href{http://arxiv.org/abs/1511.01421}{{\ttfamily arXiv:1511.01421 [hep-th]}}.
%%CITATION = ARXIV:1511.01421;%%.

\bibitem{Cole:2017kve}
A.~Cole and G.~Shiu, ``{Persistent Homology and Non-Gaussianity},''
  \href{http://dx.doi.org/10.1088/1475-7516/2018/03/025}{{\em JCAP} {\bfseries
  1803} no.~03, (2018) 025},
\href{http://arxiv.org/abs/1712.08159}{{\ttfamily arXiv:1712.08159
  [astro-ph.CO]}}.
%%CITATION = ARXIV:1712.08159;%%.

\bibitem{Spreemann:2018}
G.~Spreemann, B.~Dunn, M.~B. Botnan, and N.~A. Baas, ``Using persistent
  homology to reveal hidden covariates in systems governed by the kinetic ising
  model,'' \href{http://dx.doi.org/10.1103/PhysRevE.97.032313}{{\em Phys. Rev.
  E} {\bfseries 97} (Mar, 2018) 032313}.
  \url{https://link.aps.org/doi/10.1103/PhysRevE.97.032313}.

\bibitem{Nadler:2005:DMS:2976248.2976368}
B.~Nadler, S.~Lafon, R.~R. Coifman, and I.~G. Kevrekidis, ``Diffusion maps,
  spectral clustering and eigenfunctions of fokker-planck operators,'' in {\em
  Proceedings of the 18th International Conference on Neural Information
  Processing Systems}, NIPS'05, pp.~955--962.
\newblock MIT Press, Cambridge, MA, USA, 2005.
\newblock
  \url{https://papers.nips.cc/paper/2942-diffusion-maps-spectral-clustering-and-eigenfunctions-of-fokker-planck-operators}.

\bibitem{Coifman7426}
R.~R. Coifman, S.~Lafon, A.~B. Lee, M.~Maggioni, B.~Nadler, F.~Warner, and
  S.~W. Zucker, ``Geometric diffusions as a tool for harmonic analysis and
  structure definition of data: Diffusion maps,''
  \href{http://dx.doi.org/10.1073/pnas.0500334102}{{\em Proceedings of the
  National Academy of Sciences} {\bfseries 102} no.~21, (2005) 7426--7431},
  \href{http://arxiv.org/abs/http://www.pnas.org/content/102/21/7426.full.pdf}{{\ttfamily
  http://www.pnas.org/content/102/21/7426.full.pdf}}.
  \url{http://www.pnas.org/content/102/21/7426}.

\bibitem{Landsberg2012}
{Landsberg, J. M.}, {\em {Tensors: Geometry and Applications}}.
\newblock {American Mathematical Society, Providence}, {2012}.

\bibitem{Bocci2014}
C.~Bocci, L.~Chiantini, and G.~Ottaviani, ``Refined methods for the
  identifiability of tensors,''
  \href{http://dx.doi.org/10.1007/s10231-013-0352-8}{{\em Annali di Matematica
  Pura ed Applicata (1923 -)} {\bfseries 193} no.~6, (Dec, 2014) 1691--1702}.
  \url{https://doi.org/10.1007/s10231-013-0352-8}.

\bibitem{doi:10.1002/cem.1236}
P.~Comon, X.~Luciani, and A.~L.~F. de~Almeida, ``Tensor decompositions,
  alternating least squares and other tales,''
  \href{http://dx.doi.org/10.1002/cem.1236}{{\em Journal of Chemometrics}
  {\bfseries 23} no.~7-8, 393--405}.
  \url{https://onlinelibrary.wiley.com/doi/pdf/10.1002/cem.1236}.

\bibitem{zbMATH00773851}
J.~{Alexander} and A.~{Hirschowitz}, ``{Polynomial interpolation in several
  variables.},'' {\em {J. Algebr. Geom.}} {\bfseries 4} no.~2, (1995) 201--222.

\bibitem{doi:10.1137/060661569}
P.~Comon, G.~Golub, L.-H. Lim, and B.~Mourrain, ``Symmetric tensors and
  symmetric tensor rank,'' \href{http://dx.doi.org/10.1137/060661569}{{\em SIAM
  Journal on Matrix Analysis and Applications} {\bfseries 30} no.~3, (2008)
  1254--1279},
  \href{http://arxiv.org/abs/https://doi.org/10.1137/060661569}{{\ttfamily
  https://doi.org/10.1137/060661569}}.

\bibitem{Sakata:2016:ACA:3027777}
T.~Sakata, T.~Sumi, and M.~Miyazaki, {\em Algebraic and Computational Aspects
  of Real Tensor Ranks}.
\newblock Springer Publishing Company, Incorporated, 1st~ed., 2016.

\bibitem{TenBerge2011}
J.~{Ten Berge}, ``Simplicity and typical rank results for three-way arrays,''
  \href{http://dx.doi.org/10.1007/s11336-010-9193-1}{{\em Psychometrika}
  {\bfseries 76} no.~1, (1, 2011) 3--12}.

\bibitem{hackbusch-2012}
W.~Hackbusch, {\em Tensor Spaces And Numerical Tensor Calculus}.
\newblock Springer--Verlag, Berlin, 2012.

\bibitem{Sasakura:2005js}
N.~Sasakura, ``{An Invariant approach to dynamical fuzzy spaces with a
  three-index variable},''
  \href{http://dx.doi.org/10.1142/S0217732306020329}{{\em Mod. Phys. Lett.}
  {\bfseries A21} (2006) 1017--1028},
\href{http://arxiv.org/abs/hep-th/0506192}{{\ttfamily arXiv:hep-th/0506192
  [hep-th]}}.
%%CITATION = HEP-TH/0506192;%%.

\bibitem{Sasakura:2006pq}
N.~Sasakura, ``{Tensor model and dynamical generation of commutative
  nonassociative fuzzy spaces},''
  \href{http://dx.doi.org/10.1088/0264-9381/23/17/017}{{\em Class. Quant.
  Grav.} {\bfseries 23} (2006) 5397--5416},
\href{http://arxiv.org/abs/hep-th/0606066}{{\ttfamily arXiv:hep-th/0606066
  [hep-th]}}.
%%CITATION = HEP-TH/0606066;%%.

\bibitem{Sasakura:2011ma}
N.~Sasakura, ``{Tensor models and 3-ary algebras},''
  \href{http://dx.doi.org/10.1063/1.3654028}{{\em J. Math. Phys.} {\bfseries
  52} (2011) 103510},
\href{http://arxiv.org/abs/1104.1463}{{\ttfamily arXiv:1104.1463 [hep-th]}}.
%%CITATION = ARXIV:1104.1463;%%.

\bibitem{Sasakura:2011nj}
N.~Sasakura, ``{Tensor models and hierarchy of n-ary algebras},''
  \href{http://dx.doi.org/10.1142/S0217751X1105381X}{{\em Int. J. Mod. Phys.}
  {\bfseries A26} (2011) 3249--3258},
\href{http://arxiv.org/abs/1104.5312}{{\ttfamily arXiv:1104.5312 [hep-th]}}.
%%CITATION = ARXIV:1104.5312;%%.

\bibitem{Sasakura:2011qg}
N.~Sasakura, ``{Super tensor models, super fuzzy spaces and super n-ary
  transformations},'' \href{http://dx.doi.org/10.1142/S0217751X11054449}{{\em
  Int. J. Mod. Phys.} {\bfseries A26} (2011) 4203--4216},
\href{http://arxiv.org/abs/1106.0379}{{\ttfamily arXiv:1106.0379 [hep-th]}}.
%%CITATION = ARXIV:1106.0379;%%.

\bibitem{Madore:2000aq}
J.~Madore, ``{An introduction to noncommutative differential geometry and its
  physical applications},''
{\em Lond. Math. Soc. Lect. Note Ser.} {\bfseries 257} (2000) 1--371.
%%CITATION = LMSSD,257,1;%%.

\bibitem{matousek2002lectures}
J.~Matou{\v{s}}ek, {\em Lectures on Discrete Geometry}.
\newblock Graduate Texts in Mathematics. Springer, 2002.

\bibitem{Calcagni:2013vsa}
G.~Calcagni, A.~Eichhorn, and F.~Saueressig, ``{Probing the quantum nature of
  spacetime by diffusion},''
  \href{http://dx.doi.org/10.1103/PhysRevD.87.124028}{{\em Phys. Rev.}
  {\bfseries D87} no.~12, (2013) 124028},
\href{http://arxiv.org/abs/1304.7247}{{\ttfamily arXiv:1304.7247 [hep-th]}}.
%%CITATION = ARXIV:1304.7247;%%.

\bibitem{Ambjorn:2005db}
J.~Ambjorn, J.~Jurkiewicz, and R.~Loll, ``{Spectral dimension of the
  universe},'' \href{http://dx.doi.org/10.1103/PhysRevLett.95.171301}{{\em
  Phys. Rev. Lett.} {\bfseries 95} (2005) 171301},
\href{http://arxiv.org/abs/hep-th/0505113}{{\ttfamily arXiv:hep-th/0505113
  [hep-th]}}.
%%CITATION = HEP-TH/0505113;%%.

\bibitem{polchinski_1998}
J.~Polchinski, \href{http://dx.doi.org/10.1017/CBO9780511816079}{{\em String
  Theory}}, vol.~1 of {\em Cambridge Monographs on Mathematical Physics}.
\newblock Cambridge University Press, 1998.

\bibitem{doi:10.1063/1.527513}
A.~Higuchi, ``Symmetric tensor spherical harmonics on the n‐sphere and their
  application to the de sitter group so(n,1),''
  \href{http://dx.doi.org/10.1063/1.527513}{{\em Journal of Mathematical
  Physics} {\bfseries 28} no.~7, (1987) 1553--1566},
  \href{http://arxiv.org/abs/https://doi.org/10.1063/1.527513}{{\ttfamily
  https://doi.org/10.1063/1.527513}}.

\bibitem{packharmonic}
S.~Axler, {\em {Harmonic Function Theory and Mathematica}}.
\newblock http://www.axler.net.

\bibitem{Obster:2017pdq}
D.~Obster and N.~Sasakura, ``{Symmetric configurations highlighted by
  collective quantum coherence},''
  \href{http://dx.doi.org/10.1140/epjc/s10052-017-5355-y}{{\em Eur. Phys. J.}
  {\bfseries C77} no.~11, (2017) 783},
\href{http://arxiv.org/abs/1704.02113}{{\ttfamily arXiv:1704.02113 [hep-th]}}.
%%CITATION = ARXIV:1704.02113;%%.

\bibitem{Klitgaard:2017ebu}
N.~Klitgaard and R.~Loll, ``{Introducing Quantum Ricci Curvature},''
  \href{http://dx.doi.org/10.1103/PhysRevD.97.046008}{{\em Phys. Rev.}
  {\bfseries D97} no.~4, (2018) 046008},
\href{http://arxiv.org/abs/1712.08847}{{\ttfamily arXiv:1712.08847 [hep-th]}}.
%%CITATION = ARXIV:1712.08847;%%.

\bibitem{Klitgaard:2018snm}
N.~Klitgaard and R.~Loll, ``{Quantizing quantum Ricci curvature},''
\href{http://arxiv.org/abs/1802.10524}{{\ttfamily arXiv:1802.10524 [hep-th]}}.
%%CITATION = ARXIV:1802.10524;%%.

\end{thebibliography}
 
\providecommand{\href}[2]{#2}\begingroup\raggedright\endgroup 
 
\end{document}